\documentclass[referee, sn-basic]{sn-jnl}
\usepackage[sort]{cite}
\usepackage{url,colortbl}
\usepackage{xcolor}
\usepackage{bm,amsmath,amssymb}
\usepackage[]{graphicx}
\usepackage[normalem]{ulem}
\usepackage{bbding}
\usepackage{enumitem}
\usepackage{setspace}
\usepackage{tabularx}
\usepackage{tabularray}
\jyear{2025}%

\begin{document}

\author*[1]{\fnm{Rebecca M.} \sur{Crossley}}\email{crossley@maths.ox.ac.uk}

\author[1]{\fnm{Philip K.} \sur{Maini}}\email{maini@maths.ox.ac.uk}

\author[1]{\fnm{Ruth E.} \sur{Baker}}\email{baker@maths.ox.ac.uk}

\affil*[1]{\orgdiv{Mathematical Institute}, \orgname{University of Oxford}, \orgaddress{\street{Woodstock Road}, \city{Oxford}, \postcode{OX2 6GG}, \country{United Kingdom}}}
\title[Phenotypic heterogeneity in cell migration]{{{Modelling the impact of phenotypic heterogeneity on cell migration: a continuum framework derived from individual-based principles}}}

\abstract{
{{Collective}} cell migration plays a crucial role in numerous biological processes, including tumour growth, wound healing, and the immune response. 
{{Often, the migrating population consists of cells with various different phenotypes.}}
This study {{derives}} a general mathematical framework for modelling cell migration in the local environment, 
{{which is coarse-grained}} from an underlying individual-based model that captures the dynamics of cell migration that are influenced by {{the phenotype of the cell}}, such as random movement, proliferation, phenotypic transitions, and interactions with the local environment. 
{{The resulting, flexible, and general model provides}} a continuum, macroscopic description of cell invasion, {{which represents}} the phenotyp{{e of}} the cell as a continuous variable {{and is much more amenable to simulation and analysis than its individual-based counterpart when considering a large number of phenotypes}}. 
We {{showcase the utility of {{the}} generalised framework in}} three biological scenarios: range expansion; cell invasion into the extracellular matrix; and T cell exhaustion. 
The results highlight how phenotypic structuring impacts the spatial and temporal dynamics of cell populations, demonstrating that different environmental pressures and phenotypic transition mechanisms significantly influence migration patterns{{, a phenomenon that would be computationally very expensive to explore using an individual-based model alone}}. 
This framework provides a versatile and robust tool for understanding the role of phenotypic heterogeneity in collective cell migration, with potential applications in optimising therapeutic strategies for diseases involving cell migration.
}


\maketitle

\section{Introduction}\label{sec:intro}
Mathematical models are essential tools for helping us to understand the key features and mechanisms underpinning biological processes, such as collective cell migration. 
Various modelling techniques exist to analyse cell behaviour during migration, ranging from microscopic, individual cell-based models to {{macroscopic-level}}, continuum-based models.

{{Stochastic}} individual-based models track the dynamics of single cells, describing migration through rules that dictate cell interactions with one another and their environment \citep{anderson2007single, cornell2019unified, van2015simulating, wang2015simulating, west2023agent}.
Deterministic continuum models, on the other hand, usually focus on the collective migration of cells into {{external tissue, stroma, or the local environment,}} {{containing, for example,}} chemo-attractants, adhesive substances or other cell populations that impact cell migration \citep{merino2022unravelling, trepat2012cell}.
They utilise a variety of different mathematical approaches, such as partial differential equations (PDEs),  that describe the evolution of cell densities and are amenable to both computational and analytical exploration.
While some PDE models are adaptations of classical invasion models from other contexts, many are derived from first principles as the deterministic, continuum limit of stochastic, discrete models, such as individual-based models. 
This process of formally deriving the deterministic model ensures that the continuum equation provides a mean-field representation of the underlying dynamics of the individual cells and their environment{{, which is valid in specific parameter regimes}} \citep{macfarlane2022individual}.

Mathematical models for cell invasion often assume that the cell population is phenotypically homogeneous, meaning every cell behaves identically in terms of division, movement and interactions with the local environment. 
However, such homogeneity is rarely present in biological systems. 
During collective cell migration, it is common to observe different cell types working together to facilitate invasion. 
The observable differences in physical or biochemical characteristics present within most cell populations are known as phenotypic heterogeneity. 
Over recent years, {{the role of}} phenotypic heterogeneity in cell populations has garnered significant attention.
Models have been developed to account for distinctly different cell behaviours, using {{a}} discrete {{number of}} phenotypes \citep{carrillo2024spatial, chauviere2010model, crossley2024phenotypic, FALCO2024109209, stepien2018traveling}, as well as for a {{continuous}} spectrum of phenotypes, whereby population members exhibit a variety of behaviours to different degrees \citep{bouin2012invasion, macfarlane2022impact}. 
Variability in cell phenotypes can be incorporated into mathematical models of cell dynamics {{involving differential equations}} by introducing a variable to describe the phenotypic state of the cells. 

{{When there is a discrete set of known cell phenotypes with distinct behaviors, it may be most appropriate to model the phenotypic state as a discrete variable, typically using integer values. 
However, when there are numerous (or potentially infinite) phenotypes with incremental differences or smoothly transitioning behaviours between them, then a continuously structured phenotype model might be more pertinent.
In this case, t}}he resulting evolution equation for the cell population density often takes the form of a non-local reaction-diffusion equation \citep{arnold2012existence, berestycki2015existence, lorenzi2022trade}, or a non-local advection-reaction-diffusion equation \citep{celora2021phenotypic, celora2023spatio, lorenzi2022invasion}.
Understanding the role of different cell phenotypes during collective migration can guide experimental design, enhance understanding of cell behaviors, and aid the development of treatments for diseases where collective cell migration is crucial, such as during wound healing.
{{
However, a critical question arises when these continuum models are constructed phenomenologically, without derivation from an underlying individual-based model. 
In such cases, we may not fully understand the biological significance of the terms within the model, especially in realising their connection to the behaviours of the individual cells and their interactions with the local environment, leaving the meaning of various terms in the model ambiguous. 
{{Moreover, we may unknowingly be making intrinsic assumptions about cell behaviour that could be invalid.}}
{{This}} research, therefore, focuses on constructing a general continuum model that is explicitly derived from the individual-based behaviours of the cells and their interactions with surrounding environmental features. 
This approach ensures that each term in the resulting continuum model has a well-defined interpretation in relation to the underlying cell dynamics, providing clarity and a deeper understanding of the biological processes being modelled.
{We are not, however, concerned with a detailed quantitative comparison between individual- and population-level models as this is already well explored in the literature \citep{ardavseva2020comparative, byrne2009individual, lorenzi2022trade, lorenzi2020discrete, macfarlane2022individual, murray2009discrete, murray2011comparing, schaller2006continuum}.}
}}

Therefore, in this article, we present a methodology for deriving a continuously structured PDE model for general cell migration into the local environment that is robustly {{derived from an}} underlying individual-based {{model that takes into account the individual}} interactions between the cells and their local environment. 
Using this {{approach}}, we demonstrate the model's applicability to various biological scenarios, {{highlighting the flexibility of this general framework and the consistent, coherent connections between the microscale behaviours {{captured}} in the individual-based model and the macroscale descriptions in the resulting PDE model.
{{For the purposes of the derivation of the continuum model, we have chosen cell invasion into the local environment in general, but due to its generality, the local environment could be replaced with a variety of substances, such as neighbouring tissues or organs, a tissue engineering scaffold or a wound, during healing.}}
The applications {{in this article}} are carefully selected to illustrate a wide range of cell behaviours by employing different functional forms that describe the probabilities of movement in both physical and phenotypic spaces, as well as the {{behaviours governing growth}} at the individual-based level. However these applications were not chosen with the aim to provide detailed biological insights at this stage.}}
We present these results through examples.

In Sec.~\ref{sec:pop_struc}, we consider {{a simplified model, without phenotype-driven migration, that shows}} how {{different growth mechanisms in the population impact the cell phenotypes present throughout a population over time}}. 
Next, in Sec.~\ref{sec:gg}, we study the migration-proliferation dichotomy {{(that states that cells can either migrate or proliferate but cannot {{do}} both at the same time)}} for cells migrating into the extracellular matrix (ECM), {{by extending this {{model}} to consider a continuum of phenotypes with a trade-off between the cells ability to grow and divide and their ability to move and degrade ECM.
In this model, we consider a range of different environmental features, such as the density of ECM, which we postulate could impact the phenotypic drift of the cells, and compare the resulting phenotypic structure of the invading cell population.}} 
Finally, in Sec.~\ref{sec:Tcell}, we {{examine the results of {{the}} general macroscopic framework for depicting the}} phenotypic and spatial dynamics of T cells infiltrating into a tumour, {{as described by microscopic individual-based interactions. 
H}}ere we consider the phenotype of the T cells to describe their exhaustion levels. 
{{Then, to summarise, f}}uture research directions and concluding remarks are discussed in Sec.~\ref{sec:conc}.

\section{The individual-based model}\label{sec:IBM}
I{{n order to incorporate microscopic descriptions of the interactions occurring between cells and their local environment,}} we formulate a phenotype-structured{{, on-lattice,}} individual-based model for collective cell migration. 

In this model, the cells are represented as individual, discrete agents and we assume that the features of the local environment we are interested in, such as the ECM or another species of cells, take up space. 
In order to fit in with {{the}} individual-based framework, we therefore choose to model the local environmental as being composed of individual, discrete elements of the same finite volume as the cells. 
Depending on the phenotype {{of the individual cell}} and   {{the}} number of cells and elements of the local environment {{in the same lattice site, each individual cell has a capacity}} to  undergo random, undirected movement, heritable phenotypic changes and proliferation{{, that can be adapted to the specific biological application of interest by employing appropriate individual-based rules to describe these changes}}. 
Furthermore, we assume that {{each individual}} cell can {{also interact with the surrounding}} environment{{, and that the cell's capacity to impact their local environment depends on the phenotype of the cell}}. 

Considering a one-dimensional spatial domain, we allow the cells and the local environment to be distributed in the region $x\in[X_{\text{min}}, X_{\text{max}}].$
We describe the phenotypic state of each individual cell through a structuring variable $y\in[Y_{\text{min}}, Y_{\text{max}}].$

In this individual-based model, we discretise the time variable $t\in\mathbb{R}^{+}$, as $t_h=h\Delta_t$ with $h\in\mathbb{N}$ and $\Delta_t\in\mathbb{R}^{+}.$
We discretise the spatial variable into {{an integer number of lattice}} sites {{$x_i = X_{\text{min}} + \Delta_x(i-1)$}} for $\Delta_x\in\mathbb{R}^{+}$ and {{$i=1,\dots, N_x+1.$}} 
We discretise the phenotype variable using {{$y_j = Y_{\text{min}}+ \Delta_y(j-1)$}} for $\Delta_y\in\mathbb{R}^{+}$ and {{$j=1,\dots, N_y+1.$}}
In this case, $\Delta_t,  \Delta_x,  \Delta_y\in\mathbb{R}^{+}$ are the time-, space- and phenotype-step, respectively.

We introduce the dependent variable $n_i^j(t_h)\in\mathbb{N}_0$ to model the number of cells that occupy a position on the lattice ${x_i} \times {y_j}\in[X_{\text{min}}, X_{\text{max}}]\times[Y_{\text{min}}, Y_{\text{max}}]$ at time $t_h,$ where $\mathbb{N}_0$ represents the natural numbers, including zero.
Then, we define the total cell number at a spatial position $x_i$ at time $t_h$ as $N_i(t_h)= \sum_{j=1}^{N_y+1}n_i^j(t_h)\in\mathbb{N}_0$ and the number of elements of the local environment at spatial position $x_i$ at time $t_h$ is denoted by $e_i(t_h)\in\mathbb{N}_0.$

 \subsection{Modelling the dynamics of the cells}\label{sec:cells}
We denote by $p({\bf{n}}, {\bf{e}},  t_h)$ the joint probability that the number of cells in spatial position $x_i$ in phenotypic state $y_j$ at time $t_h$ is given by ${\bf{n}}=([n_1^1, \dots ,  n_{N_x+1}^1], \dots , [n_1^{N_y+1}, \dots , n_{N_x+1}^{N_y+1}])$ and that the number of {{discrete, constitutive}} elements of the local environment in spatial position $x_i$ is given by ${\bf{e}}=[e_1, \dots, e_i, \dots, e_{N_x+1}].$

Between a time step $t_h$ and $t_{h+1}${{ (equivalently described by ${{t_h}}+\Delta_t$)}}, each cell in phenotypic state $y_j\in[Y_{\text{min}}, Y_{\text{max}}]$ at position $x_i\in[X_{\text{min}}, X_{\text{max}}]$ can undergo random movement, heritable phenotypic changes and cell proliferation{{ independently of time and}} according to the following assumptions in this section. 
We note here that we write $n_i^j(t_h)$ as $n_i^j$ and  $e_i(t_h)$ as $e_i$ for simplicity going forward.

\subsubsection{Random cell movement}\label{sec:mvmt}
We model cell movement in space as an on-lattice, biased random walk between neighbouring lattice sites. The probability of cell movement can depend on a number of factors, such as the local environment or the phenotype of the cell, but it is easy to relax these assumptions to consider other variables of interest. 
In particular, we {{introduce the following two changes in state vector}} that describe movement {{left or right in physical space of a single cell}} in phenotypic state $y_j$ into position $x_{i\pm1}$ from position $x_i$:
\begin{align*}
    L_{i,j}^{\text{m}}&: [n_1^j, \dots, n_{i-1}^j, n_{i}^{j}, \dots, n_{N_x+1}^j] \longrightarrow [n_1^j, \dots, n_{i-1}^j+1, n_{i}^{j}-1, \dots, n_{N_x+1}^j], \nonumber \\ &\qquad\qquad\qquad\qquad\qquad\qquad \text{for} \quad i=2, \dots , N_x+1, \quad j=1,\dots, N_y+1,\\
    R_{i,j}^{\text{m}}&: [n_1^j, \dots, n_{i}^j, n_{i+1}^{j}, \dots, n_{N_x+1}^j] \longrightarrow [n_1^j, \dots, n_{i}^j-1, n_{i+1}^{j}+1, \dots, n_{N_x+1}^j], \nonumber \\ &\qquad\qquad\qquad\qquad\qquad\qquad  \text{for} \quad i=1, \dots , N_x, \quad j=1,\dots, N_y+1,
\end{align*}
where $L_{i, j}^{\text{m}},  R_{i,j}            ^{\text{m}}: \mathbb{N}^{N_x+1}\rightarrow \mathbb{N}^{N_x+1}.$
{{We assume that the probability of cell movement depends on the phenotype of the cell and the number of cells and elements of the local environment in the target site, rather than the lattice site that the cell is currently in. 
As such, w}}e define the probability of movement to the left, to spatial position $x_{i-1}${{ from $x_i$, during a single time step $\Delta_t$}}, as 
\begin{equation*}
    \beta_{-}(j, N_{i-1},e_{i-1})\in[0,1], \qquad i=2,  \dots ,  N_x+1, \quad j=1,\dots, N_y+1,
\end{equation*}
and the probability of movement to the right, to spatial position $x_{i+1}${{ from $x_i$, during a single time step $\Delta_t$}}, as described by the {{change to the state vector}} $R_{i,j}^{\text{m}}$, as 
\begin{equation*}
    \beta_{+}(j, N_{i+1},e_{i+1})\in[0,1], \qquad i=1,  \dots ,  N_x, \quad j=1,\dots, N_y+1,
\end{equation*}
which depends on the phenotypic state of the cell, $j$, the number of elements of the local environment and the total number of cells in the target site. 
{{In order to ensure that cells cannot move to a physical site ``outside of the domain" $x\in[X_{\text{min}}, X_{\text{max}}]$, we assume that cells cannot move left out of site $i=1$, or right out of site $i=N_x+1$, so that $\beta_{-}(j, N_{0}, e_{0})=0$ and $\beta_{+}(j, N_{N_x+2}, e_{N_x+2})=0.$ }}
For $i=1,  \dots ,  N_x+1, \; j=1,\dots, N_y+1$, cells remain in their current site (\textit{i.e.}, do not move) with probability
\begin{equation*}
    1-\beta_{+}(j,N_{i+1},e_{i+1})-\beta_{-}(j, N_{i-1},e_{i-1})\in[0,1]. \nonumber
\end{equation*}

\subsubsection{Cell proliferation}\label{sec:prolif}
In order to {{model}} cell proliferation, we assume that a dividing cell is instantaneously replaced by two identical cells of equal volume to one another and the parent cell, such that the daughter cells inherit the same spatial position and phenotypic state of the parent cell. 
As such, the corresponding {{change in state vector}} {{during}} a time step $\Delta_t$ can be written as 
\begin{align*}
    G_{i, j}&: [n_1^j, \dots, n_{i}^j, \dots, n_{N_x+1}^j] \longrightarrow [n_1^j, \dots, n_{i}^j -1, \dots, n_{N_x+1}^j], \nonumber \\ &\qquad\qquad\qquad\qquad\qquad\qquad \text{for} \quad i=1, \dots , N_x+1, \quad j=1,\dots, N_y+1. 
\end{align*}
To {{represent phenotype-dependent}} cell proliferation, we assume that the probability of a proliferation event is dependent on the phenotypic state of the cell, and the total number of cells and elements of the local environment in the same physical site as the cell that is dividing. 
Therefore, 
we define the probability that a cell in site $i$ with phenotype $j$ proliferates {{during time step $\Delta_t$}}, as 
\begin{equation*}
    \gamma(j, N_i,e_i)\in[0,1], \qquad\qquad i=1, \dots , N_x+1, \quad j=1,\dots, N_y+1.
\end{equation*}
The probability of a cell not undergoing proliferation during a time step $\Delta_t$ can then be written as 
\begin{equation}
    1-\gamma(j, N_i,e_i)\in[0,1], \qquad\qquad i=1, \dots , N_x+1, \quad j=1,\dots, N_y+1. \nonumber
\end{equation}

\subsubsection{Cell phenotypic changes}\label{sec:phen}
{{During}} a single time step, $\Delta_t$, we {{model}} transitions in phenotype space from state $y_j$ to $y_{j\pm1}$ via the following {{changes in state vectors}}:
\begin{align*}
    D_{i,j}^{\text{p}}&: [n_i^1, \dots, n_{i}^{j-1}, n_{i}^{j}, \dots, n_{i}^{N_y+1}] \longrightarrow [n_i^1, \dots, n_{i}^{j-1}+1, n_{i}^{j}-1, \dots, n_{i}^{N_y+1}], \nonumber \\ &\qquad\qquad\qquad\qquad\qquad\qquad \text{for} \quad i=1, \dots , N_x+1,\quad j=1, \dots , N_y+1, \\
    U_{i,j}^{\text{p}}&: [n_i^1, \dots, n_{i}^{j}, n_{i}^{j+1}, \dots, n_{i}^{N_y+1}] \longrightarrow [n_i^1, \dots, n_{i}^{j}-1, n_{i}^{j+1}+1, \dots, n_{i}^{N_y+1}], \nonumber \\ &\qquad\qquad\qquad\qquad\qquad\qquad \text{for} \quad i=1, \dots , N_x+1,\quad j=1, \dots , N_y+1, 
\end{align*}
where $D_{i,j}^{\text{p}},  U_{i,j}^{\text{p}}: \mathbb{N}^{N_y+1}\rightarrow \mathbb{N}^{N_y+1}.$
A cell in site $i$ transitions from phenotypic state $y_j$ to $y_{j+1}$ {{during time step $\Delta_t$}} with a probability that depends on the phenotypic state of the cell and the total number of cells and elements of the local environment in the site $i$. 
Therefore, we can write that this transition, described by the {{change in state vector}} $U_{i,j}^{\text{p}}$, occurs with a probability 
\begin{equation*}
    \mu_{+}(j, N_i,  e_i)\in[0,1], \qquad i=1, \dots , N_x+1, \quad j=1,  \dots ,  N_y+1.
\end{equation*} 
Similarly, a cell in site $i$ transitions from phenotypic state $y_j$ to $y_{j-1}$ {{during time step $\Delta_t$ }}with a probability that depends on the phenotypic state of the cell and the total number of cells and elements of the local environment in the site $i$. 
Therefore, we can write that this transition, described by the {{change in state vector}} $D_{i,j}^{\text{p}}$, occurs with a probability 
\begin{equation*}
    \mu_{-}(j,  N_i,  e_i)\in[0,1], \qquad i=1, \dots , N_x+1, \quad j=1,  \dots ,  N_y+1.
\end{equation*} 
{{In order to ensure cells cannot transition to phenotypic states}} {{``outside of the domain"}} $y_j\in[Y_{\text{min}}, Y_{\text{max}}]$, we take $\mu_{-}(1, N_i,e_i) = 0$ and $\mu_{+}(N_y+1, N_i, e_i) = 0.$ 
Taking this into consideration, the probability that a cell in phenotypic state $y_j$ and spatial position $x_i$ will not change phenotype during a time step $\Delta_t$ is given by 
\begin{equation*}
    1- \mu_{+}(j, N_i,  e_i) - \mu_{-}(j,  N_i,  e_i)\in[0,1], \qquad i=1, \dots , N_x+1, \quad j=1,  \dots ,  N_y+1.
\end{equation*}

\subsection{Modelling the dynamics of the local environment}\label{sec:ecm_dyn}
We model degradation of elements of the local environment through contact with cells in the same physical site. 
Other cell-environment interactions, such as haptotaxis, or environmental changes such as production by cells, could also be considered here. These extensions are trivial, so in this work we focus solely on degradation of the environment in order to keep the scope of our applications narrow.
In particular, we define the {{change in state vector}} $H_i: \mathbb{N}^{N_x+1}\rightarrow \mathbb{N}^{N_x+1}$ to describe degradation of an element of the local environment in spatial position $x_i$ as:
\begin{equation*}
    H_i:[e_1, \dots, e_i, \dots , e_{N_x+1}] \longrightarrow [e_1, \dots, e_i +1, \dots, e_{N_x+1}], \qquad i=1,  \dots ,  N_x+1.
\end{equation*}
We assume that the probability of degradation of an element of the local environment {{during time step $\Delta_t$}} depends on the number of cells in each phenotypic state $j$ in the same spatial position $x_i$. 
As such, we define the probability of a cell in site $i$ of phenotype $j$ degrading an element of the local environment during a time step $\Delta_t$ as
\begin{equation*}
    \lambda (j, n_i^j)\in[0,1],    \qquad i=1,  \dots ,  N_x+1, \quad j=1,  \dots ,  N_y+1,.
\end{equation*}

\subsection{The corresponding continuum model}\label{sec:cont}
{{
In order to derive the corresponding continuum model describing the dynamics of the entire population of cells and the local environment over time, we employ a process known as coarse-graining. 
This procedure is described in full in the Supplementary Information}}.
Assuming that the probability of two or more events occurring in time step $\Delta_t$ is sufficiently small, the master equation, which describes the evolution of the probability density over time, is given by
\begin{align}
    &\Delta_t \dfrac{\partial}{\partial t} p({\bf{n}}, {\bf{e}},  t_h) + O(\Delta_t^2)\nonumber \\ &=  \sum_{i=1}^{N_x+1}\sum_{j=1}^{N_y+1}\mu_{-}(j+1, N_i, e_i)\left\{ (n_i^{j+1}+1)p(U_{i,j}^{\text{p}}{\bf{n}}, {\bf{e}},   t_h)-n_i^{j+1}p({\bf{n}}, {\bf{e}},  t_h)\right\} \nonumber \\ 
    &\quad +  \sum_{i=1}^{N_x+1}\sum_{j=1}^{N_y+1}\mu_{+}(j-1, N_i, e_i)\left\{ (n_i^{j-1}+1)p(D_{i,j}^{\text{p}}{\bf{n}}, {\bf{e}},   t_h)-n_i^{j-1}p({\bf{n}}, {\bf{e}},  t_h)\right\} \nonumber \\
    &\quad + \sum_{i=1}^{N_x} \sum_{j=1}^{N_y+1}\beta_{-}(j, {N_i,e_i)}\left\{ (n_{i+1}^{j}+1)p(R_{i,j}^{\text{m}}{\bf{n}}, {\bf{e}},   t_h)-n_{i+1}^{j}p({\bf{n}}, {\bf{e}},  t_h)\right\} \nonumber \\ 
    &\quad + \sum_{i=2}^{N_x+1} \sum_{j=1}^{N_y+1}\beta_{+}(j, N_i,e_i)\left\{ (n_{i-1}^{j}+1)p(L_{i,j}^{\text{m}}{\bf{n}}, {\bf{e}},   t_h)-n_{i-1}^{j}p({\bf{n}}, {\bf{e}},  t_h)\right\}\nonumber \\ 
    &\quad + \sum_{i=1}^{N_x+1} \sum_{j=1}^{N_y+1}\left\{\gamma(j, N_i-1,e_i)(n_i^j-1)p(G_{i,j}{\bf{n}}, {\bf{e}},  t_h)-\gamma(j, N_i,e_i)n_i^jp({\bf{n}}, {\bf{e}},  t_h)\right\} \nonumber \\
    &\quad + \sum_{i=1}^{N_x+1}\sum_{j=1}^{N_y+1} \lambda(j, n_i^j)\left\{ (e_i+1) p({\bf{n}},  H_i{\bf{e}},  t_h) - e_ip({\bf{n}},  {\bf{e}},  t_h)\right\}.
    \label{eq:master}
\end{align}
{{Briefly, the first two lines on the right hand side correspond to changes in the phenotypic state of the cell, the second two correspond to changes in the physical position of the cell, the penultimate line describes proliferation of the cell and the final line describes degradation of the local environment.}}

\subsubsection{The coarse-grained model of the cells}

{{As is standard in the literature, we define the ensemble average for the function, $f$, of the number of cells at position {$i=1,\dots, N_x+1$} in state {$j=1,\dots, N_y+1$} and number of elements of {{local environment}} in lattice site {$i=1,\dots, N_x+1$} in the following way:
\begin{equation}
    {{\langle f(n_i^j, e_i)\rangle = \sum_{\bf{n}}\sum_{\bf{e}}f(n_i^j, e_i)p({\bf{n}}, {\bf{e}},  t_h).}}
\end{equation}
}}We can {{therefore}} formally derive (as seen in Supplementary Information Sec.~\ref{app:deriv}) the following equation describing the {{evolution of the mean number of cells in physical site {{$i=1,\dots, N_x+1$}} and phenotypic state {{$j=1,\dots, N_y+1$}} based on the rules described in Sec.~\ref{sec:cells}:
\begin{align}
    \dfrac{\partial}{\partial t} \langle n_i^j\rangle &= \dfrac{1}{\Delta_t}\langle\beta_{+}(j, N_i,e_i)n_{i-1}^j\rangle +\dfrac{1}{\Delta_t} \langle\beta_{-}(j, N_i,e_i)n_{i+1}^j\rangle \nonumber \\ 
    &\quad-\dfrac{1}{\Delta_t} \langle\beta_{-}(j, N_{i-1},e_{i-1})n_i^j\rangle -\dfrac{1}{\Delta_t}\langle\beta_{+}(j, N_{s+1},e_{i+1})n_i^j\rangle \nonumber \\ 
    &\quad+\dfrac{1}{\Delta_t}\langle\mu_{+}(j-1, N_i,e_i)n_i^{j-1}\rangle+\dfrac{1}{\Delta_t}\langle\mu_{-}(j+1, N_i,e_i)n_i^{j+1}\rangle\nonumber \\ 
    &\quad-\dfrac{1}{\Delta_t}\langle\mu_{+}(j, N_i,e_i)n_i^j\rangle-\dfrac{1}{\Delta_t}\langle\mu_{-}(j, N_i,e_i)n_i^j\rangle\nonumber\\
    &\quad+\dfrac{1}{\Delta_t} \langle\gamma(j, N_i,e_i)n_i^j\rangle. \label{eq:full_IBM_prolif_ECM}
\end{align}
{{We now derive a PDE description of Eq.~\eqref{eq:full_IBM_prolif_ECM} by taking}} limits as $\Delta_x\rightarrow 0$, $\Delta_y\rightarrow 0$ and $\Delta_t\rightarrow 0$.
{{In order to do this, the }}discrete values of $\langle{n}_i^j(t_h)\rangle$ and $\langle{e}_i(t_h)\rangle$ are written in terms of the continuous variables $n(x,  y,  t)$ and $e(x,t)$, describing the cell and local environment density, respectively. 
{{W}}e find that, correct to $\mathcal{O}(\Delta_t)$:
\begin{align}
    \dfrac{\partial n(x, y, t)}{\partial t} &= \dfrac{1}{\Delta_t}\beta_{+}(y, \rho(x,  t),e(x,  t))n(x-\Delta_x,  y,  t)\nonumber \\ &\quad +\dfrac{1}{\Delta_t} \beta_{-}(y, \rho(x,  t),e(x,  t))n (x+\Delta_x,  y,  t) \nonumber \\ 
    &\quad-\dfrac{1}{\Delta_t} \beta_{-}(y, \rho(x-\Delta_x,  t),e(x-\Delta_x,  t))n(x,y,t) \nonumber \\ &\quad-\dfrac{1}{\Delta_t}\beta_{+}(y, \rho(x+\Delta_x,  t),e(x+\Delta_x,  t))n(x,y,t) \nonumber \\ 
    &\quad+\dfrac{1}{\Delta_t}\mu_{+}(y-\Delta_y, \rho(x,  t),e(x,  t))n(x,  y-\Delta_y,  t)\nonumber \\
    &\quad+\dfrac{1}{\Delta_t}\mu_{-}(y+\Delta_y, \rho(x, t),e(x,  t))n(x,  y+\Delta_y,  t)\nonumber \\ 
    &\quad-\dfrac{1}{\Delta_t}\mu_{+}(y, \rho(x,  t), e(x,  t))n(x,y,t) \nonumber \\ &\quad -\dfrac{1}{\Delta_t}\mu_{-}(y, \rho(x, t),e(x,  t))n(x,y,t)\nonumber\\
    &\quad+\dfrac{1}{\Delta_t} \gamma(y, \rho(x,  t),e(x,  t))n(x,y,t). \label{eq:full_IBM_prolif_ECM4}
\end{align}
Employing a Taylor series expansion around $(x,y)$, rearranging and collecting terms, we obtain
\begin{align*}
    &\dfrac{\partial}{\partial t}n(x,y,t) \\ &=\dfrac{\Delta_x}{\Delta_t}\dfrac{\partial}{\partial x}\Big(\left(\beta_{-}(y, \rho(x,t), e(x,t))-\beta_{+}(y, \rho(x,t), e(x,t))\right)n\Big) \nonumber \\
    & \quad+ \dfrac{\Delta_x^2}{2\Delta_t} \dfrac{\partial}{\partial x}\Bigg(\Big(\beta_{-}\left(y, \rho(x,t), e(x,t)\right)+\beta_{+}\left(y, \rho(x,t), e(x,t)\right)\Big)\dfrac{\partial}{\partial x} n(x,y,t) \nonumber \\ &\qquad\qquad\qquad- n(x,y,t)\dfrac{\partial}{\partial x}\Big(\beta_{-}\left(y, \rho(x,t), e(x,t)\right)+\beta_{+}\left(y, \rho(x,t), e(x,t)\right)\Big)\Bigg) \nonumber \\ 
    &\quad+\dfrac{\Delta_y}{\Delta_t}\dfrac{\partial}{\partial y} \left(\Big(\mu_{-}\left(y, \rho(x,t), e(x,t)\right)-\mu_{+}\left(y, \rho(x,t), e(x,t)\right)\Big)n(x,y,t)\right)\nonumber \\ &\quad+\dfrac{\Delta_y^2}{2\Delta_t} \dfrac{\partial ^2}{\partial y^2}\left(\Big(\mu_{-}\left(y, \rho(x,t), e(x,t)\right)+\mu_{+}\left(y, \rho(x,t), e(x,t)\right)\Big)n(x,y,t)\right)\nonumber \\
    &\quad+\dfrac{1}{\Delta_t}\gamma\left(y, \rho(x,t), e(x,t)\right)n(x,y,t).
\end{align*}

We define the following functions}}
{\small
\begin{align*}
    \lim_{\Delta_x,  \Delta_t\rightarrow 0} \dfrac{\Delta_x}{\Delta_t} \Big(\beta_{-}(y, \rho(x,t), e(x,t))-\beta_{+}(y, \rho(x,t), e(x,t))\Big) &= v^m(y, \rho(x,t), e(x,t)), \\
    \lim_{\Delta_x,  \Delta_t\rightarrow 0} \dfrac{\Delta_x^2}{2\Delta_t} \Big(\beta_{-}(y, \rho(x,t), e(x,t))+\beta_{+}(y, \rho(x,t), e(x,t))\Big) & = D^m(y, \rho(x,t), e(x,t)), \\
    \lim_{\Delta_y,  \Delta_t\rightarrow 0} \dfrac{\Delta_y}{\Delta_t} \Big(\mu_{-}(y, \rho(x,t), e(x,t))-\mu_{+}\left(y, \rho(x,t), e(x,t)\right)\Big) &= v^p(y, \rho(x,t), e(x,t)), \\
    \lim_{\Delta_y,  \Delta_t\rightarrow 0} \dfrac{\Delta_y^2}{2\Delta_t} \Big(\mu_{-}(y, \rho(x,t), e(x,t))+\mu_{+}\left(y, \rho(x,t), e(x,t)\right)\Big) &= D^p(y, \rho(x,t), e(x,t)), \\
    \lim_{\Delta_t\rightarrow 0} \dfrac{1}{\Delta_t} \gamma\left(y, \rho(x,t), e(x,t)\right) &= r(y, \rho(x,t), e(x,t)),
\end{align*}
}
such that the final equation governing the dynamics of the cell population is given by
\begin{align}
    \dfrac{\partial}{\partial t}n(x,y,t) &= \dfrac{\partial}{\partial x}\Big(v^m(y, \rho(x,t), e(x,t))n\Big) \nonumber \\
    &\qquad+\dfrac{\partial}{\partial x}\Bigg(D^m\left(y, \rho(x,t), e(x,t)\right)\dfrac{\partial}{\partial x} n(x,y,t) \nonumber \\ &\qquad\qquad\qquad\qquad- n(x,y,t)\dfrac{\partial}{\partial x}D^m\left(y, \rho(x,t), e(x,t)\right)\Bigg) \nonumber \\ 
    & \qquad+\dfrac{\partial}{\partial y} \left(v^p\left(y, \rho(x,t), e(x,t)\right)n(x,y,t)\right)\nonumber \\ &\qquad+\dfrac{\partial ^2}{\partial y^2}\left(D^p\left(y, \rho(x,t), e(x,t)\right)n(x,y,t)\right)\nonumber \\
    &\qquad +r\left(y, \rho(x,t), e(x,t)\right)n(x,y,t), \label{eq:cont-n}
\end{align}
where 
\begin{equation*}
    \rho(x,t)= \int_{y=Y_{\text{min}}}^{y=Y_{\text{max}}}n(x, y,  t) \mathrm{d}y,
\end{equation*}
describes the total cell density.

{{The differential equation governing the cell population evolution over time is complemented with the initial condition \begin{equation}n(x,y,0)=n_0(x,y),\label{eq:ICn}\end{equation} and}} is subject to zero-flux boundary conditions at $x=X_{\text{min}}, X_{\text{max}}$ and $y=Y_{\text{min}}, Y_{\text{max}}$, {{which are derived in Supplementary Information Sec.~\ref{app:BCs} and}} given by
\begin{align}
    v^mn +D^m\dfrac{\partial n}{\partial x}-n\dfrac{\partial D^m}{\partial x}  = 0  \qquad\qquad \text{at} \quad x=X_{\text{min}}, \label{eq:BC1}\\
    -v^mn +D^m\dfrac{\partial n}{\partial x}-n\dfrac{\partial D^m}{\partial x} = 0  \qquad\qquad \text{at} \quad x=X_{\text{max}}.
\end{align}
on the physical domain and
\begin{align}
    v^pn +\dfrac{\partial }{\partial y}(D^pn)= 0  \qquad\qquad \text{at} \qquad y=Y_{\text{min}}, \\
    -v^pn +\dfrac{\partial }{\partial y}(D^pn)= 0  \qquad\qquad \text{at} \qquad y=Y_{\text{max}}, \label{eq:BC4}
\end{align} 
at the ends of phenotype space. 
{{The differences in the boundary conditions in phenotype and physical space are a result of the varied assumptions underlying the movement probabilities. 
In physical space, the probability of movement depends on the number of cells and elements of the local environment in the target site, whereas the probability of movement in phenotype space depends on the number of cells and elements of the local environment in the same site as the cell.}}

\subsubsection{The coarse-grained model of the local environment}
Using probabilistic approximations of the same form as those underlying Eq.~\eqref{eq:full_IBM_prolif_ECM}, we recover the following equation describing the evolution of elements of the local environment in site $i$ over time:
\begin{equation}
     \Delta_t \dfrac{\partial}{\partial t} \langle e_s\rangle=- \sum_{j=1}^{N_y}\langle\lambda (j, n_s^j)e_s\rangle. \label{eq:es}
\end{equation}
Defining 
\begin{equation*}
    \lim_{\Delta_t\rightarrow0}\dfrac{1}{\Delta_t} \lambda(y, n(x,y,t), e(x,t))= \nu(y, n(x, y, t), e(x,t)),
\end{equation*}
which we can substitute into {{Eq.~\eqref{eq:es}}}, rearrange and take limits as $\Delta_x, \Delta_y, \Delta_t \rightarrow 0$, to find that the differential equation for the {{evolution of the}} density of the local environment, $e(x,t)$, is given by
\begin{equation}
       \dfrac{\partial }{\partial t} e(x,t) = - \int_{y=Y_{\text{min}}}^{y=Y_{\text{max}}} \nu(y, n(x,  y,  t)) e(x,  t) \mathrm{d}y. \label{eq:cont-e}
\end{equation}
The corresponding initial condition is then \begin{equation}e(x,0)=e_0(x).\label{eq:ICe}\end{equation}

Now that we have derived the coarse-grained model in full {{(Eqs.~\eqref{eq:cont-n}-\eqref{eq:BC4}, \eqref{eq:cont-e} and \eqref{eq:ICe})}}, we present a series of applications that demonstrate the utility of {{this}} framework through the choice of specific functional forms for the functions $v^m\left(y, \rho(x,t), e(x,t)\right)$, $D^m\left(y, \rho(x,t), e(x,t)\right)$, $v^p\left(y, \rho(x,t), e(x,t)\right)$, $D^p\left(y, \rho(x,t), e(x,t)\right)$, $r\left(y, \rho(x,t), e(x,t)\right)$ and $\nu(y, n(x,y,t))$.
We will assume in this article that all movement in physical space is undirected, and therefore we take $\beta_{+}(y, \rho(x,t), e(x,t))=\beta_{-}(y, \rho(x,t), e(x,t)$, such that $v^m=0$ hereon in. 
{{Nevertheless, the general form of the governing equations is retained, enabling readers to readily adapt the framework to cases involving directed movement or other specific applications.}}

\section{{{Broad spectrum applications in mathematical biology}}}\label{sec:num}
In this section, we showcase the versatility of the PDE model{{ling framework}} given by Eqs.~\eqref{eq:cont-n}-\eqref{eq:BC4}, \eqref{eq:cont-e} and \eqref{eq:ICe} by applying {{it}} to several exemplar biological scenarios. 
These applications demonstrate how the PDE framework effectively captures emergent population-level dynamics across diverse biological contexts. 
By considering a range of different underlying characteristics and interaction rules (prescribed in Supplementary Information Sec.~\ref{app:IBMfcts}), we showcase the ability of these models to encode complex behaviours while maintaining analytical and computational tractability. 

\subsection{Simulation methods}\label{sec:sims}
The deterministic, continuum counterpart of the individual-based model described in Sec.~\ref{sec:IBM} is given by the PDEs in Eqs.~\eqref{eq:cont-n}~and~\eqref{eq:cont-e}, with boundary conditions given in Eqs.~\eqref{eq:BC1}-\eqref{eq:BC4} and initial conditions given in Eqs.~\eqref{eq:ICn}~and~\eqref{eq:ICe}.
To solve this {{system}} numerically, we use an {{advection-diffusion-reaction}} (A-DR) scheme that discretises the spatial variable $x$ using a central finite difference stencil modified from previous work \citep{crossley2023travelling}, employing ghost points to enforce the zero-flux boundary conditions. 
{{The full system of discretised equations can be found in the Supplementary Information Sec.~\ref{sec:SI-NM}.}}
In the phenotypic axis, $y$, we use a finite volume scheme, which divides the axis into $N_y+1$ sites of equal width. The advective component is controlled using the Koren limiter \citep{koren1993robust}. 
The resulting {{system of ordinary differential equations}} are then integrated in time using python's in-built ordinary differential equation solver {\tt{scipy.integrate.solve\_ivp}} with the explicit Runge-Kutta integration method of order 5 and time step $\Delta_t=0.1$. 
The phenotype step is $\Delta_y=0.02$ and the spatial step is $\Delta_x=0.1$, both of which were chosen to be sufficiently small to ensure that we observed convergence in the solutions.               
Code is available for all computations at the following GitHub repository:
\url{https://github.com/beckycrossley/cont_phen}.

\subsection{Phenotypic structuring during range expansion}\label{sec:pop_struc}
Understanding how cell populations expand and evolve is a fundamental question in biology, particularly in contexts such as tumour growth, microbial colony expansion, and tissue development. 
A key aspect of these processes is tracking cell lineages to uncover how phenotypic traits propagate and shape population dynamics over time. 
In this section, we demonstrate how {{this}} modelling framework provides a convenient and effective approach for studying these lineage dynamics within an evolving population. 
Specifically, we consider a phenotypically structured population of homogeneous cells (\textit{i.e.}, cells that share the same underlying behaviour but are distinguishable by a phenotypic marker) to gain deeper insights into how individual lineages contribute to the overall invasion process. By analysing the spatio-temporal evolution of the phenotypic structure as the population spreads, we highlight how {{this}} approach enables the systematic tracking of cell lineages during range expansion.

Previous studies, such as those by \citep{marculis2020inside1}, have investigated similar population dynamics using stage-structured integrodifference equations \citep{marculis2020inside}, with further extensions incorporating trade-offs between reproductive and dispersal abilities \citep{marculis2020modeling}. 
{{While these approaches offer valuable insights into structured population dynamics, they rely on discrete phenotypic stages, which may limit the resolution of evolutionary and ecological interactions.

In contrast, {{this}} work provides a more nuanced perspective by modelling the evolution of a continuously structured phenotype, $y\in[0,1]$, over space, $x\geq0${{, and time, $t\geq0.$}}
{{This allows for a finer representation of phenotypic variation and subsequent exploration of its role during population expansion.
Specifically, we describe the spatio-temporal evolution of the cell population, }}
$n(x,y,t)$, using the following governing equation: 
\begin{equation}
\dfrac{\partial}{\partial t} n(x,y,t) = \dfrac{\partial^2}{\partial x^2} n(x,y,t) + r\left(\rho(x,t)\right)n(x,y,t). \label{eq:struct}
\end{equation}
As in \citep{marculis2020inside1}, we study Eq.~\eqref{eq:struct} subject to two different functions describing net cell proliferation: 
\begin{equation}
r_K(y, \rho(x,t)) = 1-\rho(x,t), \label{eq:KPP}
\end{equation}
for {{Fisher-}}KPP type invasion (pulled waves) and 
\begin{equation}
    r_A(y, \rho(x,t)) = (1-\rho(x,t))(\rho(x,t)-p^{*}), \label{eq:Allee}
\end{equation} 
for the Allee effect, with $p^{*}\in(0, 1/2)$ (corresponding to pushed waves), where $\rho(x,t)=\int_{y=0}^{y=1}n(x,y,t)\mathrm{d}y.$
The individual-based functions underlying these continuum equations can be found in Supplementary Information Sec.~\ref{app:PSfct}.
We compliment this setup with an initial condition that ensures initial phenotypic structuring of the population, which can then be tracked over time. Specifically, we take
\begin{equation}
    u_0(x,y)=\begin{cases}
        1 \quad \text{if}\;\; {{x=5y}}, \\
        0 \quad \text{otherwise.}
    \end{cases}
\end{equation}

\begin{figure}[htbp]
    \centering
    \includegraphics[width=\linewidth]{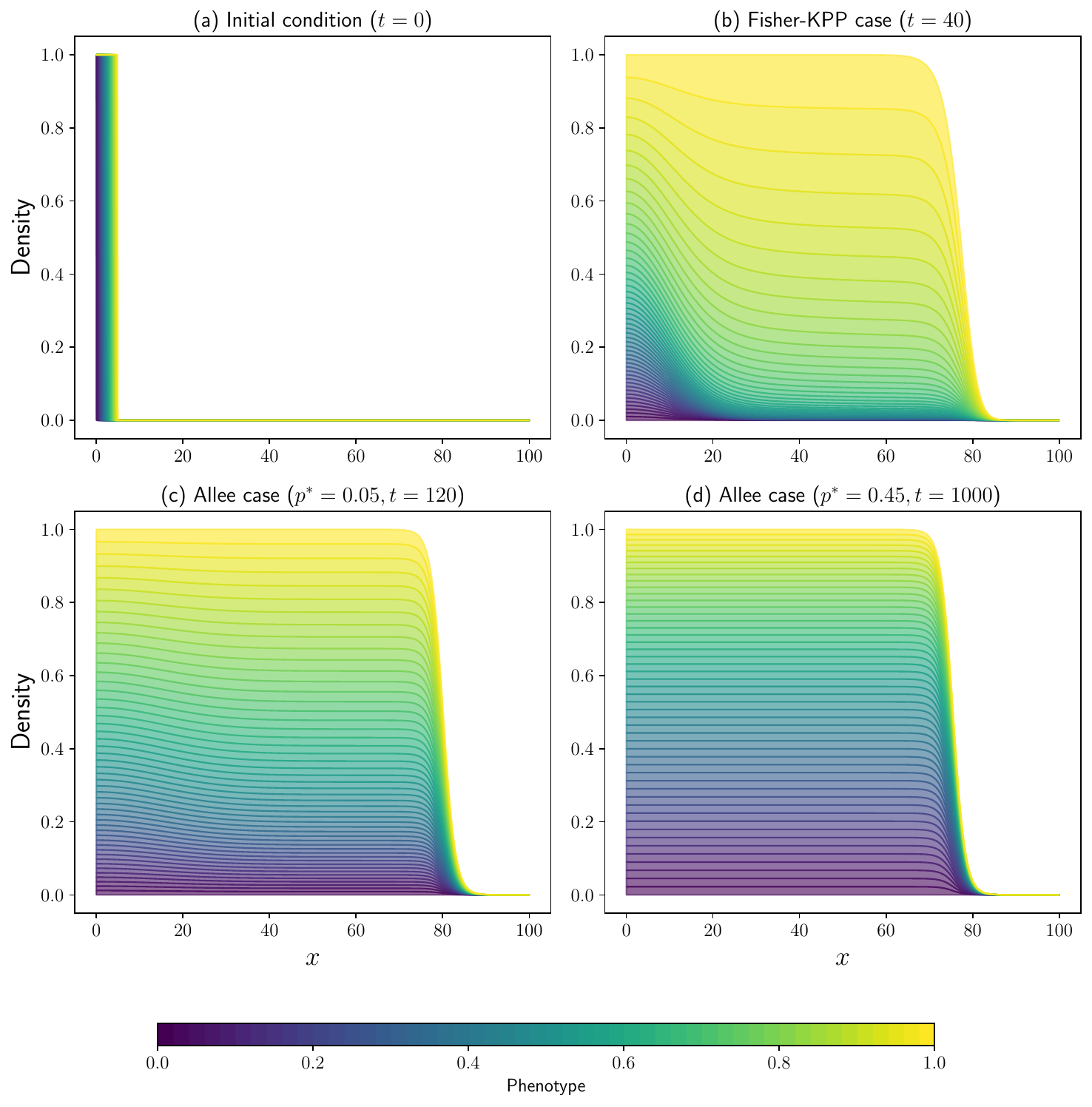}
    \caption{Evolution of the phenotypic structure of cells in Eq.~\eqref{eq:struct} subject to various growth terms. (a) The initial distribution of the cells with different phenotypes. (b) The spatial structure of the invading wave subject to the Fisher-KPP growth term (Eq.~\eqref{eq:KPP}). Results in (c) and (d) show the spatial structure of the invading wave subject to the Allee effect (Eq.~\eqref{eq:Allee}).}
    \label{fig:struct}
\end{figure}

Recalling that cells in this case are homogeneous, and thus the phenotypic variable $y\in[0,1]$ is used purely to label cells as they evolve, we can see that
Fig.~\ref{fig:struct} shows the phenotypic structure of the population of cells as they invade, subject to the aforementioned growth terms (Eq.~\eqref{eq:KPP} and Eq.~\eqref{eq:Allee}). 
Specifically, Fig.~\ref{fig:struct}(b) shows the spatial structure of Eq.~\eqref{eq:struct} with the {{Fisher-}}KPP growth term (Eq.~\eqref{eq:KPP}). 
{{As invasion progresses, the leading edge of the expanding population is dominated by cells that originated from the rightmost part of the initial distribution, with phenotypes closer to $y=1$. 
Over time, the density of these dominant phenotypes increases, and due to diffusion, cells with these traits spread backward, integrating into the bulk of the population.}}

This is a form of what is known as \emph{surfing} \citep{klopfstein2006fate}{{--}}a phenomenon well-studied in the context of drifting genetic mutations in expanding populations.
Surfing occurs {{when areas of low cell density allow space for increased growth}}, such as along the invading front (see Fig.~\ref{fig:struct}) \citep{excoffier2009genetic}.

The structure observed in the case of the {{Fisher-}}KPP type growth is often described as a `vertical pattern' \citep{marculis2020inside1}, which indicates that the total population is composed of different underlying phenotypes at different points in space, demonstrating a high level of spatial structuring. 
However, we see that even in a continuously structured cell population, it is those cells with the phenotypes that were nearest the front of the initial distribution of cells that dominate during range expansion. The lower phenotypes, which began at the rear of the invading population, primarily remain here and diminish in size over time and increasing space.

Alternatively, if we now consider the case of Allee growth (Eq.~\eqref{eq:Allee}) in Eq.~\eqref{eq:struct}, then the numerical results in Fig.~\ref{fig:struct} display a different pattern of invasion. 
As the cells invade, a much larger proportion of all of the initially present phenotypes remain present in the travelling wave front at later times, and throughout the bulk of the invading population. 
As in the {{Fisher-}}KPP case, the {{cells}} {{in the rightmost portion of the initial population, with phenotypes near $1$,}} contribute the largest portions to the wave. 
However, with the addition of the Allee effect, there now exist contributions from all initially present phenotypic states throughout the wave. 
{{This is because the Allee effect introduces a dependency on the total local population density, which means cells at the leading edge, in areas of low density, have reduced proliferation, preventing them from rapidly outcompeting other phenotypes within the cell bulk.}}
We note that by increasing the strength of the Allee effect, {{by taking $p^{*}$ closer to $0.5$,}} the proportion of all of the phenotypes present in the wave becomes closer to one another (equalises). 
The spatial pattern observed in this case is described as `horizontal' as it does not differ in space, but still shows high phenotypic variation in the front \citep{roques2012allee}. 

These structural differences at the wave front agree with those observed {{by}} \citep{roques2012allee} {{who}} consider a similar system but with discrete, rather than continuous, phenotypes. 
{{These results}} indicate that the lineage structure of an invading wave could potentially be used to distinguish between the underlying growth mechanisms of the cell populations.
It is also notable that the speed of cell invasion subject to the Allee effect (Eq.~\eqref{eq:Allee}) is significantly slower than the speed of invasion in the {{Fisher-}}KPP case (Eq.~\eqref{eq:KPP}).
As such, although the Allee effect maintains diversity across the travelling wave front, it also decreases the speed of invasion in doing so, and so a trade-off is observed between diversity maintenance and speed of population invasion, which favours faster invasion for a weaker Allee effect. 

\subsection{A go-or-grow model of cells invading the extracellular matrix (ECM)}\label{sec:gg}
In this section, we apply the aforementioned system of Eqs.~\eqref{eq:cont-n}-\eqref{eq:BC4}, \eqref{eq:cont-e} and \eqref{eq:ICe} to a general model for collective cell migration into the {{ECM}}{{, exploring how individual cell-level properties give rise to emergent population-level behaviours. 
By examining the interactions between cells and their environment at a population-level, we aim to uncover the underlying individual cell-ECM mechanisms that drive large-scale migration patterns and spatial organization, providing insight into how cell processes shape collective movement in biological systems.}}

The ECM is a complex network of proteins and carbohydrates that supports and provides structure for migrating cells \citep{crossley2024modeling, winkler2020concepts}. 
{{Its composition varies by location, making its role in collective migration difficult to define. However, a well-agreed notion is that cells need to breakdown the ECM in order to make space in which to invade \citep{jablonska2016matrix, nagase1999matrix, nagase2006structure, visse2003matrix}.}}

{{In this work, we aim to model the fundamental trade-off in energetic costs associated with motility and proliferation, known as the ``go-or-grow" hypothesis \citep{hatzikirou2012go}, while also incorporating the role of ECM degradation by migrating cells. 
Our goal is to use this framework to bridge the gap between individual-cell behaviours and emergent population-level dynamics.

Previous mathematical models have captured this trade-off by considering only two discrete phenotypes \citep{crossley2024phenotypic}, due to limitations in the available mathematical frameworks. 
However, biological evidence strongly supports the existence of a continuum of phenotypes rather than a simple dichotomy \citep{bendall2012single, campbell2021cooperation}. 
By extending prior work to a continuously structured phenotypic model, we provide a more biologically realistic representation of go-or-grow dynamics, allowing us to explore how individual-level decision-making translates into large-scale invasion patterns.}}

We model the cells, denoted $n(x,y,t)$, as able to move, proliferate and degrade the ECM, $e(x,t)$. 
To model the trade-off between ECM degradation, motility and proliferation, we {{assume}} that cells with a larger value of the phenotype variable, $y\in[0,1]$, have a higher proliferative potential but a lower motility and lower ECM degrading potential whilst, in comparison, cells with a lower value of the phenotypic variable $y$ correspond to those with a higher motility and higher ECM degrading potential, but a lower proliferative potential. 
Therefore, cells with phenotype $y=0$ degrade ECM most rapidly and are the most motile, but they are unable to proliferate.
On the other hand, cells with phenotype $y=1$ are the most proliferative, but cannot degrade ECM or move. 
{{As in \citep{crossley2024phenotypic},}} we implement a linear relationship between the phenotype of the cells and the associated ability to degrade ECM, move and proliferate.  

{{The initial functions describing the probabilities of transitions in the individual-based model underlying these continuum equations can be found in Supplementary Information Sec.~\ref{app:GGfct}. 
Hereon in, we focus on the corresponding continuum functions stated below.}}

Following the volume exclusion principles described by \citep{crossley2023travelling, crossley2024phenotypic}, we assume that the probability of a cell moving randomly in physical space linearly decreases as the occupancy of space increases.
As described earlier, we also introduce the assumption that the probability of a cell moving randomly increases as the phenotype of the cell decreases.
As such, the function describing movement in physical space can be written as 
\begin{equation*}
    D^m(y, \rho(x,t), e(x,t)) = (1-y)\left(1-\dfrac{\rho(x,t)+e(x,t)}{\kappa}\right), %\dfrac{1}{2}
\end{equation*}
where $\kappa>0$ is the total available density for cells and ECM, known as the carrying capacity.
Similarly, for all cells, the probability of cell proliferation linearly decreases as the space around the cell fills with other cells and ECM, but it also decreases as the phenotype decreases, such that 
\begin{equation*}
     r(y, \rho(x,t), e(x,t)) = y\left(1-\dfrac{\rho(x,t)+e(x,t)}{\kappa}\right).
\end{equation*}
Furthermore, we assume that degradation rate of the ECM is proportional to the density of surrounding cells and decreases as the phenotype of the cells increases. 
We write this as 
\begin{equation*}
    \nu(y,  n(x, y, t)) = (1-y)n(x,y,t).
\end{equation*}

{{One major challenge in understanding the cell phenotype dynamics during invasion is the lack of direct experimental observations, as visualizing phenotypic transitions in real time is extremely difficult. 
As a result, there is limited guidance in the literature on the appropriate mathematical forms for these transitions, leaving a key gap in {{the}} understanding of how phenotype-dependent behaviours shape collective migration.

This highlights the final undetermined functions in Eqs.~\eqref{eq:cont-n}~and~\eqref{eq:cont-e}, that describe the transitions between cell phenotypes. 
In this section, we systematically investigate the impact of different density-dependent phenotypic transition rules, $\mu_{\pm}(y, \rho(x,t), e(x,t))$, along with their associated drift and diffusion terms, $v^p(y, \rho(x,t), e(x,t))$ and $D^p(y, \rho(x,t), e(x,t))$, respectively, as summarised in Table~\ref{tab:PD}. 
By exploring how the invasion dynamics change under different assumptions, we aim to determine whether population-level (and therefore potentially more observable) behaviours can provide insight into the underlying, unobservable phenotypic structures, offering a potential approach for inferring hidden biological mechanisms from macroscopic invasion patterns.}}

\begin{table}[htbp]
\begin{center}
{\renewcommand{\arraystretch}{3}

\begin{tblr}{
  cells={valign=m,halign=c},
  row{1}={bg=lightgray,font=\bfseries,rowsep=8pt},
  colspec={QQQQ},
  hlines,
  vlines
}
\hline
Phenotypic drift & $\mu_{-}(y, \rho, e)$ & $\mu_{+}(y, \rho, e)$        & $v^p(y, \rho,e)$      & $D^p(y, \rho, e)$ \\
\hline
\hline
Cell-dependent           & $ \dfrac{\rho}{\kappa}$           & $1-\dfrac{\rho}{\kappa}$    & $2\dfrac{\rho}{\kappa}-1$   & $1$                              \\
\hline
ECM-dependent            & $ \dfrac{e}{\kappa}$              & $1-\dfrac{e}{\kappa}$       & $2\dfrac{e}{\kappa}-1$      & $1$                              \\
\hline
Space-dependent          & $ \dfrac{\rho+e}{\kappa}$         & $ 1-\dfrac{\rho+e}{\kappa}$ & $2\dfrac{\rho+e}{\kappa}-1$ & $1$           \\
\hline
\end{tblr}
% \end{tabular}
}
\end{center}
\vspace{0.4em}
\caption{Table listing the functions employed in Eq.~\eqref{eq:cont-n} describing the probabilities of transitions up and down in phenotype space, resulting in the phenotypic drift, ${{v^p}}(y, \rho(x,t), e(x,t))$, and phenotypic diffusion, ${{D^p}}(y, \rho(x,t), e(x,t))$, functions shown.}
\label{tab:PD}
\end{table}
{{We investigate three phenotypic drift mechanisms influencing cell transitions: firstly, cell-dependent drift, where cells shift toward more motile phenotypes at higher cell densities; next, ECM-dependent drift, where cells adopt more ECM-degrading phenotypes as ECM density increases; and finally, space-dependent drift, where cells become more proliferative as available space increases. 
Each drift term is bounded between zero and one, which is consistent with the constraints placed on the total cell and ECM density. 
By comparing these mechanisms for phenotype change, we assess how different environment- and density-dependent phenotypic transitions shape the structure of the migrating cell front.}}

{{
We simulate this system of Eqs.~\eqref{eq:cont-n}-\eqref{eq:BC4}, \eqref{eq:cont-e} and \eqref{eq:ICe} subject to the following initial conditions:
\begin{align} 
    n_0(x,y)&= \dfrac{{\text{exp}\left(-100(x^2 + y^2)\right)}}{\max \left( \int_{X_{\text{min}}}^{X_\text{max}} {\text{exp}\left(-100(x^2 + y^2)\right)} \, \mathrm{d}x \right)}, 
    \label{eq:IC_ggecm_cells}\\
    e_0(x)&=\begin{cases}
        0 \quad\quad \text{if} \;\; n_0(x,y)>0.001, \\
        0.5 \quad \text{otherwise.}
    \end{cases}\label{eq:IC_ggecm_ecm}
\end{align}
}}
\begin{figure}
    \centering
    \includegraphics[width=\linewidth]{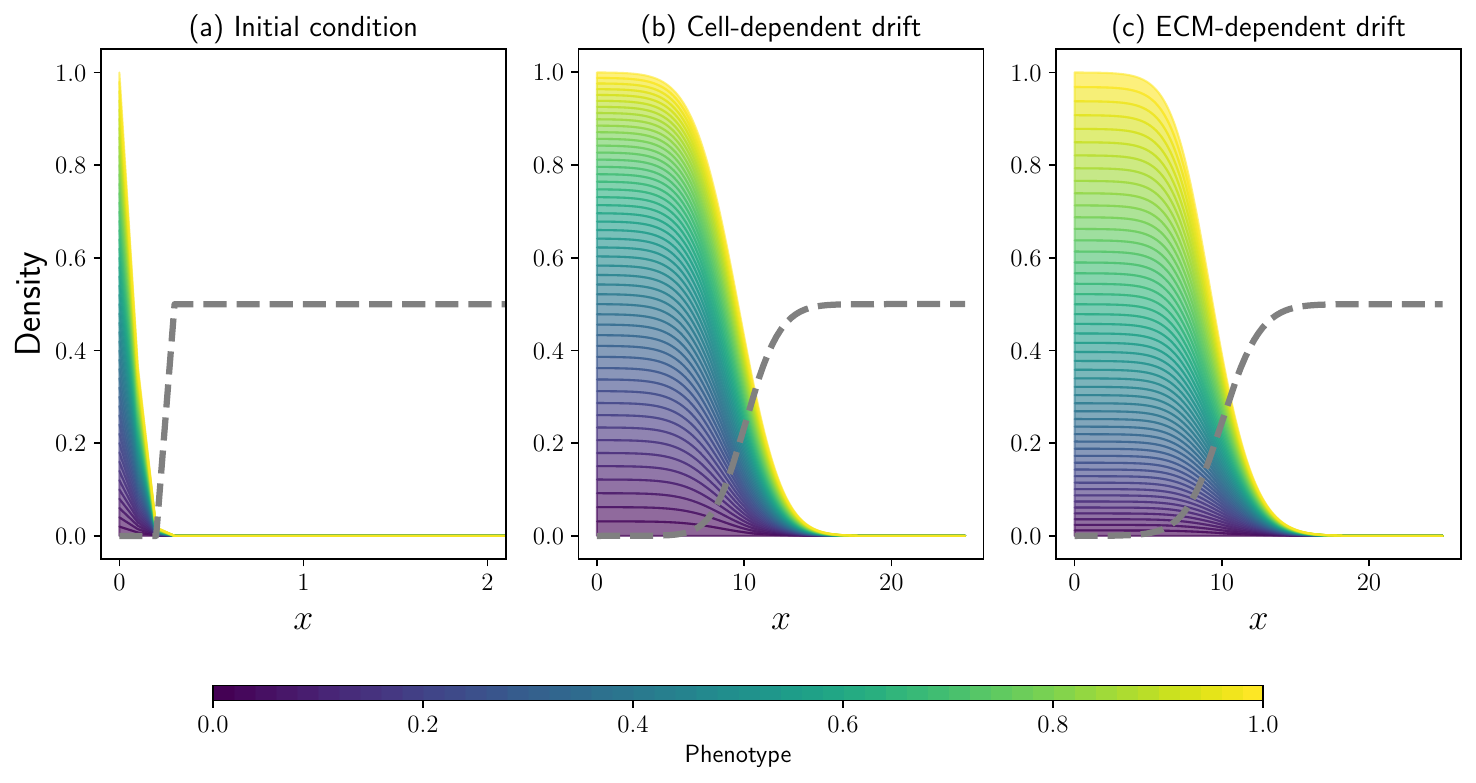}
    \caption{Evolution of the phenotypic structure of cells in Eqs.~\eqref{eq:cont-n}--\eqref{eq:cont-e} subject to various phenotypic drift terms, with the corresponding ECM density shown as a dashed grey line. (a) The initial distribution of the ECM and the cells with different phenotypes. (b) The spatial structure of the invading wave subject to cell-dependent phenotypic drift. (c)  The spatial structure of the invading wave subject to ECM-dependent phenotypic drift. 
    Results in (b) and (c) are all plotted at $t=30$ and simulations are carried out with $\kappa=1$. See Table~\ref{tab:PD} for explicit forms of the phenotypic drift terms.}
    \label{fig:gg_ECM}
\end{figure}

By examining the simulation results shown in Fig.~\ref{fig:gg_ECM} we can see that for a fully mixed initial population of cells, the phenotypic drift terms considered produce travelling wave solutions with similar{{, constant}} speeds of invasion. 
{{See Supplementary Information Fig.~\ref{fig:x} for a comparison including space-dependent phenotypic drift, which shows very similar results to cell-dependent drift.}}

In the case of cell-dependent phenotypic drift, we notice that there is a larger density of cells with lower phenotypes, which correspond to more motile and ECM degrading but less proliferative cells, in the bulk of the wave. 
At the front, we instead see a higher proportion of more proliferative and immobile cells, which are able to divide into the available space. 
As such, in the bulk of the wave the mean phenotype remains the same but {{as $x$ increases}}, we observe an initially gradual increase in mean phenotype before a much sharper increase at the very front of the wave where extremely low cell densities induce cells to switch to a proliferative phenotype.
Very similar patterning is observed in the space-dependent phenotypic drift case, without the sharp change in phenotype at the very front of the travelling wave (results in Supplementary Information).
In this context, this model predicts that the density of the ECM has minimal impact on the phenotypic structure of the invading wave.

However, when examining ECM-dependent phenotypic drift in Fig.~\ref{fig:gg_ECM}(c), we notice that the cells with higher phenotypes, corresponding to less motile and less degrading but more proliferative, now constitute a larger proportion of the population behind the invading front. 
Compared to cell- or space-dependent phenotypic drift, the mean phenotype of cells in the bulk is significantly higher. 
However, at the front of the invading wave in the ECM-dependent case, we see a larger proportion of cells with low phenotypes, namely, cells that have a higher ability to degrade the ECM and move into the subsequently available space. 

As a result of the phenotypic structures developed when the system is subject to different phenotypic drift functions, the spatial structure of the individual phenotypes within the invading wave could be used to better understand the underlying mechanisms governing phenotypic transitions during collective cell migration into the ECM. 

\subsection{T cell exhaustion}\label{sec:Tcell}
{{In this subsection, we demonstrate the broad applicability of the framework by deriving a population-level PDE model for T cell exhaustion, starting from an individual-based description of the underlying dynamics. 
This approach systematically coarse-grains the underlying cell processes, ensuring that the resulting PDEs are not merely phenomenological but instead retain a direct mechanistic link to individual cell properties. 
Specifically, we incorporate a T cell population with varying levels of exhaustion invading into a tumour, and examine the role of phenotype-dependent drift in shaping its exhaustion dynamics. 
This application highlights how {{this}} framework can be adapted to capture complex immune responses while preserving biologically meaningful connections between individual and population-level behaviour.}}

T cells are a key component of the immune system, with an important role to play in locating and attacking tumour cells \citep{weninger2001migratory}. 
When space is available, T cells will infiltrate into the tumour where they kill malignant cells by releasing cytotoxic enzymes. 
During the sustained activation of T cells required to combat and restrict further growth of a tumour, T cells will differentiate and eventually ``exhaust" \citep{yi2010t}. 
This occurs as a result of continued exposure to the antigens of the tumour cells and as T cells exhaust, their effector functions reduce \citep{blank2019defining, chow2022clinical}.
Exhaustion results in diminished cytokine production, impaired proliferation and reduced motility in the T cells \citep{miller2019subsets}.
Completely exhausted T cells are no longer able to move or grow \citep{wherry2004memory}, and have impaired toxicity, which reduces their ability to kill off tumour cells \citep{jiang2015t}. 
Furthermore, T cells will die much faster the more exhausted they are \citep{wherry2004memory}.
Understanding the dynamics of the T cells as they infiltrate a tumour and their exhaustion during this process is crucial to developing treatments for tumours, and so we develop a mathematical model to gain further insights.

{{As such, after applying the T cell specific individual-based model functions described in {{the}} Supplementary Information Sec.~\ref{app:Tfct} to the coarse-graining process, w}}e investigate the spatio-temporal evolution of T cell {{density}}, denoted $T(x,y,t)$, invading into {{a population of}} tumour cells, {{with density denoted by}} $C(x,t)$. 
In this application, the phenotype of the cells, $y\in[0,1]$, represents the exhaustion of the T cells. 
As such, T cells with phenotype $y=1$ are na\"ive, and are able to move freely and randomly in physical space, whilst also attacking nearby tumour cells and dividing \citep{reina2019antitumour, worbs2009t}. 
However, T cells with phenotype $y=0$ are considered exhausted, or terminally-differentiated memory T cells, which are considered to be in a resting state \citep{van2018full, sprent2011normal}. 
The resulting model takes the following form:
\begin{align}
    \dfrac{\partial}{\partial t}T(x,y,t) &= \dfrac{\partial}{\partial x}\bigg({{D^m}}\big(y, \rho(x,t), C(x,t)\big)\dfrac{\partial}{\partial x} T(x,y,t) \nonumber \\ &\qquad\qquad\qquad\qquad- T(x,y,t)\dfrac{\partial}{\partial x}{{D^m}}\big(y, \rho(x,t), C(x,t)\big)\bigg) \nonumber \\ 
    & \qquad+\dfrac{\partial}{\partial y} \bigg({{v^p}}\big(y, \rho(x,t), C(x,t)\big)T(x,y,t)\bigg)\nonumber \\ &\qquad+\dfrac{\partial ^2}{\partial y^2}\bigg({{D^p}}\big(y, \rho(x,t), C(x,t)\big)T(x,y,t)\bigg)\nonumber \\
    &\qquad +{{r}}\big(y, \rho(x,t), C(x,t)\big)T(x,y,t), \label{eq:Tcell-T}
\end{align}
{{where the}} net proliferation of the T cells, which depends on the exhaustion of the T cells and the available surrounding space for growth, can be described by
$${{r}}(y, \rho(x,t), C(x,t)) = \gamma_1y\left(1-\frac{\rho(x,t)+C(x,t)}{\kappa}\right)-\gamma_0(1-y),$$
where $\gamma_1\geq0$ describes the growth rate, and $\gamma_0\geq0$ describes the death rate of the T cells. $\kappa>0$ is the carrying capacity for the cells, as described in Sec.~\ref{sec:gg}.

T-cell movement in physical space is assumed to be random and undirected, but it is prevented by the presence of other T cells or tumour cells (in line with the volume-filling assumptions described in Sec.~\ref{sec:gg}). 
As such, diffusion in physical space is given by $${{D^m}}(y, \rho(x,t), C(x,t)) = y \bigg(1-\dfrac{\rho(x,t)+C(x,t)}{\kappa}\bigg),$$
which describes a higher diffusion rate for T cells with a higher phenotype.

Furthermore, we wish to model the phenotypic transitions of the T cells, or how exhausted the T cells become as they move, grow and interact with tumour cells. 
We have already assumed that the higher the phenotype of a T cell, the higher its probability of moving and growing, and this in turn will exhaust it. 
We also know that the tumour cells can be killed by the T cells, which we assume will further exhaust the T cells at a rate proportional to the number of interactions they have with the surrounding tumour cells. 
Therefore, we model the drift in phenotype space as $${{v^p}}(y, \rho(x,t), C(x,t)) = y(k_1+k_2C(x,t)),$$ where $k_1, k_2 \geq0$ describe the exhaustion rate of the T cells as a result of movement and growth, and as a result of interactions with the tumour cells, respectively.
We allow for some small randomness in the exhaustion levels of the T cells by including a diffusive term in phenotype space of the form $${{D^m}}(y, \rho(x,t), C(x,t))=\varepsilon\ll1.$$  

Now, it is well-known that tumour cells are also mobile and able to grow \citep{SURESH2007413}. 
As such, we can derive an equation similar to Eq.~\eqref{eq:cont-e} which also includes terms describing random movement and proliferation terms, and are derived in the same manner as those in Eq.~\eqref{eq:cont-n}. 
The resulting equation governing the evolution of the {{density of the}} tumour cells in space, $x\geq0$, and time, $t\geq0$, is therefore given by
\begin{align}
    \dfrac{\partial}{\partial t} C(x,t) &=  \dfrac{\partial}{\partial x}\bigg({{D^C}}\big(\rho(x,t), C(x,t)\big)\dfrac{\partial}{\partial x} C(x,t) \nonumber \\ &\qquad\qquad\qquad\qquad- C(x,t)\dfrac{\partial}{\partial x}{{D^C}}\big(\rho(x,t), C(x,t)\big)\bigg) \nonumber \\ &\qquad-\int_{y=0}^{y=1}{{\nu}}(y, T(x,y,t)) C(x,t) \mathrm{d}y + {{g}}(\rho(x, t), C(x,t))C(x,t).\label{eq:Tcell-C}
\end{align} 

Assuming that the motility and proliferation of tumour cells is restricted in regions of high cell density, we take 
\begin{align*}
    {{g}}(\rho(x,t), C(x,t))&=r_C\left(1-\dfrac{\rho(x,t)+C(x,t)}{\kappa}\right), \\
    {{D^C}}(\rho(x,t), C(x,t))&=1-\dfrac{\rho(x,t)+C(x,t)}{\kappa},
\end{align*}
with $r_C\geq0$ describing the growth rate of the tumour cells. 
Finally, 
T cells with a higher phenotype are less exhausted, and therefore kill tumour cells at a higher rate. 
As such, we write 
$${{\nu}}(y, T(x,y,t)) = \bar{\lambda}yT(x,y,t),$$ 
where $\bar{\lambda}\geq0$ describes the rate of degradation of the tumour cells by the T cells, or the toxicity of the T cells on the tumour.

{{
In the case of T-cell exhaustion, we simulate this system of Eqs.~~\eqref{eq:Tcell-T}--\eqref{eq:Tcell-C} subject to the following initial conditions:
\begin{align}
    T_0(x,y)&= \dfrac{\text{exp}\left(-100(x^2+(y-1)^2)\right)}{\max \left( \int_{X_{\text{min}}}^{X_\text{max}} \text{exp}\left(-100(x^2+(y-1)^2)\right) \, \mathrm{d}x \right)}, 
    \label{eq:IC_T_cells}\\
    C_0(x)&=\begin{cases}
        0 \quad\quad \text{if} \;\; n_0(x,y)>0.001, \\
        0.5 \quad \text{otherwise.}
    \end{cases}\label{eq:IC_T_ecm}
\end{align}
}}

Examining Fig.~\ref{fig:Tcell} it is clear that two main invasion behaviours are exhibited, which {{depend on the parameters of the system}}.
The first, which can be observed in Fig.~\ref{fig:Tcell}(b)~and~Fig.~\ref{fig:Tcell}(c), shows T cells that attack the tumour cells and produce travelling wave type profiles that invade through the tumour cells into the domain, killing off the tumour as they do so.

\begin{figure}[htbp]
    \centering
    \includegraphics[width=\linewidth]{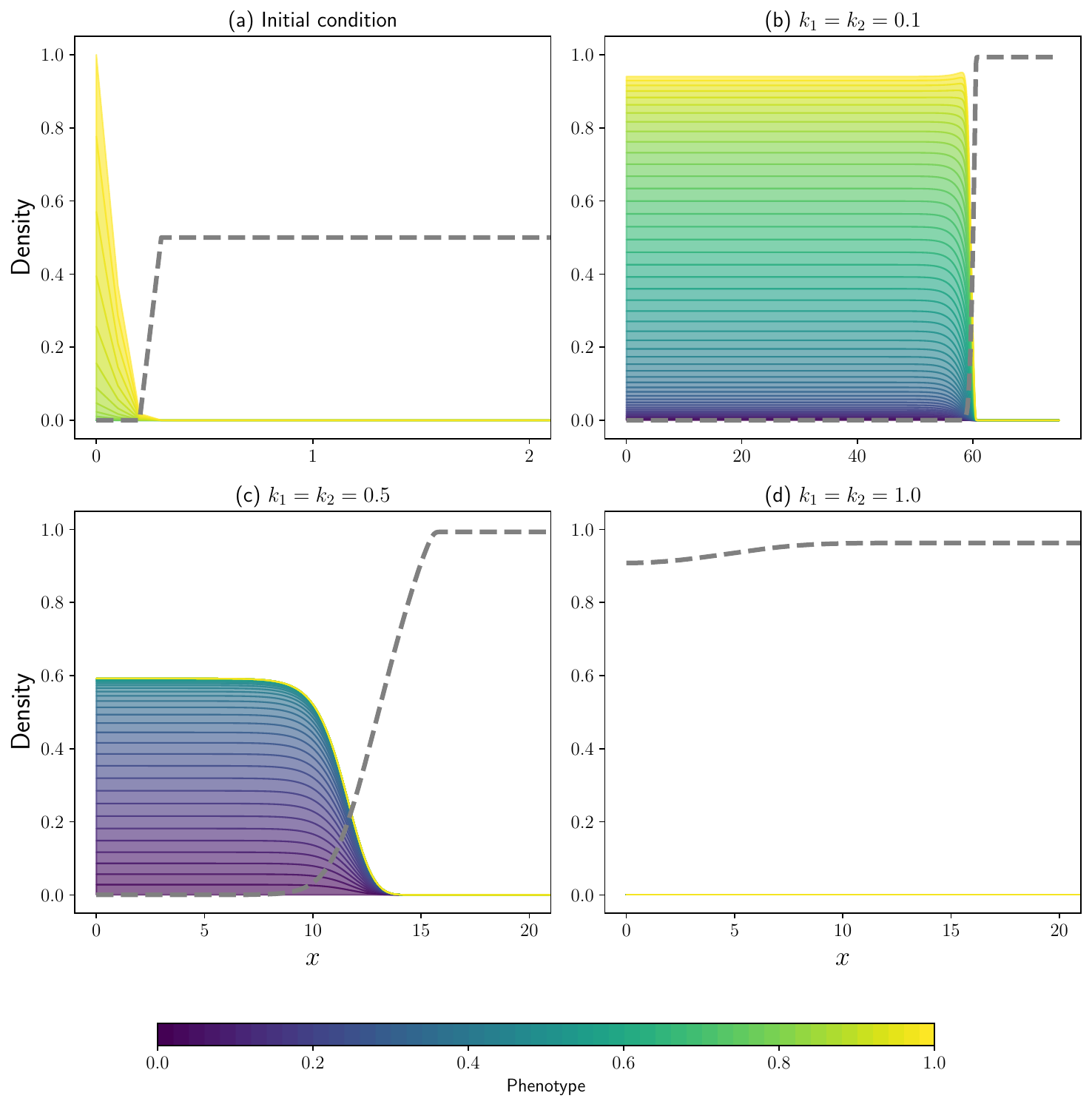}
    \caption{Evolution of the phenotypic structure of {{a population of}} T cells invading into a {{a population of}} tumour cells as described in Eqs.~\eqref{eq:Tcell-T}--\eqref{eq:Tcell-C} subject to various exhaustion rates, with the corresponding tumour cell density shown as a dashed grey line. (a) The initial distribution of the cells. Results in (b), (c) and (d) show the spatial structure of the invading wave subject to exhaustion rates $0.1$, $0.5$ and $1.0$, respectively. Solutions are plotted at $t=50$, and simulations are carried out with $\varepsilon=0.01$, $\gamma_0=1$, $\gamma_1=10$, $r_C=0.1$, $\kappa=1$, and $\bar{\lambda}=10$.}
    \label{fig:Tcell}
\end{figure}

In these simulations, {{altering the parameters of the system, namely the exhaustion rate of the T cells, leads to a range of different predicted behaviours. T}}he initial population of T cells were na\"ive. 
As such, the phenotypic structure of the invading wave of T cells with a lower exhaustion rate consisted of a larger range of phenotypes of cells (see Fig.~\ref{fig:Tcell}(b)). 
Consequently, the invading wave retains cells with a high phenotype. 
Comparatively, {{by increasing}} the exhaustion rate of the T cells, a travelling wave of invasion can still be observed (see Fig.~\ref{fig:Tcell}(c), for example), but with a much more restricted set of phenotypes in the bulk of the wave. 
The very front of the wave contains T cells with higher phenotypes that are the least exhausted, but the total density of T cells invading into the tumour cells is decreased, and subsequently, so is the invasion speed. 

In another scenario, whereby the growth of the tumour cells exceeds the degradation of the tumour cells by the T cells, results similar to those in Fig.~\ref{fig:Tcell}(d) are {{found}}, where the T cells exhaust extremely quickly as a result of a high number of interactions with tumour cells. 
This continued exposure quickly exhausts the T cells, which in turn increases death and creates available space for subsequent colonization by the tumour cells. 
In Fig.~\ref{fig:Tcell}(d), all of the T cells have died and we observe that the tumour cells (plotted as a dashed grey line) are growing to fill the entire domain, which they have already occupied in its entirety far ahead of the location of the initial population of T cells.

Due to the large number of parameters in Eqs.~\eqref{eq:Tcell-T}--\eqref{eq:Tcell-C}, complete exhaustion and death of the T cells can also be observed by decreasing the diffusion coefficient of the T cells, increasing (decreasing) the death (growth) rate of the T cells, decreasing the toxicity of the T cells or increasing the growth or motility rates of the tumour cells. 
In all of these cases, the tumour cells will eradicate the T cells and growth of the tumour to fill the available space will be observed. 
This is an example of competitive exclusion \citep{strobl2020mix}.
The opposite effects on each of these parameters provides examples of when the T cells are able to either completely eradicate or prevent further growth of the tumour cells. 
These behaviours have been observed experimentally \citep{schreiber2011cancer} and thus the modelling results shown in this work could provide useful insights into developing treatments for tumours in the future, once biologically realistic parameter sets are investigated.

\section{Discussion}\label{sec:conc}
In conclusion, numerous biological processes can be described in a simple form by phenotypically structured cell migration. 
Yet, general frameworks for modelling this, built from vastly adaptable underlying individual-based principles, are largely understudied \citep{villa2024reducing}.
In this article, we have demonstrated how to derive a general model for phenotypically structured cell migration from the underlying individual-based processes, that can be used to reproduce a variety of results in the literature, {{whilst modelling}} a continuum of phenotypes. 

Whilst we have illustrated that the general modelling framework is easily adaptable and can be applied to a {{range of}} biological systems considering spatial invasion of structured populations, we highlight the need for adaptation of the underlying assumptions to the specific modelling question of interest. 
{{For instance, the continuum model derived in this work employs distinctly different boundary conditions in the physical and phenotypic domains. These differences arise from the varying assumptions about the processes governing movement in each domain. However, these conditions can be straightforwardly modified through analogous derivations.}}
{{We therefore stress that this work represents an important first step towards}} generalising this modelling approach. 

In changing the form of the phenotypic drift terms, care must also be taken to ensure that limits taken in moving from the individual-based derivation to the continuum model are satisfied. 
For reaction-diffusion models with drift terms, the impact of these scalings and quantitative comparisons between the individual-based and resulting continuum models in such limits are well-studied in the literature \citep{lorenzi2020discrete, macfarlane2022individual}.

The continuum modelling framework presented in this work describes invasion in two dimensions: space and phenotype.  
We could extend this into higher dimensions, both spatially and phenotypically, such that two or three dimensional experiments, such as those looking at the evolution of tumour spheroids for example, can be more accurately modelled, and the results validated against data.

Beyond this, we have applied this system of Eqs.~\eqref{eq:cont-n}-\eqref{eq:BC4}, \eqref{eq:cont-e} and \eqref{eq:ICe} to several biological applications to demonstrate its versatility and utility. 
We recognise that these applications are for illustrative purposes only, and thus we employ several biological simplifications.
We therefore propose that this framework could provide a base model for continuum modelling of invasion processes derived from underlying individual-based assumptions.
The modelling framework presented would therefore benefit from expanding or adapting the forms of the functions to the specific biological application of interest, and the addition of extra details from various sub-models in the literature that may have been validated with experimental data would be needed in order to use this modelling approach to infer specific conclusions about the application of interest. 
For example, in the go-or-grow application, we have considered the volume-filling effects of both the cells and the ECM. However, if we were to consider modelling a chemoattractant, for example, the volume-filling assumptions underlying this model would no longer apply. As such, the derivation of the model would need to be adjusted accordingly, in a trivial manner that results in the removal of non-linearities from the diffusion terms. 

Additionally, in the case of T cell infiltration into 
tumour cells, where we interpret the phenotype of the cell as its exhaustion level, we are making a first step into the spatio-temporal modelling of T cell movement with phenotypic structuring, and there remains a large number of questions to be asked. For example, what are the specific functions that most accurately describe the exhaustion of the T cells, or what does it really mean to be exhausted in this context \citep{blank2019defining}?

Overall, when simulation results from a particular biological scenario present vastly different phenotypically structured solutions, this general modelling framework could provide a useful tool for developing understanding of the mechanisms underlying heterogeneity during migration, such as the bet-hedging strategies observed as a result of environmental pressures in \citep{almeida2024evolutionary}. 
The framework presented in this work has been derived from descriptions of interactions at an individual-based level and can be easily adapted to investigate a variety of biological applications. 
As a result, the versatility of this tool could help us understand the role of heterogeneity in a wide range of circumstances and its resulting insights could ultimately be used to help inform subsequent important treatment decisions. 

\backmatter

\bmhead{Acknowledgments}

R.M.C. would like to thank the Engineering and Physical Sciences Research Council (EP/T517811/1) and the Oxford-Wolfson-Marriott scholarship at Wolfson College, University of Oxford for funding.
R.E.B. also acknowledges support of a grant from the Simons Foundation (MP-SIP-00001828).

\section*{Declarations}

The authors do not have any competing interests to declare.

\bibliography{references}% common bib file

\newpage
\newcommand{\ind}{\mathbb{I}}
\begin{center}
    {\Large \textbf{Supplementary Information}}\\[1.5ex]
    \textit{Modelling the impact of phenotypic heterogeneity on cell migration: a continuum framework derived from individual-based principles}\\[3ex]
    \textbf{Rebecca M. Crossley}$^{1}$, \textbf{Philip K. Maini}$^{1}$, \textbf{Ruth E. Baker}$^{1}$\\[1ex]
    \small
    $^{1}$Mathematical Institute, University of Oxford, Woodstock Road, Oxford OX2 6GG, United Kingdom\\
    \texttt{crossley@maths.ox.ac.uk}, \texttt{maini@maths.ox.ac.uk}, \texttt{baker@maths.ox.ac.uk}
\end{center}

\date{}
\setcounter{section}{0}
\setcounter{figure}{0}
\setcounter{equation}{0}
\setcounter{table}{0}
\renewcommand{\theequation}{S\arabic{equation}}
\renewcommand{\thesection}{S\arabic{section}}
\renewcommand{\thefigure}{S\arabic{figure}}

\newpage
\section{Formal derivation of the continuum Eqs.~\eqref{eq:cont-n} and \eqref{eq:cont-e}}\label{app:deriv}

In this supplementary information, we present the full formal derivation of the continuum model~\eqref{eq:cont-n} and \eqref{eq:cont-e} from the individual-based model described in Sec.~\ref{sec:IBM}. 

When the {{dynamics}} of the cells {{and the local environment are}}  described by the rules outlined in Sec.~\ref{sec:IBM}, then 
the master {{Eq.~\eqref{eq:master}}} is given by
\begin{align}
    &\Delta_t \dfrac{\partial}{\partial t} p({\bf{n}}, {\bf{e}},  t_h) + O(\Delta_t^2)\nonumber \\ &=  \sum_{i=1}^{N_x+1}\sum_{j=1}^{N_y}\mu_{-}(j+1, N_i, e_i)\left\{ (n_i^{j+1}+1)p(U_{i,j}^{\text{p}}{\bf{n}}, {\bf{e}},   t_h)-n_i^{j+1}p({\bf{n}}, {\bf{e}},  t_h)\right\} \nonumber \\ 
    &\quad +  \sum_{i=1}^{N_x+1}\sum_{j=2}^{N_y+1}\mu_{+}(j-1, N_i, e_i)\left\{ (n_i^{j-1}+1)p(D_{i,j}^{\text{p}}{\bf{n}}, {\bf{e}},   t_h)-n_i^{j-1}p({\bf{n}}, {\bf{e}},  t_h)\right\} \nonumber \\
    &\quad + \sum_{i=1}^{N_x} \sum_{j=1}^{N_y+1}\beta_{-}(j, {N_i,e_i)}\left\{ (n_{i+1}^{j}+1)p(R_{i,j}^{\text{m}}{\bf{n}}, {\bf{e}},   t_h)-n_{i+1}^{j}p({\bf{n}}, {\bf{e}},  t_h)\right\} \nonumber \\ 
    &\quad + \sum_{i=2}^{N_x+1} \sum_{j=1}^{N_y+1}\beta_{+}(j, N_i,e_i)\left\{ (n_{i-1}^{j}+1)p(L_{i,j}^{\text{m}}{\bf{n}}, {\bf{e}},   t_h)-n_{i-1}^{j}p({\bf{n}}, {\bf{e}},  t_h)\right\}\nonumber \\ 
    &\quad + \sum_{i=1}^{N_x+1} \sum_{j=1}^{N_y+1}\left\{\gamma(j, N_i-1,e_i)(n_i^j-1)p(G_{i,j}{\bf{n}}, {\bf{e}},  t_h)-\gamma(j, N_i,e_i)n_i^jp({\bf{n}}, {\bf{e}},  t_h)\right\} \nonumber \\
    &\quad + \sum_{i=1}^{N_x+1}\sum_{j=1}^{N_y+1} \lambda(j, n_i^j)\left\{ (e_i+1) p({\bf{n}},  H_i{\bf{e}},  t_h) - e_ip({\bf{n}},  {\bf{e}},  t_h)\right\},
    \label{SIeq:master}
\end{align}

\subsection{Equation for cell density}
As is standard in the literature, we define the ensemble average for the function, $f$, of the number of cells at position $i$ in state $j$ and the number of elements of the local environment in lattice site $i$ in the following way:
\begin{equation}
    {{\langle f(n_i^j, e_i)\rangle = \sum_{\bf{n}}\sum_{\bf{e}}f(n_i^j, e_i)p({\bf{n}}, {\bf{e}},  t_h).}}
\end{equation}

Now, returning to Eq.~\eqref{SIeq:master}{{, we first take moments and multiply by}} $n_s^q$, where $s=1, \dots, N_x+1$ and $q=1, \dots, N_y+1$ and take the sum over the vectors ${\bf{n}}$ and ${\bf{e}}$. 
{{The final term in Eq.~\eqref{SIeq:master} corresponds to the evolution of the number of elements of the local environment, and does not contribute to the cell dynamics.}}
{{Then,}} to begin with, we consider only the first remaining term on the right-hand side, which we denote by $I${{,}} in the following way:
\begin{equation*}
    I= \sum_{\bf{n}}\sum_{\bf{e}}n_s^q \sum_{i=1}^{N_x+1}\sum_{j=1}^{N_y}\mu_{-}(j+1, N_i, e_i)\left\{ (n_i^{j+1}+1)p(U_j^{\text{p}}{\bf{n}}, {\bf{e}},   t_h)-n_i^{j+1}p({\bf{n}}, {\bf{e}},  t_h)\right\}. \nonumber
\end{equation*}
To continue, we must now consider the following cases in turn: 
\begin{itemize}
    \item $I_1$: $i\neq s$, $j\neq q,  q-1${{;}}
    \item $I_2$: $i\neq s$ and $j=q${{;}}
    \item $I_3$: $i\neq s$ and  $j=q-1${{;}}
    \item $I_4$: $i = s$ and $j \neq q, q-1${{;}}
    \item $I_5$: $i=s$ and $j=q${{;}}
    \item $I_6$: $i=s$ and $j=q-1.$
\end{itemize}
We begin with the case {{$I_1$}} where $i\neq s$ and $j \neq q,  q-1.$ 
In this scenario, we find
\begin{align}
    I_1&= \sum_{\bf{n}}\sum_{\bf{e}}n_s^q \sum_{\substack{i=1, \\ i\neq s}}^{N_x+1}\sum_{\substack{j=1, \\ j \neq q,  q-1}}^{N_y}\mu_{-}(j+1, N_i, e_i)\left\{ (n_i^{j+1}+1)p(U_j^{\text{p}}{\bf{n}}, {\bf{e}},   t_h)-n_i^{j+1}p({\bf{n}}, {\bf{e}},  t_h)\right\} \nonumber \\ 
    &= \sum_{\bf{n}}\sum_{\bf{e}}n_s^q\sum_{\substack{i=1, \\ i\neq s}}^{N_x+1}\sum_{\substack{j=1, \\ j \neq q,  q-1}}^{N_y}\mu_{-}(j+1, N_i, e_i)\times \nonumber \\ &\qquad\qquad\Big\{ (n_i^{j+1}+1)p([n_i^1, \dots, n_{i}^{j}-1, n_{i}^{j+1}+1, \dots, n_{i}^{N_y+1}], {\bf{e}},   t_h)-n_i^{j+1}p({\bf{n}}, {\bf{e}},  t_h)\Big\}. \nonumber
\end{align}
Under the change of variables $\bar{n}_i^j = \bar{n}_i^j -1$ and $\bar{n}_i^{j+1} = \bar{n}_i^{j+1} +1$ in the first term, we have that 
\begin{align*}
    I_1&= \sum_{\bf{n}}\sum_{\bf{e}}n_s^q\sum_{\substack{i=1, \\ i\neq s}}^{N_x+1}\sum_{\substack{j=1, \\ j \neq q,  q-1}}^{N_y}\mu_{-}(j+1, N_i, e_i)\times \nonumber \\ &\qquad\qquad\qquad\qquad\qquad\Big\{ \bar{n}_i^{j+1}p([n_i^1, \dots, \bar{n}_{i}^{j}, \bar{n}_{i}^{j+1}, \dots, n_{i}^{N_y+1}], {\bf{e}},   t_h)-n_i^{j+1}p({\bf{n}}, {\bf{e}},  t_h)\Big\} \nonumber \\ 
    &= \sum_{\bf{n}}\sum_{\bf{e}}n_s^q\sum_{\substack{i=1, \\ i\neq s}}^{N_x+1}\sum_{\substack{j=1, \\ j \neq q,  q-1}}^{N_y}\mu_{-}(j+1, N_i, e_i)\left\{ {n}_i^{j+1}p({\bf{n}}, {\bf{e}},  t_h)-n_i^{j+1}p({\bf{n}}, {\bf{e}},  t_h)\right\} \nonumber \\ 
    &= 0, 
\end{align*}
where we can arbitrarily drop the bar after the change of variables.

Next, we consider the case when $i\neq s$ and $j=q$, and apply the same change of variables argument (with $\bar{n}_i^q = \bar{n}_i^q -1$ and $\bar{n}_i^{q+1} = \bar{n}_i^{q+1} +1$) so that
\begin{align*}
    I_2 &=  \sum_{\bf{n}}\sum_{\bf{e}}n_s^q\sum_{\substack{i=1, \\ i\neq s}}^{N_x+1}\mu_{-}(q+1, N_i, e_i)\times \nonumber \\ &\qquad\qquad\Big\{ (n_i^{q+1}+1)p([n_i^1, \dots, {n}_{i}^{q}-1, {n}_{i}^{q+1}+1, \dots, n_{i}^{N_y+1}], {\bf{e}},   t_h)-n_i^{q+1}p({\bf{n}}, {\bf{e}},  t_h)\Big\} \nonumber \\
    &=  \sum_{\bf{n}}\sum_{\bf{e}}n_s^q\sum_{\substack{i=1, \\ i\neq s}}^{N_x+1}\mu_{-}(q+1, N_i, e_i)\times \nonumber \\ &\qquad\qquad\qquad\qquad\Big\{ \bar{n}_i^{q+1}p([n_i^1, \dots, \bar{n}_{i}^{q}, \bar{n}_{i}^{q+1}, \dots, n_{i}^{N_y+1}], {\bf{e}},   t_h)-n_i^{q+1}p({\bf{n}}, {\bf{e}},  t_h)\Big\} \nonumber \\
    &= \sum_{\bf{n}}\sum_{\bf{e}}n_s^q\sum_{\substack{i=1, \\ i\neq s}}^{N_x+1}\mu_{-}(q+1, N_i, e_i)\left\{ {n}_i^{q+1}p({\bf{n}}, {\bf{e}},  t_h)-n_i^{q+1}p({\bf{n}}, {\bf{e}},  t_h)\right\} \nonumber \\
    &=0.
\end{align*}

Finally we consider the last case for $i\neq s$, namely, where $j=q-1.$
Once again, by changing the variables using $\bar{n}_i^{q-1}=n_i^{q-1}-1$ and $\bar{n}_i^q=n_i^q+1,$ we have 
\begin{align*}
    I_3 &=  \sum_{\bf{n}}\sum_{\bf{e}}n_s^q\sum_{\substack{i=1, \\ i\neq s}}^{N_x+1}\mu_{-}(q, N_i, e_i)\times \nonumber \\ &\qquad\qquad\left\{ (n_i^{q}+1)p([n_i^1, \dots, {n}_{i}^{q-1}-1, {n}_{i}^{q}+1, \dots, n_{i}^{N_y+1}], {\bf{e}},   t_h)-n_i^{q}p({\bf{n}}, {\bf{e}},  t_h)\right\} \nonumber \\
    &=  \sum_{\bf{n}}\sum_{\bf{e}}n_s^q\sum_{\substack{i=1, \\ i\neq s}}^{N_x+1}\mu_{-}(q, N_i, e_i)\times \nonumber \\ &\qquad\qquad\qquad\qquad\left\{ \bar{n}_i^{q}p([n_i^1, \dots, \bar{n}_{i}^{q-1}, \bar{n}_{i}^{q}, \dots, n_{i}^{N_y+1}], {\bf{e}},   t_h)-n_i^{q}p({\bf{n}}, {\bf{e}},  t_h)\right\} \nonumber \\
    &= \sum_{\bf{n}}\sum_{\bf{e}}n_s^q\sum_{\substack{i=1, \\ i\neq s}}^{N_x+1}\mu_{-}(q, N_i, e_i)\left\{ {n}_i^{q}p({\bf{n}}, {\bf{e}},  t_h)-n_i^{q}p({\bf{n}}, {\bf{e}},  t_h)\right\} \nonumber \\
    &=0.
\end{align*}
As such, we can see that there are no contributions arising from the terms for which $i\neq s.$

Now we can consider cases where $i = s.$ First off, we consider $ j \neq q,  q-1$, and apply the change of variables $\bar{n}_s^j = n_s^j -1$ and $\bar{n}_s^{j+1} =  n_s^{j+1}+1$:
\begin{align*}
    I_4 &= \sum_{\bf{n}}\sum_{\bf{e}}n_s^q\sum_{\substack{j=1, \\ j \neq q,  q-1}}^{N_y}\mu_{-}(j+1, N_{s}, e_s)\times \nonumber \\ &\qquad\qquad\left\{ (n_s^{j+1}+1)p([n_s^1, \dots, n_{s}^{j}-1, n_{s}^{j+1}+1, \dots, n_{s}^{N_y+1}], {\bf{e}},   t_h)-n_s^{j+1}p({\bf{n}}, {\bf{e}},  t_h)\right\} \nonumber \\
    &= \sum_{\bf{n}}\sum_{\bf{e}}n_s^q\sum_{\substack{j=1, \\ j \neq q,  q-1}}^{N_y}\mu_{-}(j+1, N_{s}, e_s)\times \nonumber \\ &\qquad\qquad\left\{ \bar{n}_s^{j+1}p([n_s^1, \dots, \bar{n}_{s}^{j}, \bar{n}_{s}^{j+1}, \dots, n_{s}^{N_y+1}], {\bf{e}},   t_h)-n_s^{j+1}p({\bf{n}}, {\bf{e}},  t_h)\right\} \nonumber \\
    &= \sum_{\bf{n}}\sum_{\bf{e}}n_s^q\sum_{\substack{j=1, \\ j \neq q,  q-1}}^{N_y}\mu_{-}(j+1, N_{s}, e_s)\left\{ {n}_s^{j+1}p({\bf{n}}, {\bf{e}},  t_h)-n_s^{j+1}p({\bf{n}}, {\bf{e}},  t_h)\right\} \nonumber \\
    &=0.
\end{align*}
Then, considering $i = s$ and $j =q$, we see that 
\begin{align*}
    I_5 &= \sum_{\bf{n}}\sum_{\bf{e}}n_s^q\mu_{-}(q+1, N_{s}, e_s)\times \nonumber \\ &\qquad\qquad\left\{ (n_s^{q+1}+1)p([n_s^1, \dots, n_{s}^{q}-1, n_{s}^{q+1}+1, \dots, n_{s}^{N_y+1}], {\bf{e}},   t_h)-n_s^{q+1}p({\bf{n}}, {\bf{e}},  t_h)\right\}. \nonumber 
\end{align*}
Now we want to separate the terms and apply the change of variables $\bar{n}_s^q = n_s^q-1$ and $\bar{n}_s^{q+1}=n_s^{q+1}+1$ in the first term to see 
{\small{
\begin{align*}
    I_5 &= \sum_{\bf{n}}\sum_{\bf{e}}\mu_{-}(q+1, N_{s}, e_s)\times \nonumber \\ &\qquad\qquad\left\{ n_s^q(n_s^{q+1}+1)p([n_s^1, \dots, n_{s}^{q}-1, n_{s}^{q+1}+1, \dots, n_{s}^{N_y+1}], {\bf{e}},   t_h)-n_s^qn_s^{q+1}p({\bf{n}}, {\bf{e}},  t_h)\right\} \nonumber \\
    &= \sum_{\bf{n}}\sum_{\bf{e}}\mu_{-}(q+1, N_{s}, e_s)\times \nonumber \\ &\qquad\qquad\left\{ (\bar{n}_s^q+1) \bar{n}_s^{q+1}p([n_s^1, \dots, \bar{n}_{s}^{q}, \bar{n}_{s}^{q+1}, \dots, n_{s}^{N_y+1}], {\bf{e}},   t_h)-n_s^q n_s^{q+1}p({\bf{n}}, {\bf{e}},  t_h)\right\} \nonumber \\
    &= \sum_{\bf{n}}\sum_{\bf{e}}\mu_{-}(q+1, N_{s}, e_s)\left\{ {n}_s^{q+1}({n}_s^q+1) p({\bf{n}}, {\bf{e}},  t_h)-n_s^qn_s^{q+1}p({\bf{n}}, {\bf{e}},  t_h)\right\} \nonumber \\
    &= \sum_{\bf{n}}\sum_{\bf{e}}\mu_{-}(q+1, N_{s}, e_s) {n}_s^{q+1}p({\bf{n}}, {\bf{e}},  t_h) \nonumber \\
    &= \langle \mu_{-}(q+1, N_{s}, e_s) n_s^{q+1} \rangle.
\end{align*}
}}
Finally, considering the case when $i=s$ and $j=q-1$ we have 
\begin{align}
    I_6 &= \sum_{\bf{n}}\sum_{\bf{e}}n_s^q\mu_{-}(q, N_{s}, e_s)\times \nonumber \\ &\qquad\qquad\left\{ (n_s^{q}+1)p([n_s^1, \dots, n_{s}^{q-1}-1, n_{s}^{q}+1, \dots, n_{s}^{N_y+1}], {\bf{e}},   t_h)-n_s^{q}p({\bf{n}}, {\bf{e}},  t_h)\right\}. \nonumber
\end{align}
Separating out the two terms and applying the change of variables $\bar{n}_s^q = n_s^q +1$ and $\bar{n}_s^{q-1}=n_s^{q-1}-1$ and then dropping the bar{{s}} in the first term, we see 
\begin{align*}
    I_6 &= \sum_{\bf{n}}\sum_{\bf{e}}\mu_{-}(q, N_{s}, e_s)\times \nonumber \\ &\qquad\qquad\left\{n_s^q (n_s^{q}+1)p([n_s^1, \dots, n_{s}^{q-1}-1, n_{s}^{q}+1, \dots, n_{s}^{N_y+1}], {\bf{e}},   t_h)-n_s^q n_s^{q}p({\bf{n}}, {\bf{e}},  t_h)\right\} \nonumber \\
    &= \sum_{\bf{n}}\sum_{\bf{e}}\mu_{-}(q, N_{s}, e_s)\times \nonumber \\ &\qquad\qquad\left\{(\bar{n}_s^q-1) \bar{n}_s^{q}p([n_s^1, \dots, \bar{n}_{s}^{q-1}, \bar{n}_{s}^{q}, \dots, n_{s}^{N_y+1}], {\bf{e}},   t_h)-n_s^q n_s^{q}p({\bf{n}}, {\bf{e}},  t_h)\right\} \nonumber \\
    &= \sum_{\bf{n}}\sum_{\bf{e}}\mu_{-}(q, N_{s}, e_s)\left\{{n}_s^{q}({n}_s^q-1) p({\bf{n}}, {\bf{e}},  t_h)-(n_s^q)^2p({\bf{n}}, {\bf{e}},  t_h)\right\} \nonumber \\
    &= -\sum_{\bf{n}}\sum_{\bf{e}}\mu_{-}(q, N_{s}, e_s){n}_s^{q}
    p({\bf{n}}, {\bf{e}},  t_h) \nonumber \\
    &= - \langle \mu_{-}(q, N_{s}, e_s){n}_s^{q} \rangle.
\end{align*}
Putting these all back together, we find that 
\begin{align*}
    I&=I_1 + I_2 +I_3 + I_4 +I_5 +I_6 \nonumber \\
    &=\langle \mu_{-}(q+1, N_{s}, e_s) n_s^{q+1} \rangle - \langle \mu_{-}(q, N_{s}, e_s){n}_s^{q} \rangle.
\end{align*}
Now we repeat this process for the second term in Eq.~\eqref{SIeq:master}, which we call $J$:
\begin{align*}
    J &= \sum_{\bf{n}}\sum_{\bf{e}}n_s^q\sum_{i=1}^{N_x+1}\sum_{j=2}^{N_y+1}\mu_{+}(j-1, N_i, e_i)\left\{ (n_i^{j-1}+1)p(D_{i,j}^{\text{p}}{\bf{n}}, {\bf{e}},   t_h)-n_i^{j-1}p({\bf{n}}, {\bf{e}},  t_h)\right\}, \\
    &= {{J_1 + J_2 + J_3 + J_4 + J_5 + J_6,}}
\end{align*}
and consider the six cases: $J_1,  \dots,  J_6$ (defined in the same way as before) where we have $i=s,  i\neq s,  j=q,  j=q+1, $ and $j\neq q, q+1.$ 
Starting with $i \neq s$ and $j \neq q,  q+1$ we see, using the change of variables $\bar{n}_i^j = n_i^j -1$ and $\bar{n}_i^{j-1}=n_i^{j-1}+1$, that we have
\begin{align*}
    J_1&= \sum_{\bf{n}}\sum_{\bf{e}}n_s^q\sum_{\substack{i=1, \\ i\neq s}}^{N_x+1}\sum_{\substack{j=2, \\ j \neq q,  q-1}}^{N_y+1}\mu_{+}(j-1, N_i, e_i)\left\{ (n_i^{j-1}+1)p(D_{i,j}^{\text{p}}{\bf{n}}, {\bf{e}},   t_h)-n_i^{j-1}p({\bf{n}}, {\bf{e}},  t_h)\right\} \nonumber \\ 
    &= \sum_{\bf{n}}\sum_{\bf{e}}n_s^q\sum_{\substack{i=1, \\ i\neq s}}^{N_x+1}\sum_{\substack{j=2, \\ j \neq q,  q-1}}^{N_y+1}\mu_{+}(j-1, N_i, e_i)\times \nonumber \\ &\qquad\qquad\Big\{ (n_i^{j-1}+1)p([n_i^1, \dots, n_{i}^{j-1}+1, n_{i}^{j}-1, \dots, n_{i}^{N_y+1}], {\bf{e}},   t_h)-n_i^{j-1}p({\bf{n}}, {\bf{e}},  t_h)\Big\} \nonumber \\
    &= \sum_{\bf{n}}\sum_{\bf{e}}n_s^q\sum_{\substack{i=1, \\ i\neq s}}^{N_x+1}\sum_{\substack{j=2, \\ j \neq q,  q-1}}^{N_y+1}\mu_{+}(j-1, N_i, e_i)\times \nonumber \\ 
    &\qquad\qquad\qquad\qquad\left\{ \bar{n}_i^{j-1}p([n_i^1, \dots, \bar{n}_{i}^{j-1}, \bar{n}_{i}^{j}, \dots, n_{i}^{N_y+1}], {\bf{e}},   t_h)-n_i^{j-1}p({\bf{n}}, {\bf{e}},  t_h)\right\} \nonumber \\
    &= \sum_{\bf{n}}\sum_{\bf{e}}n_s^q\sum_{\substack{i=1, \\ i\neq s}}^{N_x+1}\sum_{\substack{j=2, \\ j \neq q,  q-1}}^{N_y+1}\mu_{+}(j-1, N_i, e_i)\left\{ {n}_i^{j-1}p({\bf{n}}, {\bf{e}},   t_h)-n_i^{j-1}p({\bf{n}}, {\bf{e}},  t_h)\right\} \nonumber \\
    &= 0. 
\end{align*}
Next, we consider the case when $i\neq s$ and $j=q$, and apply the same change of variables argument (with $\bar{n}_i^q = \bar{n}_i^q -1$ and $\bar{n}_i^{q-1} = \bar{n}_i^{q-1} +1$) to the first term:
\begin{align*}
    J_2 &=  \sum_{\bf{n}}\sum_{\bf{e}}n_s^q\sum_{\substack{i=1, \\ i\neq s}}^{N_x+1}\mu_{+}(q-1, N_i, e_i)\times \nonumber \\ &\qquad\qquad\left\{ (n_i^{q-1}+1)p([n_i^1, \dots, {n}_{i}^{q-1}+1, {n}_{i}^{q}-1, \dots, n_{i}^{N_y+1}], {\bf{e}},   t_h)-n_i^{q-1}p({\bf{n}}, {\bf{e}},  t_h)\right\} \nonumber \\
    &=  \sum_{\bf{n}}\sum_{\bf{e}}n_s^q\sum_{\substack{i=1, \\ i\neq s}}^{N_x+1}\mu_{+}(q-1, N_i, e_i)\times \nonumber \\ &\qquad\qquad\left\{ \bar{n}_i^{q-1}p([n_i^1, \dots, \bar{n}_{i}^{q-1}, \bar{n}_{i}^{q}, \dots, n_{i}^{N_y+1}], {\bf{e}},   t_h)-n_i^{q-1}p({\bf{n}}, {\bf{e}},  t_h)\right\} \nonumber \\
    &= \sum_{\bf{n}}\sum_{\bf{e}}n_s^q\sum_{\substack{i=1, \\ i\neq s}}^{N_x+1}\mu_{+}(q-1, N_i, e_i)\left\{ {n}_i^{q-1}p({\bf{n}}, {\bf{e}},  t_h)-n_i^{q-1}p({\bf{n}}, {\bf{e}},  t_h)\right\} \nonumber \\
    &=0.
\end{align*}
Finally we consider the last remaining situation when $i\neq s$, where $j=q+1.$
Once again, by changing the variables using $\bar{n}_i^{q+1}=n_i^{q+1}-1$ and $\bar{n}_i^q=n_i^q+1,$ we have 
{\small{
\begin{align*}
    J_3 &=  \sum_{\bf{n}}\sum_{\bf{e}}n_s^q\sum_{\substack{i=1, \\ i\neq s}}^{N_x+1}\mu_{+}(q, N_i, e_i)\times \nonumber \\ &\qquad\qquad\left\{ (n_i^{q}+1)p([n_i^1, \dots, {n}_{i}^{q}+1, {n}_{i}^{q+1}-1, \dots, n_{i}^{N_y+1}], {\bf{e}},  t_h)-n_i^{q}p({\bf{n}}, {\bf{e}},  t_h)\right\} \nonumber \\
    &=  \sum_{\bf{n}}\sum_{\bf{e}}n_s^q\sum_{\substack{i=1, \\ i\neq s}}^{N_x+1}\mu_{+}(q, N_i, e_i)\left\{ \bar{n}_i^{q}p([n_i^1, \dots, \bar{n}_{i}^{q}, \bar{n}_{i}^{q+1}, \dots, n_{i}^{N_y+1}], {\bf{e}},  t_h)-n_i^{q}p({\bf{n}}, {\bf{e}},  t_h)\right\} \nonumber \\
    &= \sum_{\bf{n}}\sum_{\bf{e}}n_s^q\sum_{\substack{i=1, \\ i\neq s}}^{N_x+1}\mu_{+}(q, N_i, e_i)\left\{ {n}_i^{q}p({\bf{n}}, {\bf{e}},  t_h)-n_i^{q}p({\bf{n}}, {\bf{e}},  t_h)\right\} \nonumber \\
    &=0.
\end{align*}
}}
As such, we can see that there are no contributions arising from the terms when $i\neq s.$

Now we can consider cases where $i = s.$ 
First, we consider $ j \neq q,  q+1$, and apply the change of variables $\bar{n}_s^j = n_s^j -1$ and $\bar{n}_s^{j-1} =  n_s^{j-1}+1$:
\begin{align*}
    J_4 &= \sum_{\bf{n}}\sum_{\bf{e}}n_s^q\sum_{\substack{j=2, \\ j \neq q,  q-1}}^{N_y+1}\mu_{+}(j-1, N_{s}, e_s)\times \nonumber \\ &\qquad\qquad\Big\{ (n_s^{j-1}+1)p([n_s^1, \dots, n_{s}^{j-1}+1, n_{s}^{j}-1, \dots, n_{s}^{N_y+1}], {\bf{e}},  t_h)-n_s^{j-1}p({\bf{n}}, {\bf{e}},  t_h)\Big\} \nonumber \\
    &= \sum_{\bf{n}}\sum_{\bf{e}}n_s^q\sum_{\substack{j=2, \\ j \neq q,  q-1}}^{N_y+1}\mu_{+}(j-1, N_{s}, e_s)\times \nonumber \\ &\qquad\qquad\left\{ \bar{n}_s^{j-1}p([n_s^1, \dots, \bar{n}_{s}^{j-1}, \bar{n}_{s}^{j}, \dots, n_{s}^{N_y+1}], {\bf{e}},  t_h)-n_s^{j-1}p({\bf{n}}, {\bf{e}},  t_h)\right\} \nonumber \\
    &= \sum_{\bf{n}}\sum_{\bf{e}}n_s^q\sum_{\substack{j=2, \\ j \neq q,  q-1}}^{N_y+1}\mu_{+}(j-1, N_{s}, e_s)\left\{ {n}_s^{j-1}p({\bf{n}}, {\bf{e}},  t_h)-n_s^{j-1}p({\bf{n}}, {\bf{e}},  t_h)\right\} \nonumber \\
    &=0.
\end{align*}
Then, considering the case when $i = s$ and $j = q$, we see that applying the change of variables $\bar{n}_s^q = n_s^q-1$ and $\bar{n}_s^{q-1}=n_s^{q-1}+1$ in the first term, we have 
{\small{
\begin{align*}
    J_5 &= \sum_{\bf{n}}\sum_{\bf{e}}n_s^q\mu_{+}(q-1, N_{s}, e_s)\times \nonumber \\ &\qquad\qquad\left\{ (n_s^{q-1}+1)p([n_s^1, \dots, n_{s}^{q-1}+1, n_{s}^{q}-1, \dots, n_{s}^{N_y+1}], {\bf{e}},  t_h)-n_s^{q-1}p({\bf{n}}, {\bf{e}},  t_h)\right\} \nonumber \\
    &= \sum_{\bf{n}}\sum_{\bf{e}}\mu_{+}(q-1, N_{s}, e_s)\times \nonumber \\ &\qquad\qquad\left\{ n_s^q(n_s^{q-1}+1)p([n_s^1, \dots, n_{s}^{q-1}+1, n_{s}^{q}-1, \dots, n_{s}^{N_y+1}], {\bf{e}},  t_h)-n_s^qn_s^{q-1}p({\bf{n}}, {\bf{e}},  t_h)\right\} \nonumber \\
    &= \sum_{\bf{n}}\sum_{\bf{e}}\mu_{+}(q-1, N_{s}, e_s)\times \nonumber \\ &\qquad\qquad\left\{ (\bar{n}_s^q+1) \bar{n}_s^{q-1}p([n_s^1, \dots, \bar{n}_{s}^{q-1}, \bar{n}_{s}^{q}, \dots, n_{s}^{N_y+1}], {\bf{e}},  t_h)-n_s^q n_s^{q-1}p({\bf{n}}, {\bf{e}},  t_h)\right\} \nonumber \\
    &= \sum_{\bf{n}}\sum_{\bf{e}}\mu_{+}(q-1, N_{s}, e_s)\left\{ {n}_s^{q-1}({n}_s^q+1) p({\bf{n}}, {\bf{e}},  t_h)-n_s^qn_s^{q-1}p({\bf{n}}, {\bf{e}},  t_h)\right\} \nonumber \\
    &= \sum_{\bf{n}}\sum_{\bf{e}}\mu_{+}(q-1, N_{s}, e_s) {n}_s^{q-1}p({\bf{n}}, {\bf{e}},  t_h) \nonumber \\
    &= \langle \mu_{+}(q-1, N_{s}, e_s) n_s^{q-1} \rangle.
\end{align*}
}}
Finally, considering the case when $i=s$ and $j=q+1$ we consider term $J_6$.
Using the change of variables $\bar{n}_s^q = n_s^q +1$ and $\bar{n}_s^{q+1}=n_s^{q+1}-1$ and then dropping the bar{{s}} in the first term, we see that
\begin{align*}
    J_6 &= \sum_{\bf{n}}\sum_{\bf{e}}n_s^q\mu_{+}(q, N_{s}, e_s)\times \nonumber \\ &\qquad\qquad\left\{ (n_s^{q}+1)p([n_s^1, \dots, n_{s}^{q}+1, n_{s}^{q+1}-1, \dots, n_{s}^{N_y+1}], {\bf{e}},  t_h)-n_s^{q}p({\bf{n}}, {\bf{e}},  t_h)\right\} \nonumber \\
    &= \sum_{\bf{n}}\sum_{\bf{e}}\mu_{+}(q, N_{s}, e_s)\times \nonumber \\ &\qquad\qquad\left\{n_s^q (n_s^{q}+1)p([n_s^1, \dots, n_{s}^{q}+1, n_{s}^{q+1}-1, \dots, n_{s}^{N_y+1}], {\bf{e}},  t_h)-n_s^q n_s^{q}p({\bf{n}}, {\bf{e}},  t_h)\right\} \nonumber \\
    &= \sum_{\bf{n}}\sum_{\bf{e}}\mu_{+}(q, N_{s}, e_s)\times \nonumber \\ &\qquad\qquad\left\{(\bar{n}_s^q-1) \bar{n}_s^{q}p([n_s^1, \dots, \bar{n}_{s}^{q}, \bar{n}_{s}^{q+1}, \dots, n_{s}^{N_y+1}], {\bf{e}},  t_h)-n_s^q n_s^{q}p({\bf{n}}, {\bf{e}},  t_h)\right\} \nonumber \\
    &= \sum_{\bf{n}}\sum_{\bf{e}}\mu_{+}(q, N_{s}, e_s)\left\{{n}_s^{q}({n}_s^q-1) p({\bf{n}}, {\bf{e}},  t_h)-(n_s^q)^2p({\bf{n}}, {\bf{e}},  t_h)\right\} \nonumber \\
    &= -\sum_{\bf{n}}\sum_{\bf{e}}\mu_{+}(q, N_{s}, e_s){n}_s^{q}
    p({\bf{n}}, {\bf{e}},  t_h) \nonumber \\
    &= - \langle \mu_{+}(q, N_{s}, e_s){n}_s^{q} \rangle.
\end{align*}
Putting these all back together, it is clear that 
\begin{align*}
    J&=J_1 + J_2 +J_3 + J_4 +J_5 +J_6 \nonumber \\
    &=\langle \mu_{+}(q-1, N_{s}, e_s) n_s^{q-1} \rangle - \langle \mu_{+}(q, N_{s}, e_s){n}_s^{q} \rangle.
\end{align*}

Next, we consider the terms modelling movement in physical space in time step $\Delta_t.$ 
Returning to Eq.~\eqref{SIeq:master}, we look first at the third term on the right-hand side, which we call $K$. 
For this term, we once again consider the cases $j=q,  j\neq q,  i=s,  i=s-1$ and $i\neq s, s-1$ in turn, so 
\begin{align*}
    K &= \sum_{\bf{n}}\sum_{\bf{e}}n_s^q\sum_{i=1}^{N_x}\sum_{j=1}^{N_y+1}\beta_{-}(j, {N_i,e_i)}\left\{ (n_{i+1}^{j}+1)p(R_{i,j}^{\text{m}}{\bf{n}}, {\bf{e}},  t_h)-n_{i+1}^{j}p({\bf{n}}, {\bf{e}},  t_h)\right\} \nonumber \\ 
    &= K_1 + K_2 +K_3 + K_4 + K_5 + K_6.
\end{align*}
Considering first the case where $j\neq q $ and $i \neq s, s-1$, and applying the change of variables $\bar{n}_i^j = {n}_i^j -1$ and $\bar{n}_{i+1}^j = {n}_{i+1}^j +1$ and dropping the bars, we have 
\begin{align*}
    K_1 &= \sum_{\bf{n}}\sum_{\bf{e}}n_s^q\sum_{\substack{i=1, \\ i\neq s,  s-1}}^{N_x}\sum_{\substack{j=1,  \\j\neq q}}^{N_y+1}\beta_{-}(j, {N_i,e_i)}\times \nonumber \\ &\qquad\qquad\Big\{ (n_{i+1}^{j}+1)p([n_1^j, \dots, n_{i}^j-1, n_{i+1}^{j}+1, \dots, n_{N_x+1}^j], {\bf{e}},  t_h)-n_{i+1}^{j}p({\bf{n}}, {\bf{e}},  t_h)\Big\} \nonumber \\ 
    &= \sum_{\bf{n}}\sum_{\bf{e}}n_s^q\sum_{\substack{i=1, \\ i\neq s,  s-1}}^{N_x}\sum_{\substack{j=1,  \\j\neq q}}^{N_y+1}\beta_{-}(j, {N_i,e_i)}\times \nonumber \\ &\qquad\qquad\left\{ \bar{n}_{i+1}^{j}p([n_1^j, \dots, \bar{n}_{i}^j, \bar{n}_{i+1}^{j}, \dots, n_{N_x+1}^j], {\bf{e}},  t_h)-n_{i+1}^{j}p({\bf{n}}, {\bf{e}},  t_h)\right\} \nonumber \\ 
    &= \sum_{\bf{n}}\sum_{\bf{e}}n_s^q\sum_{\substack{i=1, \\ i\neq s,  s-1}}^{N_x}\sum_{\substack{j=1,  \\j\neq q}}^{N_y+1}\beta_{-}(j, {N_i,e_i)}\left\{ {n}_{i+1}^{j}p({\bf{n}}, {\bf{e}},  t_h)-n_{i+1}^{j}p({\bf{n}}, {\bf{e}},  t_h)\right\} \nonumber \\ 
    &= 0. 
\end{align*}
Now consider $ j \neq q$ with $i = s.$
Applying the change of variables $\bar{n}_s^j = {n}_s^j -1$ and $\bar{n}_{s+1}^j = {n}_{s+1}^j +1$ and dropping the bars, we have 
\begin{align*}
    K_2 &= \sum_{\bf{n}}\sum_{\bf{e}}n_s^q\sum_{\substack{j=1,  \\j\neq q}}^{N_y+1}\beta_{-}(j, {N_{s},e_s)}\times \nonumber \\ &\qquad\qquad\left\{ (n_{s+1}^{j}+1)p([n_1^j, \dots, n_{s}^j-1, n_{s+1}^{j}+1, \dots, n_{N_x+1}^j], {\bf{e}},  t_h)-n_{s+1}^{j}p({\bf{n}}, {\bf{e}},  t_h)\right\} \nonumber \\ 
    &= \sum_{\bf{n}}\sum_{\bf{e}}n_s^q\sum_{\substack{j=1,  \\j\neq q}}^{N_y+1}\beta_{-}(j, {N_{s},e_s)}\times \nonumber \\ &\qquad\qquad\left\{ \bar{n}_{s+1}^{j}p([n_1^j, \dots, \bar{n}_{s}^j, \bar{n}_{s+1}^{j}, \dots, n_{N_x+1}^j], {\bf{e}},  t_h)-n_{s+1}^{j}p({\bf{n}}, {\bf{e}},  t_h)\right\} \nonumber \\ 
    &= \sum_{\bf{n}}\sum_{\bf{e}}n_s^q\sum_{\substack{j=1,  \\j\neq q}}^{N_y+1}\beta_{-}(j, {N_{s},e_s)}\left\{ {n}_{s+1}^{j}p({\bf{n}}, {\bf{e}},  t_h)-n_{s+1}^{j}p({\bf{n}}, {\bf{e}},  t_h)\right\} \nonumber \\ 
    &= 0. 
\end{align*}
Next we look at $j \neq q$ with $i = s-1.$ 
If we apply the change of variables in the first term and drop the bars ($\bar{n}_s^j = {n}_s^j +1$ and $\bar{n}_{s-1}^j = {n}_{s-1}^j -1$), we have
\begin{align*}
    K_3 &=  \sum_{\bf{n}}\sum_{\bf{e}}n_s^q\sum_{\substack{j=1,  \\j\neq q}}^{N_y+1}\beta_{-}(j, {N_{s-1},e_{s-1})}\times \nonumber \\ &\qquad\qquad\left\{ (n_{s}^{j}+1)p([n_1^j, \dots, n_{s-1}^j-1, n_{s}^{j}+1, \dots, n_{N_x+1}^j], {\bf{e}},  t_h)-n_{s}^{j}p({\bf{n}}, {\bf{e}},  t_h)\right\} \nonumber \\ 
    &= \sum_{\bf{n}}\sum_{\bf{e}}n_s^q\sum_{\substack{j=1,  \\j\neq q}}^{N_y+1}\beta_{-}(j, {N_{s-1},e_{s-1})}\times \nonumber \\ &\qquad\qquad\left\{ \bar{n}_{s}^{j}p([n_1^j, \dots, \bar{n}_{s-1}^j, \bar{n}_{s}^{j}, \dots, n_{N_x+1}^j], {\bf{e}},  t_h)-n_{s}^{j}p({\bf{n}}, {\bf{e}},  t_h)\right\} \nonumber \\ 
    &= \sum_{\bf{n}}\sum_{\bf{e}}n_s^q\sum_{\substack{j=1,  \\j\neq q}}^{N_y+1}\beta_{-}(j,  {N_{s-1},e_{s-1})}\left\{ {n}_{s}^{j}p({\bf{n}}, {\bf{e}},  t_h)-n_{s}^{j}p({\bf{n}}, {\bf{e}},  t_h)\right\} \nonumber \\ 
    &= 0. 
\end{align*}
Now we consider the three cases when $j = q$. 
First, when $i \neq s, s-1$, we apply the change of variables $\bar{n}_i^q = {n}_i^q -1$ and $\bar{n}_{i+1}^q = {n}_{i+1}^q +1$ and dropping the bars, we have 
\begin{align*}
    K_4 &= \sum_{\bf{n}}\sum_{\bf{e}}n_s^q\sum_{\substack{i=1, \\ i\neq s,  s-1}}^{N_x}\beta_{-}(q, {N_i,e_i)}\times \nonumber \\ &\qquad\qquad\left\{ (n_{i+1}^{q}+1)p([n_1^q, \dots, n_{i}^q-1, n_{i+1}^{q}+1, \dots, n_{N_x+1}^q], {\bf{e}},  t_h)-n_{i+1}^{q}p({\bf{n}}, {\bf{e}},  t_h)\right\} \nonumber \\ 
    &= \sum_{\bf{n}}\sum_{\bf{e}}n_s^q\sum_{\substack{i=1, \\ i\neq s,  s-1}}^{N_x}\beta_{-}(q, {N_i,e_i)}\times \nonumber \\ &\qquad\qquad\left\{ \bar{n}_{i+1}^{q}p([n_1^q, \dots, \bar{n}_{i}^q, \bar{n}_{i+1}^{q}, \dots, n_{N_x+1}^q], {\bf{e}},  t_h)-n_{i+1}^{q}p({\bf{n}}, {\bf{e}},  t_h)\right\} \nonumber \\ 
    &= \sum_{\bf{n}}\sum_{\bf{e}}n_s^q\sum_{\substack{i=1, \\ i\neq s,  s-1}}^{N_x}\beta_{-}(q, {N_i,e_i)}\left\{ {n}_{i+1}^{q}p({\bf{n}}, {\bf{e}},  t_h)-n_{i+1}^{q}p({\bf{n}}, {\bf{e}},  t_h)\right\} \nonumber \\ 
    &= 0. 
\end{align*}
Now consider $i = s$ and the change of variables $\bar{n}_s^q = {n}_s^q -1$ and $\bar{n}_{s+1}^q = {n}_{s+1}^q +1$ to give
\begin{align*}
    K_5 &= \sum_{\bf{n}}\sum_{\bf{e}}n_s^q\beta_{-}(q, {N_{s},e_s)}\times \nonumber \\ &\qquad\qquad\left\{ (n_{s+1}^{q}+1)p([n_1^q, \dots, n_{s}^q-1, n_{s+1}^{q}+1, \dots, n_{N_x+1}^q], {\bf{e}},  t_h)-n_{s+1}^{q}p({\bf{n}}, {\bf{e}},  t_h)\right\} \nonumber \\ 
    &= \sum_{\bf{n}}\sum_{\bf{e}}\beta_{-}(q, {N_{s},e_s)}\times \nonumber \\ &\qquad\qquad\left\{ (\bar{n}_s^q+1)\bar{n}_{s+1}^{q}p([n_1^q, \dots, \bar{n}_{s}^q, \bar{n}_{s+1}^{q}, \dots, n_{N_x+1}^q], {\bf{e}},  t_h)-n_s^qn_{s+1}^{q}p({\bf{n}}, {\bf{e}},  t_h)\right\} \nonumber \\ 
    &= \sum_{\bf{n}}\sum_{\bf{e}}\beta_{-}(q, {N_{s},e_s)}\left\{ ({n}_s^q+1){n}_{s+1}^{q}p({\bf{n}}, {\bf{e}},  t_h)-n_s^q n_{s+1}^{q}p({\bf{n}}, {\bf{e}},  t_h)\right\} \nonumber \\ 
    &= \sum_{\bf{n}}\sum_{\bf{e}}\beta_{-}(q, {N_{s},e_s)}{n}_{s+1}^{q}p({\bf{n}}, {\bf{e}},  t_h) \nonumber \\ 
    &= \langle \beta_{-}(q, {N_{s},e_s)}{n}_{s+1}^{q} \rangle.
\end{align*}
Finally, consider $i = s -1$ with the change of variables $\bar{n}_{s-1}^q = {n}_{s-1}^q -1$ and $\bar{n}_{s}^q = {n}_{s}^q +1$ to give
\begin{align*}
    K_6 &= \sum_{\bf{n}}\sum_{\bf{e}}n_s^q\beta_{-}(q, {N_{s-1}, e_{s-1})}\times \nonumber \\ &\qquad\qquad\left\{ (n_{s}^{q}+1)p([n_1^q, \dots, n_{s-1}^q-1, n_{s}^{q}+1, \dots, n_{N_x+1}^q], {\bf{e}},  t_h)-n_{s}^{q}p({\bf{n}}, {\bf{e}},  t_h)\right\} \nonumber \\ 
    &= \sum_{\bf{n}}\sum_{\bf{e}}\beta_{-}(q, {N_{s-1}, e_{s-1})}\times \nonumber \\ &\qquad\qquad\left\{ (\bar{n}_s^q-1)\bar{n}_{s}^{q}p([n_1^q, \dots, \bar{n}_{s-1}^q, \bar{n}_{s}^{q}, \dots, n_{N_x+1}^q], {\bf{e}},  t_h)-n_s^qn_{s}^{q}p({\bf{n}}, {\bf{e}},  t_h)\right\} \nonumber \\ 
    &= \sum_{\bf{n}}\sum_{\bf{e}}\beta_{-}(q, {N_{s-1}, e_{s-1})}\left\{ ({n}_s^q-1){n}_{s}^{q}p({\bf{n}}, {\bf{e}},  t_h)-n_s^q n_{s}^{q}p({\bf{n}}, {\bf{e}},  t_h)\right\} \nonumber \\ 
    &= -\sum_{\bf{n}}\sum_{\bf{e}}\beta_{-}(q, {N_{s-1}, e_{s-1})}{n}_{s}^{q}p({\bf{n}}, {\bf{e}},  t_h) \nonumber \\ 
    &= -\langle \beta_{-}(q, {N_{s-1}, e_{s-1})}{n}_{s}^{q} \rangle.
\end{align*}
Bringing back together these terms, we have
\begin{align*}
    K &= K_1 + K_2 + K_3 + K_4 + K_5 + K_6 \nonumber \\
    &= \langle \beta_{-}(q, {N_{s},e_s)}{n}_{s+1}^{q} \rangle -\langle \beta_{-}(q, {N_{s-1},e_{s-1})}{n}_{s}^{q} \rangle.
\end{align*}
For the second movement term, we repeat this process, by breaking it down into six cases, defined by $j=q,  j\neq q,  i=s,  i=s+1$ and $i\neq s, s+1$:
\begin{align*}
    \mathcal{L} &= \sum_{\bf{n}}\sum_{\bf{e}}n_s^q\sum_{i=2}^{N_x+1}\sum_{j=1}^{N_y+1}\beta_{+}(j, N_i,e_i)\left\{ (n_{i-1}^{j}+1)p(L_{i,j}^{\text{m}}{\bf{n}}, {\bf{e}},  t_h)-n_{i-1}^{j}p({\bf{n}}, {\bf{e}},  t_h)\right\}\nonumber \\ 
    &= \mathcal{L}_1 + \mathcal{L}_2 + \mathcal{L}_3 +\mathcal{L}_4 + \mathcal{L}_5 + \mathcal{L}_6.
\end{align*}

Then we start by considering $j \neq q $ and $ i \neq s,  s+1.$ 
Employing the change of variables $\bar{n}_i^j = {n}_i^j -1$ and $\bar{n}_{i-1}^j = {n}_{i-1}^j +1$ and dropping the bars, we have 
\begin{align*}
    \mathcal{L}_1 &= \sum_{\bf{n}}\sum_{\bf{e}}n_s^q\sum_{\substack{i=1,\\  i\neq s,  s+1}}^{N_x}\sum_{\substack{j=1,  \\j\neq q}}^{N_y+1}\beta_{+}(j, {N_i,e_i)}\times \nonumber \\ &\qquad\qquad\left\{ (n_{i-1}^{j}+1)p([n_1^j, \dots, n_{i-1}^j+1, n_{i}^{j}-1, \dots, n_{N_x+1}^j], {\bf{e}},  t_h)-n_{i-1}^{j}p({\bf{n}}, {\bf{e}},  t_h)\right\} \nonumber \\ 
    &= \sum_{\bf{n}}\sum_{\bf{e}}n_s^q\sum_{\substack{i=1,\\  i\neq s,  s+1}}^{N_x}\sum_{\substack{j=1,  \\j\neq q}}^{N_y+1}\beta_{+}(j, {N_i,e_i)}\times \nonumber \\ &\qquad\qquad\left\{ \bar{n}_{i-1}^{j}p([n_1^j, \dots, \bar{n}_{i-1}^j, \bar{n}_{i}^{j}, \dots, n_{N_x+1}^j], {\bf{e}},  t_h)-n_{i-1}^{j}p({\bf{n}}, {\bf{e}},  t_h)\right\} \nonumber \\ 
    &= \sum_{\bf{n}}\sum_{\bf{e}}n_s^q\sum_{\substack{i=1,\\  i\neq s,  s+1}}^{N_x}\sum_{\substack{j=1,  \\j\neq q}}^{N_y+1}\beta_{+}(j, {N_i,e_i)}\left\{ {n}_{i-1}^{j}p({\bf{n}}, {\bf{e}},  t_h)-n_{i-1}^{j}p({\bf{n}}, {\bf{e}},  t_h)\right\} \nonumber \\ 
    &= 0. 
\end{align*}
Now consider $ j \neq q$ with $i = s.$
Then, applying the change of variables $\bar{n}_s^j = {n}_s^j -1$ and $\bar{n}_{s-1}^j = {n}_{s-1}^j +1$ and dropping the bars, we have 
\begin{align*}
    \mathcal{L}_2 &= \sum_{\bf{n}}\sum_{\bf{e}}n_s^q\sum_{\substack{j=1,  \\j\neq q}}^{N_y+1}\beta_{+}(j, {N_{s},e_s)}\times \nonumber \\ &\qquad\qquad\left\{ (n_{s-1}^{j}+1)p([n_1^j, \dots, n_{s-1}^j+1, n_{s}^{j}-1, \dots, n_{N_x+1}^j], {\bf{e}},  t_h)-n_{s+1}^{j}p({\bf{n}}, {\bf{e}},  t_h)\right\} \nonumber \\ 
    &= \sum_{\bf{n}}\sum_{\bf{e}}n_s^q\sum_{\substack{j=1,  \\j\neq q}}^{N_y+1}\beta_{+}(j, {N_{s},e_s)}\times \nonumber \\ &\qquad\qquad\left\{ \bar{n}_{s-1}^{j}p([n_1^j, \dots, \bar{n}_{s-1}^j, \bar{n}_{s}^{j}, \dots, n_{N_x+1}^j], {\bf{e}},  t_h)-n_{s-1}^{j}p({\bf{n}}, {\bf{e}},  t_h)\right\} \nonumber \\ 
    &= \sum_{\bf{n}}\sum_{\bf{e}}n_s^q\sum_{\substack{j=1,  \\j\neq q}}^{N_y+1}\beta_{+}(j, {N_{s},e_s)}\left\{ {n}_{s+1}^{j}p({\bf{n}}, {\bf{e}},  t_h)-n_{s-1}^{j}p({\bf{n}}, {\bf{e}},  t_h)\right\} \nonumber \\ 
    &= 0. 
\end{align*}
Next we look at $j \neq q$ with $i = s+1.$ 
If we apply the change of variables in the first term and drop the bars ($\bar{n}_s^j = {n}_s^j +1$ and $\bar{n}_{s+1}^j = {n}_{s+1}^j -1$), we have
\begin{align*}
    \mathcal{L}_3 &=  \sum_{\bf{n}}\sum_{\bf{e}}n_s^q\sum_{\substack{j=1,  \\j\neq q}}^{N_y+1}\beta_{+}(j, {N_{s-1},e_{s-1})}\times \nonumber \\ &\qquad\qquad\left\{ (n_{s}^{j}+1)p([n_1^j, \dots, n_{s}^j+1, n_{s+1}^{j}-1, \dots, n_{N_x+1}^j], {\bf{e}},  t_h)-n_{s}^{j}p({\bf{n}}, {\bf{e}},  t_h)\right\} \nonumber \\ 
    &= \sum_{\bf{n}}\sum_{\bf{e}}n_s^q\sum_{\substack{j=1,  \\j\neq q}}^{N_y+1}\beta_{+}(j, {N_{s-1},e_{s-1})}\times \nonumber \\ &\qquad\qquad\left\{ \bar{n}_{s}^{j}p([n_1^j, \dots, \bar{n}_{s}^j, \bar{n}_{s+1}^{j}, \dots, n_{N_x+1}^j], {\bf{e}},  t_h)-n_{s}^{j}p({\bf{n}}, {\bf{e}},  t_h)\right\} \nonumber \\ 
    &= \sum_{\bf{n}}\sum_{\bf{e}}n_s^q\sum_{\substack{j=1,  \\j\neq q}}^{N_y+1}\beta_{+}(j,  {N_{s-1},e_{s-1})}\left\{ {n}_{s}^{j}p({\bf{n}}, {\bf{e}},  t_h)-n_{s}^{j}p({\bf{n}}, {\bf{e}},  t_h)\right\} \nonumber \\ 
    &= 0. 
\end{align*}
Now we consider the three cases when $j = q$. 
First, when $i \neq s, s+1$, applying the change of variables $\bar{n}_i^q = {n}_i^q -1$ and $\bar{n}_{i-1}^q = {n}_{i-1}^q +1$ and dropping the bars, we have 
\begin{align*}
    \mathcal{L}_4 &= \sum_{\bf{n}}\sum_{\bf{e}}n_s^q\sum_{\substack{i=1, \\ i\neq s,  s-1}}^{N_x}\beta_{+}(q, {N_i,e_i)}\times \nonumber \\ &\qquad\qquad\left\{ (n_{i-1}^{q}+1)p([n_1^q, \dots, n_{i-1}^q+1, n_{i}^{q}-1, \dots, n_{N_x+1}^q], {\bf{e}},  t_h)-n_{i-1}^{q}p({\bf{n}}, {\bf{e}},  t_h)\right\} \nonumber \\ 
    &= \sum_{\bf{n}}\sum_{\bf{e}}n_s^q\sum_{\substack{i=1, \\ i\neq s,  s-1}}^{N_x}\beta_{+}(q, {N_i,e_i)}\times \nonumber \\ &\qquad\qquad\left\{ \bar{n}_{i-1}^{q}p([n_1^q, \dots, \bar{n}_{i-1}^q, \bar{n}_{i}^{q}, \dots, n_{N_x+1}^q], {\bf{e}},  t_h)-n_{i-1}^{q}p({\bf{n}}, {\bf{e}},  t_h)\right\} \nonumber \\ 
    &= \sum_{\bf{n}}\sum_{\bf{e}}n_s^q\sum_{\substack{i=1, \\ i\neq s,  s-1}}^{N_x}\beta_{+}(q, {N_i,e_i)}\left\{ {n}_{i-1}^{q}p({\bf{n}}, {\bf{e}},  t_h)-n_{i-1}^{q}p({\bf{n}}, {\bf{e}},  t_h)\right\} \nonumber \\ 
    &= 0. 
\end{align*}
Now for $i = s$ and the change of variables $\bar{n}_s^q = {n}_s^q -1$ and $\bar{n}_{s-1}^q = {n}_{s-1}^q +1$, we have
\begin{align*}
    \mathcal{L}_5 &= \sum_{\bf{n}}\sum_{\bf{e}}n_s^q\beta_{+}(q, {N_{s},e_s)}\times \nonumber \\ &\qquad\qquad\left\{ (n_{s-1}^{q}+1)p([n_1^q, \dots, n_{s-1}^q+1, n_{s}^{q}-1, \dots, n_{N_x+1}^q], {\bf{e}},  t_h)-n_{s-1}^{q}p({\bf{n}}, {\bf{e}},  t_h)\right\} \nonumber \\ 
    &= \sum_{\bf{n}}\sum_{\bf{e}}\beta_{+}(q, {N_{s},e_s)}\times \nonumber \\ &\qquad\qquad\left\{ (\bar{n}_s^q+1)\bar{n}_{s-1}^{q}p([n_1^q, \dots, \bar{n}_{s-1}^q, \bar{n}_{s}^{q}, \dots, n_{N_x+1}^q], {\bf{e}},  t_h)-n_s^qn_{s}^{q}p({\bf{n}}, {\bf{e}},  t_h)\right\} \nonumber \\ 
    &= \sum_{\bf{n}}\sum_{\bf{e}}\beta_{+}(q, {N_{s},e_s)}\left\{ ({n}_s^q+1){n}_{s-1}^{q}p({\bf{n}}, {\bf{e}},  t_h)-n_s^q n_{s-1}^{q}p({\bf{n}}, {\bf{e}},  t_h)\right\} \nonumber \\ 
    &= \sum_{\bf{n}}\sum_{\bf{e}}\beta_{+}(q, {N_{s},e_s)}{n}_{s-1}^{q}p({\bf{n}}, {\bf{e}},  t_h) \nonumber \\ 
    &= \langle \beta_{+}(q, {N_{s},e_s)}{n}_{s-1}^{q} \rangle.
\end{align*}
Finally, for $i = s + 1$ with the change of variables $\bar{n}_{s+1}^q = {n}_{s+1}^q -1$ and $\bar{n}_{s}^q = {n}_{s}^q +1$, we have
\begin{align*}
    \mathcal{L}_6 &= \sum_{\bf{n}}\sum_{\bf{e}}n_s^q\beta_{+}(q, {N_{s+1},e_{s+1})}\times \nonumber \\ &\qquad\qquad\left\{ (n_{s}^{q}+1)p([n_1^q, \dots, n_{s}^q+1, n_{s+1}^{q}-1, \dots, n_{N_x+1}^q], {\bf{e}},  t_h)-n_{s}^{q}p({\bf{n}}, {\bf{e}},  t_h)\right\} \nonumber \\ 
    &= \sum_{\bf{n}}\sum_{\bf{e}}\beta_{+}(q, {N_{s+1},e_{s+1})}\times \nonumber \\ &\qquad\qquad\left\{ (\bar{n}_s^q-1)\bar{n}_{s}^{q}p([n_1^q, \dots, \bar{n}_{s}^q, \bar{n}_{s+1}^{q}, \dots, n_{N_x+1}^q], {\bf{e}},  t_h)-n_s^qn_{s}^{q}p({\bf{n}}, {\bf{e}},  t_h)\right\} \nonumber \\ 
    &= \sum_{\bf{n}}\sum_{\bf{e}}\beta_{+}(q, {N_{s+1},e_{s+1})}\left\{ ({n}_s^q-1){n}_{s}^{q}p({\bf{n}}, {\bf{e}},  t_h)-n_s^q n_{s}^{q}p({\bf{n}}, {\bf{e}},  t_h)\right\} \nonumber \\ 
    &= -\sum_{\bf{n}}\sum_{\bf{e}}\beta_{+}(q, {N_{s+1},e_{s+1})}{n}_{s}^{q}p({\bf{n}}, {\bf{e}},  t_h) \nonumber \\ 
    &= -\langle \beta_{+}(q, {N_{s+1},e_{s+1})}{n}_{s}^{q} \rangle.
\end{align*}
Bringing back together these terms, we have
\begin{align*}
    \mathcal{L} &= \mathcal{L}_1 + \mathcal{L}_2 + \mathcal{L}_3 + \mathcal{L}_4 + \mathcal{L}_5 + \mathcal{L}_6 \nonumber \\
    &= \langle \beta_{+}(q, {N_{s},e_s)}{n}_{s-1}^{q} \rangle -\langle \beta_{+}(q, {N_{s+1},e_{s+1})}{n}_{s}^{q} \rangle.
\end{align*}
Finally, we must consider the last term in Eq.~\eqref{SIeq:master}
\begin{align*}
    M&=\sum_{\bf{n}}\sum_{\bf{e}}n_s^q\sum_{i=1}^{N_x+1}\sum_{j=1}^{N_y+1}\Big\{\gamma(j, N_i-1,e_i)(n_i^j-1)p(G_{i,j}{\bf{n}}, {\bf{e}},  t_h)\nonumber\\&\qquad\qquad\qquad\qquad\qquad\qquad\qquad\qquad\qquad\qquad\qquad\qquad\qquad-\gamma(j, N_i, e_i)n_i^jp({\bf{n}}, {\bf{e}},  t_h)\Big\} \nonumber \\
    &= \sum_{\bf{n}}\sum_{\bf{e}}n_s^q\sum_{i=1}^{N_x+1}\sum_{j=1}^{N_y+1}\Big\{\gamma(j, N_i-1, e_i)(n_i^j-1)p([n_1^j, \dots, n_{i}^j -1, \dots, n_{N_x+1}^j], {\bf{e}},  t_h)\nonumber\\&\qquad\qquad\qquad\qquad\qquad\qquad\qquad\qquad\qquad\qquad\qquad\qquad-\gamma(j, N_i,e_i)n_i^jp({\bf{n}}, {\bf{e}},  t_h)\Big\}, \\
    &= {{M_1 + M_2 + M_3 + M_4 +M_5 + M_6}}.
\end{align*}
Again we consider the cases $i=s,  i\neq s,  j=q,  j\neq q.$
First, begin with $i\neq s$ and $j \neq q$, and use the change of variable $\bar{n}_i^j = n_i^j-1$, noticing that $N_i-1$ becomes $\bar{N_i}$ when we change the variable. 
Dropping the bar{{s}} gives
\begin{align*}
     M_1&= \sum_{\bf{n}}\sum_{\bf{e}}n_s^q\sum_{i=1}^{N_x+1}\sum_{j=1}^{N_y+1}\Big\{\gamma(j, N_i-1,e_i)(n_i^j-1)p([n_1^j, \dots, n_{i}^j -1, \dots, n_{N_x+1}^j], {\bf{e}},  t_h)\nonumber\\&\qquad\qquad\qquad\qquad\qquad\qquad\qquad\qquad\qquad\qquad\qquad\qquad-\gamma(j, N_i, e_i)n_i^jp({\bf{n}}, {\bf{e}},  t_h)\Big\} \nonumber \\
     &= \sum_{\bf{n}}\sum_{\bf{e}}n_s^q\sum_{\substack{i=1,\\  i \neq s}}^{N_x+1}\sum_{\substack{j=1,  \\j\neq q}}^{N_y+1}\Big\{\gamma(j, \bar{N_i},e_i)\bar{n_i}^jp([n_1^j, \dots, \bar{n}_{i}^j, \dots, n_{N_x+1}^j], {\bf{e}},  t_h)\nonumber\\&\qquad\qquad\qquad\qquad\qquad\qquad\qquad\qquad\qquad\qquad\qquad\qquad-\gamma(j, {N_i}, e_i){n}_i^jp({\bf{n}}, {\bf{e}},  t_h)\Big\} \nonumber \\
     &= \sum_{\bf{n}}\sum_{\bf{e}}n_s^q\sum_{\substack{i=1,\\  i \neq s}}^{N_x+1}\sum_{\substack{j=1,  \\j\neq q}}^{N_y+1}\gamma(j, N_i,e_i)\left\{n_i^jp({\bf{n}}, {\bf{e}},  t_h)-{n}_i^jp({\bf{n}}, {\bf{e}},  t_h)\right\} \nonumber \\
     &=0.
\end{align*}
Now consider $i\neq s$ and $j =q.$
Using the change of variable $\bar{n}_i^q = n_i^q-1$, along with the newly defined $\bar{N_i}$, before dropping the bar{{s}} we have
\begin{align*}
     M_2&= \sum_{\bf{n}}\sum_{\bf{e}}n_s^q\sum_{\substack{i=1,\\  i \neq s}}^{N_x+1}\Big\{\gamma(q, N_i-1,e_i)(n_i^q-1)p([n_1^q, \dots, n_{i}^q -1, \dots, n_{N_x+1}^q], {\bf{e}},  t_h)\nonumber\\&\qquad\qquad\qquad\qquad\qquad\qquad\qquad\qquad\qquad\qquad-\gamma(q, N_i,e_i)n_i^qp({\bf{n}}, {\bf{e}},  t_h)\Big\} \nonumber \\
     &= \sum_{\bf{n}}\sum_{\bf{e}}n_s^q\sum_{\substack{i=1,\\  i \neq s}}^{N_x+1}\Big\{\gamma(q, \bar{N_i},e_i)\bar{n}_i^qp([n_1^q, \dots, \bar{n}_{i}^q, \dots, n_{N_x+1}^q], {\bf{e}},  t_h)\nonumber\\&\qquad\qquad\qquad\qquad\qquad\qquad\qquad\qquad\qquad\qquad-\gamma(q, N_i,e_i)n_i^qp({\bf{n}}, {\bf{e}},  t_h)\Big\} \nonumber \\
     &= \sum_{\bf{n}}\sum_{\bf{e}}n_s^q\sum_{\substack{i=1,\\  i \neq s}}^{N_x+1}\gamma(q, N_i,e_i)\left\{n_i^qp({\bf{n}}, {\bf{e}},  t_h)-{n}_i^qp({\bf{n}}, {\bf{e}},  t_h)\right\} \nonumber \\
     &=0.
\end{align*}
Now take $i=s$ and consider $j \neq q$. 
Using the following change of variables, $\bar{n}_s^j = n_s^j-1$ and $\bar{N_s}=N_s-1$, {{we}} see that
\begin{align*}
     M_3&= \sum_{\bf{n}}\sum_{\bf{e}}n_s^q\sum_{\substack{j=1,  \\j\neq q}}^{N_y+1}\Big\{\gamma(j, N_{s}-1,e_s)(n_s^j-1)p([n_1^j, \dots, n_{s}^j -1, \dots, n_{N_x+1}^j], {\bf{e}},  t_h)\nonumber\\&\qquad\qquad\qquad\qquad\qquad\qquad\qquad\qquad\qquad\qquad-\gamma(j, N_{s},e_s)n_s^jp({\bf{n}}, {\bf{e}},  t_h)\Big\} \nonumber \\
     &= \sum_{\bf{n}}\sum_{\bf{e}}n_s^q\sum_{\substack{j=1,  \\j\neq q}}^{N_y+1}\Big\{\gamma(j, \bar{N_{s}},e_s)\bar{n}_s^jp([n_1^j, \dots, \bar{n}_{s}^j, \dots, n_{N_x+1}^j], {\bf{e}},  t_h)\nonumber\\&\qquad\qquad\qquad\qquad\qquad\qquad\qquad\qquad\qquad\qquad-\gamma(j, N_{s},e_s)n_s^jp({\bf{n}}, {\bf{e}},  t_h)\Big\} \nonumber \\
     &= \sum_{\bf{n}}\sum_{\bf{e}}n_s^q\sum_{\substack{j=1,  \\j\neq q}}^{N_y+1}\gamma(j, N_{s},e_s)\left\{n_s^jp({\bf{n}}, {\bf{e}},  t_h)-{n}_s^jp({\bf{n}}, {\bf{e}},  t_h)\right\} \nonumber \\
     &=0.
\end{align*}
Finally, consider $i=s$ and $j=q$ with the change of variable  $\bar{n}_s^q = n_s^q-1$ along with $\bar{N_s}$ in the second term, and then drop the bar to give 
\begin{align*}
     M_4&= \sum_{\bf{n}}\sum_{\bf{e}}n_s^q\Big\{\gamma(q, N_{s}-1,e_s)(n_s^q-1)p([n_1^q, \dots, n_{s}^q -1, \dots, n_{N_x+1}^q], {\bf{e}},  t_h)\nonumber\\&\qquad\qquad\qquad\qquad\qquad\qquad\qquad\qquad\qquad\qquad-\gamma(q, N_{s},e_s)n_s^qp({\bf{n}}, {\bf{e}},  t_h)\Big\} \nonumber \\
     &= \sum_{\bf{n}}\sum_{\bf{e}}\Big\{\gamma(q, \bar{N_{s}},e_s)\bar{n}_s^q(\bar{n}_s^q+1)p([n_1^q, \dots, \bar{n}_{s}^q, \dots, n_{N_x+1}^q], {\bf{e}},  t_h)\nonumber\\&\qquad\qquad\qquad\qquad-\gamma(q, N_{s},e_s)(n_s^q)^2p({\bf{n}}, {\bf{e}},  t_h)\Big\} \nonumber \\
     &= \sum_{\bf{n}}\sum_{\bf{e}}n_s^q\gamma(q, N_{s},e_s)p({\bf{n}}, {\bf{e}},  t_h) \nonumber \\
     &=\langle \gamma(q, N_{s},e_s) n_s^q \rangle.
\end{align*}
Combining all of these calculations, we can rewrite the master equation, Eq.~\eqref{SIeq:master}, as
\begin{align}
    \dfrac{\partial}{\partial t} \langle n_s^q\rangle &= \dfrac{1}{\Delta_t}\langle\beta_{+}(q, N_{s},e_s)n_{s-1}^q\rangle +\dfrac{1}{\Delta_t} \langle\beta_{-}(q, N_{s},e_s)n_{s+1}^q\rangle \nonumber \\ 
    &\quad-\dfrac{1}{\Delta_t} \langle\beta_{-}(q, N_{s-1},e_{s-1})n_{s}^q\rangle -\dfrac{1}{\Delta_t}\langle\beta_{+}(q, N_{s+1},e_{s+1})n_{s}^q\rangle \nonumber \\ 
    &\quad+\dfrac{1}{\Delta_t}\langle\mu_{+}(q-1, N_{s},e_s)n_{s}^{q-1}\rangle+\dfrac{1}{\Delta_t}\langle\mu_{-}(q+1, N_{s},e_s)n_{s}^{q+1}\rangle\nonumber \\ 
    &\quad-\dfrac{1}{\Delta_t}\langle\mu_{+}(q, N_{s},e_s)n_{s}^{q}\rangle-\dfrac{1}{\Delta_t}\langle\mu_{-}(q, N_{s},e_s)n_{s}^{q}\rangle\nonumber\\
    &\quad+\dfrac{1}{\Delta_t} \langle\gamma(q, N_{s},e_s)n_s^q\rangle. \label{SIeq:full_IBM_prolif_ECM}
\end{align}
This mean equation is related to a partial differential equation model in the appropriate limits as $\Delta_x\rightarrow 0$, $\Delta_y\rightarrow 0$ and $\Delta_t\rightarrow 0$ simultaneously, and the discrete values of $\langle{n}_i^j(t_h)\rangle$ and $\langle{e}_i(t_h)\rangle$ are written in terms of the continuous variables $n(x,  y,  t)$ and $e(x,t)$, respectively. 
Eq.~\eqref{SIeq:full_IBM_prolif_ECM} can be rewritten as follows, using mean-field approximations,
\begin{align}
    \dfrac{\partial n(x, y, t)}{\partial t} &= \dfrac{1}{\Delta_t}\beta_{+}(y, \rho(x,  t),e(x,  t))n(x-\Delta_x,  y,  t)\nonumber \\ &\quad +\dfrac{1}{\Delta_t} \beta_{-}(y, \rho(x,  t),e(x,  t))n (x+\Delta_x,  y,  t) \nonumber \\ 
    &\quad-\dfrac{1}{\Delta_t} \beta_{-}(y, \rho(x-\Delta_x,  t),e(x-\Delta_x,  t))n(x,y,t) \nonumber \\ &\quad-\dfrac{1}{\Delta_t}\beta_{+}(y, \rho(x+\Delta_x,  t),e(x+\Delta_x,  t))n(x,y,t) \nonumber \\ 
    &\quad+\dfrac{1}{\Delta_t}\mu_{+}(y-\Delta_y, \rho(x,  t),e(x,  t))n(x,  y-\Delta_y,  t)\nonumber \\
    &\quad+\dfrac{1}{\Delta_t}\mu_{-}(y+\Delta_y, \rho(x, t),e(x,  t))n(x,  y+\Delta_y,  t)\nonumber \\ 
    &\quad-\dfrac{1}{\Delta_t}\mu_{+}(y, \rho(x,  t), e(x,  t))n(x,y,t) \nonumber \\ &\quad -\dfrac{1}{\Delta_t}\mu_{-}(y, \rho(x, t),e(x,  t))n(x,y,t)\nonumber\\
    &\quad+\dfrac{1}{\Delta_t} \gamma(y, \rho(x,  t),e(x,  t))n(x,y,t). \label{SIeq:full_IBM_prolif_ECM4}
\end{align}
Then, we can {{employ Taylor series expansions in}}  Eq.~\eqref{SIeq:full_IBM_prolif_ECM4} and take limits to give 
\begin{align*}
    \dfrac{\partial n(x,y,t)}{\partial t} &= \dfrac{1}{\Delta_t}\beta_{+}(y, \rho(x,  t),e(x,  t))\Bigg[n(x,y,t) - \Delta_x \dfrac{\partial}{\partial x}n(x,y,t)+ \dfrac{\Delta_x^2}{2}\dfrac{\partial^2}{\partial x^2}n(x,y,t)\Bigg]\nonumber \\ &\quad+\dfrac{1}{\Delta_t} \beta_{-}(y, \rho(x,  t),e(x,  t))\Bigg[n(x,y,t) + \Delta_x \dfrac{\partial}{\partial x}n(x,y,t)+ \dfrac{\Delta_x^2}{2}\dfrac{\partial^2}{\partial x^2}n(x,y,t)\Bigg] \nonumber \\ 
    &\quad-\dfrac{1}{\Delta_t} \Bigg[\beta_{-}(y, \rho,  e) - \Delta_x \dfrac{\partial }{\partial x}\beta_{-}(y, \rho,  e)+\dfrac{\Delta_x^2}{2}\dfrac{\partial^2}{\partial x^2}\beta_{-}(y, \rho,  e)\Bigg]n(x,y,t) \nonumber \\ 
    &\quad -\dfrac{1}{\Delta_t}\Bigg[\beta_{+}(y, \rho,  e) + \Delta_x \dfrac{\partial }{\partial x}\beta_{+}(y, \rho,  e)+\dfrac{\Delta_x^2}{2}\dfrac{\partial^2}{\partial x^2}\beta_{+}(y, \rho,  e) \Bigg]n(x,y,t) \nonumber \\ 
    &\quad+\dfrac{1}{\Delta_t}\Bigg[\mu_{+}(y, \rho, e) - \Delta_y \dfrac{\partial }{\partial y} \mu_{+}(y, \rho, e) + \dfrac{\Delta_y^2}{2}\dfrac{\partial ^2}{\partial y^2}\mu_{+}(y, \rho, e)\Bigg]\times\nonumber \\ &\qquad\qquad\qquad\Bigg[ n(x,y,t) - \Delta_y \dfrac{\partial}{\partial y}n(x,y,t)+ \dfrac{\Delta_y^2}{2}\dfrac{\partial^2}{\partial y^2}n(x,y,t)\Bigg]\nonumber \\
    &\quad+\dfrac{1}{\Delta_t}\Bigg[\mu_{-}(y, \rho, e) + \Delta_y \dfrac{\partial }{\partial y} \mu_{-}(y, \rho, e) + \dfrac{\Delta_y^2}{2}\dfrac{\partial ^2}{\partial y^2}\mu_{-}(y, \rho, e)\Bigg]\times\nonumber \\ &\qquad\qquad\qquad\Bigg[ n(x,y,t) + \Delta_y \dfrac{\partial}{\partial y}n(x,y,t)+ \dfrac{\Delta_y^2}{2}\dfrac{\partial^2}{\partial y^2}n(x,y,t)\Bigg]\nonumber \\ 
    &\quad-\dfrac{1}{\Delta_t}\mu_{+}(y, \rho(x,  t), e(x,  t))n(x,y,t)-\dfrac{1}{\Delta_t}\mu_{-}(y, \rho(x, t),e(x,  t))n(x,y,t)\nonumber\\
    &\quad+\dfrac{1}{\Delta_t} \gamma(y, \rho(x,  t),e(x,  t))n(x,y,t) +\text{O}(\Delta_x^2) +\text{O}(\Delta_y^2) +\text{O}(\Delta_t^2). 
\end{align*}
Rearranging and collecting terms, we obtain
{\small{
\begin{align*}
    \dfrac{\partial}{\partial t}n(x,y,t) &= \dfrac{\Delta_x}{\Delta_t}\dfrac{\partial}{\partial x}\Big(\left(\beta_{-}(y, \rho(x,t), e(x,t))-\beta_{+}(y, \rho(x,t), e(x,t))\right)n\Big) \nonumber \\
    &\qquad + \dfrac{\Delta_x^2}{2\Delta_t} \dfrac{\partial}{\partial x}\Bigg(\Big(\beta_{-}\left(y, \rho(x,t), e(x,t)\right)+\beta_{+}\left(y, \rho(x,t), e(x,t)\right)\Big)\dfrac{\partial}{\partial x} n(x,y,t) \nonumber \\ &\qquad\qquad\qquad\qquad- n(x,y,t)\dfrac{\partial}{\partial x}\Big(\beta_{-}\left(y, \rho(x,t), e(x,t)\right)+\beta_{+}\left(y, \rho(x,t), e(x,t)\right)\Big)\Bigg) \nonumber \\ 
    & \qquad+\dfrac{\Delta_y}{\Delta_t}\dfrac{\partial}{\partial y} \left(\Big(\mu_{-}\left(y, \rho(x,t), e(x,t)\right)-\mu_{+}\left(y, \rho(x,t), e(x,t)\right)\Big)n(x,y,t)\right)\nonumber \\ &\qquad+\dfrac{\Delta_y^2}{2\Delta_t} \dfrac{\partial ^2}{\partial y^2}\left(\Big(\mu_{-}\left(y, \rho(x,t), e(x,t)\right)+\mu_{+}\left(y, \rho(x,t), e(x,t)\right)\Big)n(x,y,t)\right)\nonumber \\
    &\qquad +\dfrac{1}{\Delta_t}\gamma\left(y, \rho(x,t), e(x,t)\right)n(x,y,t). 
\end{align*}
}}
{{W}}e take the parabolic limit as $\Delta_x, \, \Delta_y, \, \Delta_t \rightarrow 0$ simultaneously (assuming $n(x,y,t)\sim O(1)$), {{and}} define
\begin{align*}
    \lim_{\Delta_x,  \Delta_t\rightarrow 0} \dfrac{\Delta_x}{\Delta_t} \Big(\beta_{-}(y, \rho(x,t), e(x,t))-\beta_{+}(y, \rho(x,t), e(x,t))\Big) &= v^m(y, \rho(x,t), e(x,t)), \\
    \lim_{\Delta_x,  \Delta_t\rightarrow 0} \dfrac{\Delta_x^2}{2\Delta_t} \Big(\beta_{-}(y, \rho(x,t), e(x,t))+\beta_{+}(y, \rho(x,t), e(x,t))\Big) &= D^m(y, \rho(x,t), e(x,t)), \\
    \lim_{\Delta_y,  \Delta_t\rightarrow 0} \dfrac{\Delta_y}{\Delta_t} \Big(\mu_{-}\left(y, \rho(x,t), e(x,t)\right)-\mu_{+}\left(y, \rho(x,t), e(x,t)\right)\Big) &= v^p(y, \rho(x,t), e(x,t)), \\
    \lim_{\Delta_y,  \Delta_t\rightarrow 0} \dfrac{\Delta_y^2}{2\Delta_t} \Big(\mu_{-}\left(y, \rho(x,t), e(x,t)\right)+\mu_{+}\left(y, \rho(x,t), e(x,t)\right)\Big) &= D^p(y, \rho(x,t), e(x,t)), \\
    \lim_{\Delta_t\rightarrow 0} \dfrac{1}{\Delta_t} \gamma\left(y, \rho(x,t), e(x,t)\right) &= r(y, \rho(x,t), e(x,t)).
\end{align*}

The final equation for the evolution of the cell density is therefore given by
\begin{align*}
    \dfrac{\partial}{\partial t}n(x,y,t) &= \dfrac{\partial}{\partial x}\Big(v^m(y, \rho(x,t), e(x,t))n\Big) \nonumber \\
    &\qquad+\dfrac{\partial}{\partial x}\Bigg(D^m\left(y, \rho(x,t), e(x,t)\right)\dfrac{\partial}{\partial x} n(x,y,t) \nonumber \\ &\qquad\qquad\qquad\qquad- n(x,y,t)\dfrac{\partial}{\partial x}D^m\left(y, \rho(x,t), e(x,t)\right)\Bigg) \nonumber \\ 
    & \qquad+\dfrac{\partial}{\partial y} \left(v^p\left(y, \rho(x,t), e(x,t)\right)n(x,y,t)\right)\nonumber \\ &\qquad+\dfrac{\partial ^2}{\partial y^2}\left(D^p\left(y, \rho(x,t), e(x,t)\right)n(x,y,t)\right)\nonumber \\
    &\qquad +r\left(y, \rho(x,t), e(x,t)\right)n(x,y,t). 
\end{align*}

\subsubsection{Model equations on the boundaries}\label{app:BCs}
We can repeat the above analysis for the boundaries of physical and phenotype space in order to retrieve the boundary equations. 
We note here that when looking at the boundaries for the cell equation, all terms involving change in the number of elements of the local environment provide no contribution and can be ignored hereon in.

\paragraph{Boundary condition at $x=X_{\text{min}}$.}
Revisiting Eq.~\eqref{SIeq:master} to derive the equation on the left-most lattice {{site}}, we multiply by $n_1^q$ and sum over all possible states ${\bf{n}}$ and ${\bf{e}}$:
{\small{
\begin{align}
    &\Delta_t \sum_{\bf{n}}\sum_{\bf{e}}n_1^q \dfrac{\partial}{\partial t} p({\bf{n}}, {\bf{e}},  t_h) +O(\Delta_t^2) \nonumber \\
    &\quad= \sum_{\bf{n}}\sum_{\bf{e}}n_1^q \sum_{i=1}^{N_x+1}\sum_{j=1}^{N_y}\mu_{-}(j+1, N_i, e_i)\left\{ (n_i^{j+1}+1)p(U_{i,j}^{\text{p}}{\bf{n}}, {\bf{e}},  t_h)-n_i^{j+1}p({\bf{n}}, {\bf{e}},  t_h)\right\} \nonumber \\ 
    &\quad + \sum_{\bf{n}}\sum_{\bf{e}}n_1^q \sum_{i=1}^{N_x+1}\sum_{j=2}^{N_y+1}\mu_{+}(j-1, N_i, e_i)\left\{ (n_i^{j-1}+1)p(D_{i,j}^{\text{p}}{\bf{n}}, {\bf{e}},  t_h)-n_i^{j-1}p({\bf{n}}, {\bf{e}},  t_h)\right\} \nonumber \\
    &\quad + \sum_{\bf{n}}\sum_{\bf{e}}n_1^q\sum_{i=1}^{N_x} \sum_{j=1}^{N_y+1}\beta_{-}(j, {N_i,e_i)}\left\{ (n_{i+1}^{j}+1)p(R_{i,j}^{\text{m}}{\bf{n}}, {\bf{e}},  t_h)-n_{i+1}^{j}p({\bf{n}}, {\bf{e}},  t_h)\right\} \nonumber \\ 
    &\quad + \sum_{\bf{n}}\sum_{\bf{e}}n_1^q\sum_{i=2}^{N_x+1} \sum_{j=1}^{N_y+1}\beta_{+}(j, N_i,e_i)\left\{ (n_{i-1}^{j}+1)p(L_{i,j}^{\text{m}}{\bf{n}}, {\bf{e}},  t_h)-n_{i-1}^{j}p({\bf{n}}, {\bf{e}},  t_h)\right\}\nonumber \\ 
    &\quad + \sum_{\bf{n}}\sum_{\bf{e}}n_1^q\sum_{i=1}^{N_x+1} \sum_{j=1}^{N_y+1}\left\{\gamma(j, N_i-1,e_i)(n_i^j-1)p(G_{i,j}{\bf{n}}, {\bf{e}},  t_h)-\gamma(j, N_i,e_i)n_i^jp({\bf{n}}, {\bf{e}},  t_h)\right\}.\label{SIeq:master_mult_x1}
\end{align}
}}
From earlier working, we know that all terms describing a movement in physical space give non-zero contributions when  $j\neq q$, and hence can be ignored. 
Now, starting by considering the first term on the right-hand side of Eq.~\eqref{SIeq:master_mult_x1}, we consider the non-zero contributions from the cases $j=q$ and $j=q-1$. 
Using $j=q$:
\begin{align}
    &\sum_{\bf{n}}\sum_{\bf{e}}n_1^q \sum_{i=1}^{N_x+1}\mu_{-}(q+1, N_i, e_i)\left\{ (n_i^{q+1}+1)p(U_{i,q}^{\text{p}}{\bf{n}}, {\bf{e}},  t_h)-n_i^{q+1}p({\bf{n}}, {\bf{e}},  t_h)\right\} \nonumber \\
    &\quad= \sum_{\bf{n}}\sum_{\bf{e}}n_1^q \sum_{i=1}^{N_x+1}\mu_{-}(q+1, N_i, e_i)\times \nonumber \\ 
    &\qquad\quad\left\{ (n_i^{q+1}+1)p([n_i^1, \dots, n_{i}^{q}-1, n_{i}^{q+1}+1, \dots, n_{i}^{N_y+1}], {\bf{e}},  t_h)-n_i^{q+1}p({\bf{n}}, {\bf{e}},  t_h)\right\}.\label{SIeq:master_mult_x12}
\end{align}
As per previous calculations, the only contributions here will come from terms where $i=1$.
Looking at these, we see that if we employ the change of variables $\bar{n}_1^{q+1}=n_1^{q+1}+1$ and $\bar{n}_1^q=n_1^q-1$ in the first term, then we can rewrite the right-hand side of Eq.~\eqref{SIeq:master_mult_x12} as
\begin{align*}
&\sum_{\bf{n}}\sum_{\bf{e}} \mu_{-}(q+1, N_1, e_1)\times \nonumber \\ 
    &\qquad\quad\left\{ \bar{n}_1^{q+1}(\bar{n}_1^q+1)p([n_1^1, \dots, \bar{n}_{1}^{q}, \bar{n}_{1}^{q+1}, \dots, n_{1}^{N_y+1}], {\bf{e}},  t_h)-n_1^q n_1^{q+1}p({\bf{n}}, {\bf{e}},  t_h)\right\} \nonumber \\
&\quad=\sum_{\bf{n}}\sum_{\bf{e}} \mu_{-}(q+1, N_1, e_1)\left\{ {n}_1^{q+1}(n_1^q+1)p({\bf{n}}, {\bf{e}},  t_h)-n_1^q n_1^{q+1}p({\bf{n}}, {\bf{e}},  t_h)\right\} \nonumber \\
&\quad=\sum_{\bf{n}}\sum_{\bf{e}} \mu_{-}(q+1, N_1, e_1)n_1^{q+1}p({\bf{n}},{\bf{e}},  t_h) \nonumber \\
    &\quad=\langle\mu_{-}(q+1, N_1, e_1) {n}_1^{q+1}\rangle.
\end{align*}
Using $j=q-1$ and $i=1$, with the change of variables $\bar{n}_1^{q-1}=n_1^{q-1}-1$ and $\bar{n}_1^q=n_1^q+1$ in the first term of Eq.~\eqref{SIeq:master_mult_x1}, then we have
\begin{align*}
&\sum_{\bf{n}}\sum_{\bf{e}}n_1^q \mu_{-}(q, N_1, e_1)\times \nonumber \\ 
    &\qquad\quad\left\{ (n_1^{q}+1)p([n_1^1, \dots, n_{1}^{q-1}-1, n_{1}^{q}+1, \dots, n_{1}^{N_y+1}], {\bf{e}},  t_h)-n_1^{q}p({\bf{n}}, {\bf{e}},  t_h)\right\} \nonumber \\
&\quad=\sum_{\bf{n}}\sum_{\bf{e}} \mu_{-}(q, N_1, e_1)\times \nonumber \\ 
    &\qquad\quad\left\{ \bar{n}_1^{q}(\bar{n}_1^q-1)p([n_1^1, \dots, \bar{n}_{1}^{q-1}, \bar{n}_{1}^{q}, \dots, n_{1}^{N_y+1}], {\bf{e}},  t_h)-n_1^q n_1^{q}p({\bf{n}}, {\bf{e}},  t_h)\right\} \nonumber \\
&\quad=\sum_{\bf{n}}\sum_{\bf{e}} \mu_{-}(q, N_1, e_1)\left\{ {n}_1^{q}(n_1^{q}-1)p({\bf{n}}, {\bf{e}},  t_h)-n_1^q n_1^{q}p({\bf{n}}, {\bf{e}},  t_h)\right\} \nonumber \\
&\quad=-\sum_{\bf{n}}\sum_{\bf{e}} \mu_{-}(q, N_1, e_1)n_1^{q}p({\bf{n}},  {\bf{e}},  t_h) \nonumber \\
    &\quad=-\langle\mu_{-}(q, N_1, e_1) {n}_1^{q}\rangle.
\end{align*}
Considering the second term of Eq.~\eqref{SIeq:master_mult_x1}, we need to consider the cases$j=q$ and $j=q+1$, for $i=1$. 
For $i=1$ and $j=q$, using the change of variables $\bar{n}_1^{q-1}=n_1^{q-1}+1$ and $\bar{n}_1^q=n_1^q-1$ we have 
\begin{align*}
    &\sum_{\bf{n}}\sum_{\bf{e}}n_1^q \mu_{+}(q-1, N_1, e_1)\left\{ (n_1^{q-1}+1)p(D_{1,q}^{\text{p}}{\bf{n}}, {\bf{e}},  t_h)-n_1^{q-1}p({\bf{n}}, {\bf{e}},  t_h)\right\} \nonumber \\
    &\quad=\sum_{\bf{n}}\sum_{\bf{e}}n_1^q \mu_{+}(q-1, N_1, e_1)\times \nonumber \\ 
    &\qquad\quad\left\{ (n_1^{q-1}+1)p([n_1^1, \dots, n_{1}^{q-1}+1, n_{1}^{q}-1, \dots, n_{1}^{N_y+1}], {\bf{e}},  t_h)-n_1^{q-1}p({\bf{n}}, {\bf{e}},  t_h)\right\} \nonumber \\
    &\quad=\sum_{\bf{n}}\sum_{\bf{e}}\mu_{+}(q-1, N_1, e_1)\times \nonumber \\ 
    &\qquad\quad\left\{ \bar{n}_1^{q-1}(\bar{n}_1^q+1)p([n_1^1, \dots, \bar{n}_{1}^{q-1}, \bar{n}_{1}^{q}, \dots, n_{1}^{N_y+1}], {\bf{e}},  t_h)-n_1^qn_1^{q-1}p({\bf{n}}, {\bf{e}},  t_h)\right\} \nonumber \\
    &\quad=\sum_{\bf{n}}\sum_{\bf{e}}\mu_{+}(q-1, N_1, e_1)\left\{ {n}_1^{q-1}({n}_1^q+1)p({\bf{n}}, {\bf{e}},  t_h)-n_1^qn_1^{q-1}p({\bf{n}}, {\bf{e}},  t_h)\right\} \nonumber \\
    &\quad= \sum_{\bf{n}}\sum_{\bf{e}}\mu_{+}(q-1, N_1, e_1) {n}_1^{q-1}p({\bf{n}}, {\bf{e}},  t_h) \nonumber \\
    &\quad=\langle\mu_{+}(q-1, N_1, e_1) {n}_1^{q-1}\rangle.
\end{align*}
For the case $j=q+1$ and $i=1$, the change of variables $\bar{n}_1^{q+1}=n_1^{q+1}-1$ and $\bar{n}_1^q=n_1^q+1$ gives
\begin{align*}
    &\sum_{\bf{n}}\sum_{\bf{e}}n_1^q \mu_{+}(q, N_1, e_1)\left\{ (n_1^{q}+1)p(D_{1,q}^{\text{p}}{\bf{n}}, {\bf{e}},  t_h)-n_1^{q}p({\bf{n}}, {\bf{e}},  t_h)\right\} \nonumber \\
    &\quad=\sum_{\bf{n}}\sum_{\bf{e}}n_1^q \mu_{+}(q, N_1, e_1)\times \nonumber \\ 
    &\qquad\quad\left\{ (n_1^{q}+1)p([n_1^1, \dots, n_{1}^{q}+1, n_{1}^{q+1}-1, \dots, n_{1}^{N_y+1}], {\bf{e}},  t_h)-n_1^{q}p({\bf{n}}, {\bf{e}},  t_h)\right\} \nonumber \\
    &\quad=\sum_{\bf{n}}\sum_{\bf{e}}\mu_{+}(q, N_1, e_1)\times \nonumber \\ 
    &\qquad\quad\left\{ \bar{n}_1^{q}(\bar{n}_1^q-1)p([n_1^1, \dots, \bar{n}_{1}^{q}, \bar{n}_{1}^{q+1}, \dots, n_{1}^{N_y+1}], {\bf{e}},  t_h)-n_1^qn_1^{q}p({\bf{n}}, {\bf{e}},  t_h)\right\} \nonumber \\
    &\quad=\sum_{\bf{n}}\sum_{\bf{e}}\mu_{+}(q, N_1, e_1)\left\{ {n}_1^{q}({n}_1^q-1)p({\bf{n}}, {\bf{e}},  t_h)-n_1^qn_1^{q}p({\bf{n}}, {\bf{e}},  t_h)\right\} \nonumber \\
    &\quad= -\sum_{\bf{n}}\sum_{\bf{e}}\mu_{+}(q, N_1, e_1) {n}_1^{q}p({\bf{n}}, {\bf{e}},  t_h) \nonumber \\
    &\quad=-\langle\mu_{+}(q, N_1, e_1) {n}_1^{q}\rangle.
\end{align*}
Now, looking at the third term of Eq.~\eqref{SIeq:master_mult_x1}, which governs movement in physical space, for $j=q$ we have
\begin{align*}
    &\sum_{\bf{n}}\sum_{\bf{e}}n_1^q\sum_{i=1}^{N_x} \beta_{-}(q, {N_i,e_i)}\left\{ (n_{i+1}^{q}+1)p(R_{i,q}^{\text{m}}{\bf{n}}, {\bf{e}},  t_h)-n_{i+1}^{q}p({\bf{n}}, {\bf{e}},  t_h)\right\} \nonumber \\
    &\quad=\sum_{\bf{n}}\sum_{\bf{e}}n_1^q\sum_{i=1}^{N_x} \beta_{-}(q, {N_i,e_i)}\times \nonumber \\ 
    &\qquad\quad\left\{ (n_{i+1}^{q}+1)p([n_1^q, \dots, n_{i}^q-1, n_{i+1}^{q}+1, \dots, n_{N_x+1}^q], {\bf{e}},  t_h)-n_{i+1}^{q}p({\bf{n}}, {\bf{e}},  t_h)\right\}, 
\end{align*}
which only produces non-zero contributions when $i=1$, namely,
\begin{align*}
    &\sum_{\bf{n}}\sum_{\bf{e}}n_1^q \beta_{-}(q, {N_1,e_1)}\left\{ (n_{2}^{q}+1)p([n_1^q-1, n_{2}^{q}+1, \dots, n_{N_x+1}^q], {\bf{e}},  t_h)-n_{2}^{q}p({\bf{n}}, {\bf{e}},  t_h)\right\}  \nonumber \\
    &\quad=\sum_{\bf{n}}\sum_{\bf{e}} \beta_{-}(q, {N_1,e_1)}\left\{ (\bar{n}_1^q+1)\bar{n}_{2}^{q}p([\bar{n}_1^q, \bar{n}_{2}^{q}, \dots, n_{N_x+1}^q], {\bf{e}},  t_h)-n_1^qn_{2}^{q}p({\bf{n}}, {\bf{e}},  t_h)\right\} \nonumber \\
    &\quad=\sum_{\bf{n}}\sum_{\bf{e}} \beta_{-}(q, {N_1,e_1)}\left\{ ({n}_1^q+1){n}_{2}^{q}p({\bf{n}}, {\bf{e}},  t_h)-n_1^qn_{2}^{q}p({\bf{n}}, {\bf{e}},  t_h)\right\} \nonumber \\
    &\quad=\sum_{\bf{n}}\sum_{\bf{e}} \beta_{-}(q, {N_1,e_1)} {n}_{2}^{q}p({\bf{n}}, {\bf{e}},  t_h) \nonumber \\
    &\quad=\langle\beta_{-}(q, {N_1,e_1)} {n}_{2}^{q}\rangle, 
\end{align*}
where we used the change of variables $\bar{n}_1^q=n_1^q-1$ and $\bar{n}_2^q=n_2^q+1$, and then dropped the bar.
Next, this argument can be repeated for $j=q$ and $i=2$ in the fourth term of Eq.~\eqref{SIeq:master_mult_x1} (chosen such that $i-1=1$, and recalling that terms with $i>2$ will give non-zero contributions), using the change of variable $\bar{n}_1^q=n_1^q+1$ and $\bar{n}_2^q=n_2^q-1$:
\begin{align*}
    & \sum_{\bf{n}}\sum_{\bf{e}}n_1^q\beta_{+}(q, N_2,e_2)\left\{ (n_{1}^{q}+1)p(L_{2,q}^{\text{m}}{\bf{n}}, {\bf{e}},  t_h)-n_{1}^{q}p({\bf{n}}, {\bf{e}}, t_h)\right\}  \nonumber \\
    &\quad=  \sum_{\bf{n}}\sum_{\bf{e}}n_1^q\beta_{+}(q, N_2,e_2)\times \nonumber \\ 
    &\qquad\qquad\qquad\qquad\left\{ (n_{1}^{q}+1)p([ n_{1}^q+1, n_{2}^{q}-1, \dots, n_{N_x+1}^q], {\bf{e}},  t_h)-n_{1}^{q}p({\bf{n}}, {\bf{e}},  t_h)\right\} \nonumber \\ 
    &\quad=  \sum_{\bf{n}}\sum_{\bf{e}}\beta_{+}(q, N_2,e_2)\left\{ \bar{n}_{1}^{q}(\bar{n}_1^q-1)p([\bar{n}_{1}^q, \bar{n}_{2}^{q}, \dots, n_{N_x+1}^q], {\bf{e}},  t_h)-n_1^qn_{1}^{q}p({\bf{n}}, {\bf{e}},  t_h)\right\} \nonumber \\ 
    &\quad=  \sum_{\bf{n}}\sum_{\bf{e}}\beta_{+}(q, N_2,e_2)\left\{ {n}_{1}^{q}({n}_1^q-1)p({\bf{n}}, {\bf{e}},  t_h)-n_1^qn_{1}^{q}p({\bf{n}}, {\bf{e}},  t_h)\right\} \nonumber \\  
    &\quad=  -\sum_{\bf{n}}\sum_{\bf{e}}\beta_{+}(q, N_2,e_2) {n}_{1}^{q}p({\bf{n}}, {\bf{e}},  t_h) \nonumber \\ 
    &\quad=  -\langle\beta_{+}(q, N_2,e_2) {n}_{1}^{q}\rangle. 
\end{align*}
Now, finally, we look at the last term on the right-hand side of Eq.~\eqref{SIeq:master_mult_x1} which has only non-zero contributions when $j=q$ and $i=1$:
{\small{
\begin{align}
    & \sum_{\bf{n}}\sum_{\bf{e}}n_1^q\left\{\gamma(q, N_1-1,e_1)(n_1^q-1)p(G_{1,q}{\bf{n}}, {\bf{e}},  t_h)-n_1^q\gamma(q, N_1,e_1)p({\bf{n}}, {\bf{e}},  t_h)\right\} \nonumber \\ 
    &\quad= \sum_{\bf{n}}\sum_{\bf{e}}n_1^q\left\{\gamma(q, N_1-1,e_1)(n_1^q-1)p([ n_{1}^q -1, \dots, n_{N_x+1}^q], {\bf{e}},  t_h)-n_1^q\gamma(q, N_1,e_1)p({\bf{n}}, {\bf{e}},  t_h)\right\}. \label{SIeq:2gamBC}
\end{align}
}}
Using the change of variable $\bar{n}_1^q= n_1^q-1$ in the second term of Eq.~\eqref{SIeq:2gamBC} we get 
\begin{align*}
&\sum_{\bf{n}}\sum_{\bf{e}}\left\{\bar{n}_1^q(\bar{n}_1^q+1)\gamma(q, \bar{N_1},e_1){\bar{n}}_1^qp([ \bar{n}_{1}^q, \dots, n_{N_x+1}^q], {\bf{e}},  t_h)-(n_1^q)^2\gamma(q, N_1,e_1)p({\bf{n}}, {\bf{e}},  t_h)\right\} \nonumber \\ 
 &\quad= \sum_{\bf{n}}\sum_{\bf{e}}\left\{{n}_1^q({n}_1^q+1)\gamma(q, N_1,e_1){{n}}_1^qp( {\bf{n}}, {\bf{e}},  t_h)-(n_1^q)^2\gamma(q, N_1,e_1)p({\bf{n}}, {\bf{e}},  t_h)\right\} \nonumber \\ 
 &\quad= \sum_{\bf{n}}\sum_{\bf{e}}\gamma(q, N_1,e_1)n_1^qp({\bf{n}}, {\bf{e}},  t_h) \nonumber \\ 
 &\quad= \langle\gamma(q, N_1,e_1)n_1^q\rangle. 
\end{align*}
Recompiling these simplified terms, the equation for cell evolution on the left-hand boundary in physical space becomes: 
\begin{align}
    \dfrac{\partial}{\partial t} \langle n_1^q\rangle &=\dfrac{1}{\Delta_t}\langle\mu_{-}(q+1, N_1, e_1) {n}_1^{q+1}\rangle + \dfrac{1}{\Delta_t}\langle\mu_{+}(q-1, N_1, e_1) {n}_1^{q-1}\rangle \nonumber \\ &\quad-\dfrac{1}{\Delta_t}\langle\mu_{-}(q, N_1, e_1) {n}_1^{q}\rangle -\dfrac{1}{\Delta_t}\langle\mu_{+}(q, N_1, e_1) {n}_1^{q}\rangle\nonumber \\ &\quad+  \dfrac{1}{\Delta_t}\langle\beta_{-}(q, N_1,e_1)n_{2}^q\rangle -\dfrac{1}{\Delta_t}\langle\beta_{+}(q, N_2,e_2)n_{1}^q\rangle +\dfrac{1}{\Delta_t} \langle\gamma(q, N_1,e_1)n_1^q\rangle . \label{SIeq:BC_x1}
\end{align}

Now we wish to find the continuum equivalent of this equation. 
Recalling the {{continuum equivalents of the dependent variables}}, and employing Taylor series expansions around $x=X_{\text{min}}$,  we can rewrite Eq.~\eqref{SIeq:BC_x1} as (dropping the dependent variables for simplicity)
\begin{align*}
    \Delta_t \dfrac{\partial n}{\partial t} &= \left(\mu_{-}+\Delta_y\dfrac{\partial \mu_{-}}{\partial y}+\dfrac{\Delta_y^2}{2}\dfrac{\partial^2 \mu_{-}}{\partial y^2}+\dots\right)\left(n+\Delta_y\dfrac{\partial n}{\partial y}+\dfrac{\Delta_y^2}{2}\dfrac{\partial^2 n}{\partial y^2}+\dots\right) \nonumber \\ 
    &\quad +\left(\mu_{+}+\Delta_y\dfrac{\partial \mu_{+}}{\partial y}+\dfrac{\Delta_y^2}{2}\dfrac{\partial^2 \mu_{+}}{\partial y^2}+\dots\right)\left(n+\Delta_y\dfrac{\partial n}{\partial y}+\dfrac{\Delta_y^2}{2}\dfrac{\partial^2 n}{\partial y^2}+\dots\right) \nonumber \\ 
    &\quad -\mu_{+}n-\mu_{-}n +\gamma n + \beta_{-} \left(n+\Delta_x\dfrac{\partial n}{\partial x}+\dfrac{\Delta_x^2}{2}\dfrac{\partial^2 n}{\partial x^2}+\dots\right) \nonumber \\ 
    &\quad -n\left(\beta_{+}+\Delta_x\dfrac{\partial \beta_{+}}{\partial x}+\dfrac{\Delta_x^2}{2}\dfrac{\partial^2 \beta_{+}}{\partial x^2}+\dots\right),
\end{align*}
at $x=X_{\text{min}}$, so that we have
\begin{align}
    \dfrac{\partial n}{\partial t}&= \dfrac{\Delta_y}{\Delta_t}\dfrac{\partial}{\partial y}\left(\left(\mu_{-}-\mu_{+}\right)n\right)+\dfrac{\Delta_y^2}{2\Delta_t}\dfrac{\partial^2}{\partial y^2}\left(\left(\mu_{-}+\mu_{+}\right)n\right) + \dfrac{1}{\Delta_t}n\left(\beta_{-}-\beta_{+}\right) \nonumber \\ &\quad +\dfrac{\Delta_x}{\Delta_t} \left(\beta_{-} \dfrac{\partial n}{\partial x} - n \dfrac{\partial \beta_{+}}{\partial x}\right) +\dfrac{\Delta_x^2}{2\Delta_t}\left( \beta_{-} \dfrac{\partial^2 n}{\partial x^2}- n \dfrac{\partial^2 \beta_{+}}{\partial x^2}\right)+\dfrac{1}{\Delta_t}\gamma n . \label{SIeq:BC_xmin_cont}
\end{align}
Recalling that
{\small{
\begin{align*}
    \lim_{\Delta_x,  \Delta_t\rightarrow 0} \dfrac{\Delta_x}{\Delta_t} \Big(\beta_{-}(y, \rho(x,t), e(x,t))-\beta_{+}(y, \rho(x,t), e(x,t))\Big) &= v^m(y, \rho(x,t), e(x,t)), \\
    \lim_{\Delta_x,  \Delta_t\rightarrow 0} \dfrac{\Delta_x^2}{2\Delta_t} \Big(\beta_{-}(y, \rho(x,t), e(x,t))+\beta_{+}(y, \rho(x,t), e(x,t))\Big) &= D^m(y, \rho(x,t), e(x,t)), \\
    \lim_{\Delta_y,  \Delta_t\rightarrow 0} \dfrac{\Delta_y}{\Delta_t} \Big(\mu_{-}\left(y, \rho(x,t), e(x,t)\right)-\mu_{+}\left(y, \rho(x,t), e(x,t)\right)\Big) &= v^p(y, \rho(x,t), e(x,t)), \\
    \lim_{\Delta_y,  \Delta_t\rightarrow 0} \dfrac{\Delta_y^2}{2\Delta_t} \Big(\mu_{-}\left(y, \rho(x,t), e(x,t)\right)+\mu_{+}\left(y, \rho(x,t), e(x,t)\right)\Big) &= D^p(y, \rho(x,t), e(x,t)), \\
    \lim_{\Delta_t\rightarrow 0} \dfrac{1}{\Delta_t} \gamma\left(y, \rho(x,t), e(x,t)\right) &= r(y, \rho(x,t), e(x,t)),
\end{align*}
}}
such that 
\begin{align*}
    \beta_{\pm}&=\dfrac{D^m\Delta_t}{\Delta_x^2}\mp\dfrac{v^m\Delta_t}{2\Delta_x},
\end{align*}
then Eq.~\eqref{SIeq:BC_xmin_cont} can be rewritten as
\begin{align*}
    \dfrac{\partial n}{\partial t}&= \dfrac{\partial}{\partial y}\left(v^pn\right)+\dfrac{\partial^2}{\partial y^2}\left(D^pn\right) + \dfrac{1}{\Delta_x}v^mn +\dfrac{1}{2}v^m\dfrac{\partial n}{\partial x} + \dfrac{1}{\Delta_x}D^m\dfrac{\partial n}{\partial x} \nonumber \\ &\quad -\dfrac{1}{\Delta_x}n\dfrac{\partial}{\partial x}D^m-\dfrac{1}{2}n\dfrac{\partial}{\partial x}v^m+\dfrac{1}{2}D^m\dfrac{\partial^2n}{\partial x^2}-\dfrac{1}{2}n\dfrac{\partial^2}{\partial x^2}D^m+rn .
\end{align*}
In order to prevent blow-up of terms in the limit $\Delta_x\rightarrow 0$, we require
\begin{equation*}
    v^mn +D^m\dfrac{\partial n}{\partial x}-n\dfrac{\partial}{\partial x}D^m  = 0  \qquad\qquad \text{at} \quad x=X_{\text{min}}.
\end{equation*}
As such, we have no flux of cells out of the physical space boundary at $x=X_{\text{min}}.$

\paragraph{Boundary condition at $x=X_{\text{max}}$.}
Revisiting Eq.~\eqref{SIeq:master} to find the equation for the right-hand lattice site in physical space
we multiply by $n_{N_x+1}^q$ and sum over all possible states ${\bf{n}}$ and ${\bf{e}}$:
{\small{
\begin{align}
    &\Delta_t \sum_{\bf{n}}\sum_{\bf{e}}n_{N_x+1}^q \dfrac{\partial}{\partial t} p({\bf{n}}, {\bf{e}},  t_h) = \nonumber \\ 
    &\quad\sum_{\bf{n}}\sum_{\bf{e}}n_{N_x+1}^q \sum_{i=1}^{N_x+1}\sum_{j=1}^{N_y}\mu_{-}(j+1, N_i, e_i)\left\{ (n_i^{j+1}+1)p(U_{i,j}^{\text{p}}{\bf{n}}, {\bf{e}},  t_h)-n_i^{j+1}p({\bf{n}}, {\bf{e}},  t_h)\right\} \nonumber \\ 
    &\quad + \sum_{\bf{n}}\sum_{\bf{e}}n_{N_x+1}^q \sum_{i=1}^{N_x+1}\sum_{j=2}^{N_y+1}\mu_{+}(j-1, N_i, e_i)\left\{ (n_i^{j-1}+1)p(D_{i,j}^{\text{p}}{\bf{n}}, {\bf{e}},  t_h)-n_i^{j-1}p({\bf{n}}, {\bf{e}},  t_h)\right\} \nonumber \\
    &\quad + \sum_{\bf{n}}\sum_{\bf{e}}n_{N_x+1}^q\sum_{i=1}^{N_x} \sum_{j=1}^{N_y+1}\beta_{-}(j, {N_i,e_i)}\left\{ (n_{i+1}^{j}+1)p(R_{i,j}^{\text{m}}{\bf{n}}, {\bf{e}},  t_h)-n_{i+1}^{j}p({\bf{n}}, {\bf{e}},  t_h)\right\} \nonumber \\ 
    &\quad + \sum_{\bf{n}}\sum_{\bf{e}}n_{N_x+1}^q\sum_{i=2}^{N_x+1} \sum_{j=1}^{N_y+1}\beta_{+}(j, N_i,e_i)\left\{ (n_{i-1}^{j}+1)p(L_{i,j}^{\text{m}}{\bf{n}}, {\bf{e}},  t_h)-n_{i-1}^{j}p({\bf{n}}, {\bf{e}},  t_h)\right\}\nonumber \\ 
    &\quad + \sum_{\bf{n}}\sum_{\bf{e}}n_{N_x+1}^q\sum_{i=1}^{N_x+1} \sum_{j=1}^{N_y+1}\left\{\gamma(j, N_i-1,e_i)(n_i^j-1)p(G_{i,j}{\bf{n}}, {\bf{e}},  t_h)-\gamma(j, N_i,e_i)n_i^jp({\bf{n}}, {\bf{e}},  t_h)\right\}.\label{SIeq:master_mult_xNx}
\end{align}
}}
Now we can repeat the analysis from the previous section, where we considered the action on the boundary $x=X_{\text{min}}$, for $x=X_{\text{max}}.$
In the first two terms on the right-hand side of Eq.~\eqref{SIeq:master_mult_xNx}, we are interested in the $i=N_x$ terms only. 
In the first term, we need to consider the cases $j=q$ and $j=q-1$. 
First, looking at $j=q$ and using the change of variables $\bar{n}_{N_x+1}^{q+1}=n_{N_x+1}^{q+1}+1$ and $\bar{n}_{N_x+1}^q=n_{N_x+1}^q-1$ :
{\small{
\begin{align*}
    &\sum_{\bf{n}}\sum_{\bf{e}}n_{N_x+1}^q\mu_{-}(q+1, N_{N_x+1}, e_{N_x+1})\Big\{ (n_{N_x+1}^{q+1}+1)p(U_{N_x+1,q}^{\text{p}}{\bf{n}}, {\bf{e}},  t_h)\nonumber \\ &\qquad\qquad\qquad\qquad\qquad\qquad\qquad\qquad\qquad\qquad\qquad\qquad-n_{N_x+1}^{q+1}p({\bf{n}}, {\bf{e}},  t_h)\Big\} \nonumber \\
    &\quad= \sum_{\bf{n}}\sum_{\bf{e}}n_{N_x+1}^q \mu_{-}(q+1, N_{N_x+1}, e_{N_x+1})\times \nonumber \\ 
    &\qquad\qquad\Big\{ (n_{N_x+1}^{q+1}+1)p([n_{N_x+1}^1, \dots, n_{N_x+1}^{q}-1, n_{N_x+1}^{q+1}+1, \dots, n_{N_x+1}^{N_y+1}], {\bf{e}},  t_h)\nonumber \\ &\qquad\qquad\qquad\qquad\qquad\qquad\qquad\qquad\qquad\qquad\qquad\qquad-n_{N_x+1}^{q+1}p({\bf{n}}, {\bf{e}},  t_h)\Big\} \nonumber \\
&\quad=\sum_{\bf{n}}\sum_{\bf{e}} \mu_{-}(q+1, N_{N_x+1}, e_{N_x+1})\times \nonumber \\ 
    &\qquad\qquad\Big\{ \bar{n}_{N_x+1}^{q+1}(\bar{n}_{N_x+1}^q+1)p([n_{N_x+1}^1, \dots, \bar{n}_{N_x+1}^{q}, \bar{n}_{N_x+1}^{q+1}, \dots, n_{N_x+1}^{N_y+1}], {\bf{e}},  t_h)\nonumber \\ &\qquad\qquad\qquad\qquad\qquad\qquad\qquad\qquad\qquad\qquad\qquad\qquad-n_{N_x+1}^q n_{N_x+1}^{q+1}p({\bf{n}}, {\bf{e}},  t_h)\Big\} \nonumber \\
&\quad=\sum_{\bf{n}}\sum_{\bf{e}} \mu_{-}(q+1, N_{N_x+1}, e_{N_x+1})\Big\{ {n}_{N_x+1}^{q+1}(n_{N_x+1}^q+1)p({\bf{n}}, {\bf{e}},  t_h)\nonumber \\ &\qquad\qquad\qquad\qquad\qquad\qquad\qquad\qquad\qquad\qquad\qquad\qquad-n_{N_x+1}^q n_{N_x+1}^{q+1}p({\bf{n}}, {\bf{e}},  t_h)\Big\} \nonumber \\
&\quad=\sum_{\bf{n}}\sum_{\bf{e}} \mu_{-}(q+1, N_{N_x+1}, e_{N_x+1})n_{N_x+1}^{q+1}p({\bf{n}}, {\bf{e}},  t_h) \nonumber \\
    &\quad=\langle\mu_{-}(q+1, N_{N_x+1}, e_{N_x+1}) {n}_{N_x+1}^{q+1}\rangle.
\end{align*}
}}
Using $j=q$ and $i={N_x+1}$, with the change of variables $\bar{n}_{N_x+1}^{q-1}=n_{N_x+1}^{q-1}-1$ and $\bar{n}_{N_x+1}^q=n_{N_x+1}^q+1$ in the first term, then we have
{\small{
\begin{align*}
&\sum_{\bf{n}}\sum_{\bf{e}}n_{N_x+1}^q \mu_{-}(q, N_{N_x+1}, e_{N_x+1})\times \nonumber \\ 
    &\qquad\qquad\Big\{ (n_{N_x+1}^{q}+1)p([n_{N_x+1}^1, \dots, n_{N_x+1}^{q-1}-1, n_{N_x+1}^{q}+1, \dots, n_{N_x+1}^{N_y+1}], {\bf{e}},  t_h)\nonumber \\ &\qquad\qquad\qquad\qquad\qquad\qquad\qquad\qquad\qquad\qquad\qquad\qquad-n_{N_x+1}^{q}p({\bf{n}}, {\bf{e}},  t_h)\Big\} \nonumber \\
&\quad=\sum_{\bf{n}}\sum_{\bf{e}} \mu_{-}(q, N_{N_x+1}, e_{N_x+1})\times \nonumber \\ 
    &\qquad\qquad\Big\{ \bar{n}_{N_x+1}^{q}(\bar{n}_{N_x+1}^q-1)p([n_{N_x+1}^1, \dots, \bar{n}_{N_x+1}^{q-1}, \bar{n}_{N_x+1}^{q}, \dots, n_{N_x+1}^{N_y+1}], {\bf{e}},  t_h)\nonumber \\ &\qquad\qquad\qquad\qquad\qquad\qquad\qquad\qquad\qquad\qquad\qquad\qquad-n_{N_x+1}^q n_{N_x+1}^{q}p({\bf{n}}, {\bf{e}},  t_h)\Big\} \nonumber \\
&\quad=\sum_{\bf{n}}\sum_{\bf{e}} \mu_{-}(q, N_{N_x+1}, e_{N_x+1})\Big\{ {n}_{N_x+1}^{q}(n_{N_x+1}^{q}-1)p({\bf{n}}, {\bf{e}},  t_h)\nonumber \\ &\qquad\qquad\qquad\qquad\qquad\qquad\qquad\qquad\qquad\qquad\qquad\qquad-n_{N_x+1}^q n_{N_x+1}^{q}p({\bf{n}}, {\bf{e}},  t_h)\Big\} \nonumber \\
&\quad=-\sum_{\bf{n}}\sum_{\bf{e}} \mu_{-}(q, N_{N_x+1}, e_{N_x+1})n_{N_x+1}^{q}p({\bf{n}}, {\bf{e}}, t_h) \nonumber \\
    &\quad=-\langle\mu_{-}(q, N_{N_x+1}, e_{N_x+1}) {n}_{N_x+1}^{q}\rangle.
\end{align*}
}}
For the second term, we need to consider $j=q$ and $j=q+1$ whilst maintaining $i={N_x+1}$. 
For $i={N_x+1}$ and $j=q$, whilst using the change of variables $\bar{n}_{N_x+1}^{q-1}=n_{N_x+1}^{q-1}+1$ and $\bar{n}_{N_x+1}^q=n_{N_x+1}^q-1$ we have 
{\small{
\begin{align*}
    &\sum_{\bf{n}}\sum_{\bf{e}}n_{N_x+1}^q \mu_{+}(q-1, N_{N_x+1}, e_{N_x+1})\Big\{ (n_{N_x+1}^{q-1}+1)p(D_{N_x+1,q}^{\text{p}}{\bf{n}}, {\bf{e}},  t_h)\nonumber \\ &\qquad\qquad\qquad\qquad\qquad\qquad\qquad\qquad\qquad\qquad\qquad\qquad-n_{N_x+1}^{q-1}p({\bf{n}}, {\bf{e}},  t_h)\Big\} \nonumber \\
    &\quad=\sum_{\bf{n}}\sum_{\bf{e}}n_{N_x+1}^q \mu_{+}(q-1, N_{N_x+1}, e_{N_x+1})\times \nonumber \\ 
    &\qquad\qquad\Big\{ (n_{N_x+1}^{q-1}+1)p([n_{N_x+1}^1, \dots, n_{N_x+1}^{q-1}+1, n_{N_x+1}^{q}-1, \dots, n_{N_x+1}^{N_y+1}], {\bf{e}},  t_h)\nonumber \\ &\qquad\qquad\qquad\qquad\qquad\qquad\qquad\qquad\qquad\qquad\qquad\qquad-n_{N_x+1}^{q-1}p({\bf{n}}, {\bf{e}},  t_h)\Big\} \nonumber \\
    &\quad=\sum_{\bf{n}}\sum_{\bf{e}}\mu_{+}(q-1, N_{N_x+1}, e_{N_x+1})\times \nonumber \\ 
    &\qquad\qquad\Big\{ \bar{n}_{N_x+1}^{q-1}(\bar{n}_{N_x+1}^q+1)p([n_{N_x+1}^1, \dots, \bar{n}_{N_x+1}^{q-1}, \bar{n}_{N_x+1}^{q}, \dots, n_{N_x+1}^{N_y+1}], {\bf{e}},  t_h)\nonumber \\ &\qquad\qquad\qquad\qquad\qquad\qquad\qquad\qquad\qquad\qquad\qquad\qquad-n_{N_x+1}^qn_{N_x+1}^{q-1}p({\bf{n}}, {\bf{e}},  t_h)\Big\} \nonumber \\
    &\quad=\sum_{\bf{n}}\sum_{\bf{e}}\mu_{+}(q-1, N_{N_x+1}, e_{N_x+1})\Big\{ {n}_{N_x+1}^{q-1}({n}_{N_x+1}^q+1)p({\bf{n}}, {\bf{e}},  t_h)\nonumber \\ &\qquad\qquad\qquad\qquad\qquad\qquad\qquad\qquad\qquad\qquad\qquad\qquad-n_{N_x+1}^qn_{N_x+1}^{q-1}p({\bf{n}}, {\bf{e}},  t_h)\Big\} \nonumber \\
    &\quad= \sum_{\bf{n}}\sum_{\bf{e}}\mu_{+}(q-1, N_{N_x+1}, e_{N_x+1}) {n}_{N_x+1}^{q-1}p({\bf{n}}, {\bf{e}},  t_h) \nonumber \\
    &\quad=\langle\mu_{+}(q-1, N_{N_x+1}, e_{N_x+1}) {n}_{N_x+1}^{q-1}\rangle.
\end{align*}
}}
Now looking at $j=q+1$ and $i={N_x}$, the change of variables $\bar{n}_{N_x+1}^{q+1}=n_{N_x+1}^{q+1}-1$ and $\bar{n}_{N_x+1}^q=n_{N_x+1}^q+1$ gives
{\small{
\begin{align*}
    &\sum_{\bf{n}}\sum_{\bf{e}}n_{N_x+1}^q \mu_{+}(q, N_{N_x+1}, e_{N_x+1})\left\{ (n_{N_x+1}^{q}+1)p(D_{N_x+1,q}^{\text{p}}{\bf{n}}, {\bf{e}},  t_h)-n_{N_x+1}^{q}p({\bf{n}}, {\bf{e}},  t_h)\right\} \nonumber \\
    &\quad=\sum_{\bf{n}}\sum_{\bf{e}}n_{N_x+1}^q \mu_{+}(q, N_{N_x+1}, e_{N_x+1})\times \nonumber \\ 
    &\qquad\qquad\Big\{ (n_{N_x+1}^{q}+1)p([n_{N_x+1}^1, \dots, n_{N_x+1}^{q}+1, n_{N_x+1}^{q+1}-1, \dots, n_{N_x+1}^{N_y+1}], {\bf{e}},  t_h)\nonumber \\ &\qquad\qquad\qquad\qquad\qquad\qquad\qquad\qquad\qquad\qquad\qquad\qquad-n_{N_x+1}^{q}p({\bf{n}}, {\bf{e}},  t_h)\Big\} \nonumber \\
    &\quad=\sum_{\bf{n}}\sum_{\bf{e}}\mu_{+}(q, N_{N_x+1}, e_{N_x+1})\times \nonumber \\ 
    &\qquad\qquad\Big\{ \bar{n}_{N_x+1}^{q}(\bar{n}_{N_x+1}^q-1)p([n_{N_x+1}^1, \dots, \bar{n}_{N_x+1}^{q}, \bar{n}_{N_x+1}^{q+1}, \dots, n_{N_x+1}^{N_y+1}], {\bf{e}},  t_h)\nonumber \\ &\qquad\qquad\qquad\qquad\qquad\qquad\qquad\qquad\qquad\qquad\qquad\qquad-n_{N_x+1}^qn_{N_x+1}^{q}p({\bf{n}}, {\bf{e}},  t_h)\Big\} \nonumber \\
    &\quad=\sum_{\bf{n}}\sum_{\bf{e}}\mu_{+}(q, N_{N_x+1}, e_{N_x+1})\left\{ {n}_{N_x+1}^{q}({n}_{N_x+1}^q-1)p({\bf{n}}, {\bf{e}},  t_h)-n_{N_x+1}^qn_{N_x+1}^{q}p({\bf{n}}, {\bf{e}},  t_h)\right\} \nonumber \\
    &\quad= -\sum_{\bf{n}}\sum_{\bf{e}}\mu_{+}(q, N_{N_x+1}, e_{N_x+1}) {n}_{N_x+1}^{q}p({\bf{n}}, {\bf{e}},  t_h) \nonumber \\
    &\quad=-\langle\mu_{+}(q, N_{N_x+1}, e_{N_x+1}) {n}_{N_x+1}^{q}\rangle.
\end{align*}
}}
Next, we want to consider the movement terms in physical space. 
In these cases, we will only have non-zero contributions when $j=q$.
Looking at the first term, we will have contributions only when $i=N_x.$
Then, using the change of variables $\bar{n}_{N_x+1}^q=n_{N_x+1}^q+1$ and $\bar{n}_{N_x}^q=n_{N_x}^q-1$ and dropping the bar{{s}}, we get
\begin{align*}
    &\sum_{\bf{n}}\sum_{\bf{e}}n_{N_x+1}^q\beta_{-}(q, {N_{N_x},e_{N_x})}\left\{ (n_{N_x+1}^{q}+1)p(R_{N_x,q}^{\text{m}}{\bf{n}}, {\bf{e}},  t_h)-n_{N_x+1}^{q}p({\bf{n}}, {\bf{e}},  t_h)\right\} \nonumber \\
    &\quad=\sum_{\bf{n}}\sum_{\bf{e}}n_{N_x+1}^q\beta_{-}(q, {N_{N_x},e_{N_x})}\times \nonumber \\ 
    &\qquad\qquad\left\{ (n_{N_x+1}^{q}+1)p([n_1^q, \dots, n_{N_x}^q-1, n_{N_x+1}^{q}+1], {\bf{e}},  t_h)-n_{N_x+1}^{q}p({\bf{n}}, {\bf{e}},  t_h)\right\} \nonumber \\
    &\quad=\sum_{\bf{n}}\sum_{\bf{e}}n_{N_x}^q \beta_{-}(q, {N_{N_x},e_{N_x})}\times \nonumber \\ 
    &\qquad\qquad\left\{ (n_{N_x+1}^{q}+1)p([n_1^q, \dots, n_{N_x}^q-1, n_{N_x+1}^{q}+1], {\bf{e}},  t_h)-n_{N_x+1}^{q}p({\bf{n}}, {\bf{e}},  t_h)\right\},  \nonumber \\
    &\quad=\sum_{\bf{n}}\sum_{\bf{e}} \beta_{-}(q, {N_{N_x},e_{N_x})}\times \nonumber \\ 
    &\qquad\qquad\left\{ (\bar{n}_{N_x}^q-1)\bar{n}_{N_x+1}^{q}p([n_1^q, \dots, \bar{n}_{N_x}^q, \bar{n}_{N_x+1}^{q}], {\bf{e}},  t_h)-n_{N_x}^qn_{N_x+1}^{q}p({\bf{n}}, {\bf{e}},  t_h)\right\} \nonumber \\
    &\quad=\sum_{\bf{n}}\sum_{\bf{e}} \beta_{-}(q, {N_{N_x},e_{N_x})}\times \nonumber \\ 
    &\qquad\qquad\left\{ ({n}_{N_x}^q-1){n}_{N_x+1}^{q}p({\bf{n}}, {\bf{e}},  t_h)-n_{N_x}^qn_{N_x+1}^{q}p({\bf{n}}, {\bf{e}},  t_h)\right\} \nonumber \\
    &\quad=-\sum_{\bf{n}}\sum_{\bf{e}} \beta_{-}(q, {N_{N_x},e_{N_x})} {n}_{N_x+1}^{q}p({\bf{n}}, {\bf{e}},  t_h) \nonumber \\
    &\quad=-\langle\beta_{-}(q, {N_{N_x},e_{N_x})} {n}_{N_x+1}^{q}\rangle, 
\end{align*}
Now we can repeat this for $j=q$ and $i=N_x+1$ in the fourth term of Eq.~\eqref{SIeq:master_mult_xNx}, using the change of variable $\bar{n}_{N_x+1}^q=n_{N_x+1}^q-1$ and $\bar{n}_{N_x}^q=n_{N_x}^q+1$:
\begin{align*}
    & \sum_{\bf{n}}\sum_{\bf{e}}n_{N_x+1}^q\beta_{+}(q, N_{N_x+1},e_{N_x+1})\left\{ (n_{N_x}^{q}+1)p(L_{N_x, q}^{\text{m}}{\bf{n}}, {\bf{e}},  t_h)-n_{N_x}^{q}p({\bf{n}}, {\bf{e}},  t_h)\right\},  \nonumber \\
    &\quad=  \sum_{\bf{n}}\sum_{\bf{e}}n_{N_x+1}^q\beta_{+}(q, N_{N_x+1},e_{N_x+1})\times \nonumber \\ 
    &\qquad\qquad\left\{ (n_{N_x}^{q}+1)p([ n_1^q, \dots, n_{N_x}^q+1, n_{N_x+1}^{q}-1], {\bf{e}},  t_h)-n_{N_x}^{q}p({\bf{n}}, {\bf{e}},  t_h)\right\} \nonumber \\ 
    &\quad=  \sum_{\bf{n}}\sum_{\bf{e}}\beta_{+}(q, N_{N_x+1},e_{N_x+1})\times \nonumber \\ 
    &\qquad\qquad\left\{ \bar{n}_{N_x}^{q}(\bar{n}_{N_x+1}^q+1)p([n_1^q, \dots , \bar{n}_{N_x}^q, \bar{n}_{N_x+1}^{q}], {\bf{e}},  t_h)-n_{N_x+1}^qn_{N_x}^{q}p({\bf{n}}, {\bf{e}},  t_h)\right\} \nonumber \\ 
    &\quad=  \sum_{\bf{n}}\sum_{\bf{e}}\beta_{+}(q, N_{N_x+1},e_{N_x+1})\left\{ {n}_{N_x}^{q}({n}_{N_x+1}^q+1)p({\bf{n}}, {\bf{e}},  t_h)-n_{N_x}^qn_{N_x+1}^{q}p({\bf{n}}, {\bf{e}},  t_h)\right\} \nonumber \\  
    &\quad=  \sum_{\bf{n}}\sum_{\bf{e}}\beta_{+}(q, N_{N_x+1},e_{N_x+1}) {n}_{N_x}^{q}p({\bf{n}}, {\bf{e}},  t_h) \nonumber \\ 
    &\quad=  \langle\beta_{+}(q, N_{N_x+1},e_{N_x+1}) {n}_{N_x}^{q}\rangle. 
\end{align*}
The final term on right-hand side of Eq.~\eqref{SIeq:master_mult_xNx} has non-zero contributions when $j=q$ and $i=N_x+1$ only. 
We employ the change of variable $\bar{n}_{N_x+1}^q= n_{N_x+1}^q-1$ in the second term to get
{\small{
\begin{align*}
    & \sum_{\bf{n}}\sum_{\bf{e}}n_{N_x+1}^q\Big\{\gamma(q, N_{N_x+1}-1,e_{N_x+1})(n_{N_x+1}^q-1)p(G_{N_x+1,q}{\bf{n}}, {\bf{e}},  t_h)\nonumber \\ &\qquad\qquad\qquad\qquad\qquad\qquad\qquad\qquad\qquad\qquad-n_{N_x+1}^q\gamma(q, N_{N_x+1},e_{N_x+1})p({\bf{n}}, {\bf{e}},  t_h)\Big\} \nonumber \\ 
    &\quad= \sum_{\bf{n}}\sum_{\bf{e}}n_{N_x+1}^q\big\{\gamma(q, N_{N_x+1}-1,e_{N_x+1})(n_{N_x+1}^q-1)p([ n_{1}^q, \dots, n_{N_x+1}^q-1], {\bf{e}},  t_h) \nonumber \\ 
    &\qquad\qquad\qquad\qquad\qquad-n_{N_x+1}^q\gamma(q, N_{N_x+1},e_{N_x+1})p({\bf{n}}, {\bf{e}},  t_h)\big\}\nonumber \\
    &\quad= \sum_{\bf{n}}\sum_{\bf{e}}\big\{\gamma(q, \bar{N_{N_x+1}},e_{N_x+1})(\bar{n}_{N_x+1}^q-1)\bar{n}_{N_x+1}^qp([n_1^q, \dots , \bar{n}_{N_x+1}^q], {\bf{e}},  t_h)\nonumber \\
    &\qquad\qquad\qquad\qquad\qquad -\gamma(q, N_{N_x+1},e_{N_x+1})(n_{N_x+1}^q)^2p({\bf{n}}, {\bf{e}},  t_h)\big\}\nonumber \\ 
 &\quad= \sum_{\bf{n}}\sum_{\bf{e}}\gamma(q, N_{N_x+1},e_{N_x+1})\left\{({n}_{N_x+1}^q-1){n}_{N_x+1}^qp({\bf{n}}, {\bf{e}},  t_h)-(n_{N_x+1}^q)^2p({\bf{n}}, {\bf{e}},  t_h)\right\} \nonumber \\ 
 &\quad= \sum_{\bf{n}}\sum_{\bf{e}}\gamma(q, N_{N_x+1},e_{N_x+1})n_{N_x+1}^qp({\bf{n}}, {\bf{e}},  t_h) \nonumber \\ 
 &\quad= \langle\gamma(q, N_{N_x+1},e_{N_x+1})n_{N_x+1}^q\rangle. 
\end{align*}
}}
Putting these together, the equation for cell evolution on the right-hand boundary in physical space becomes: 
\begin{align}
    \dfrac{\partial}{\partial t} \langle n_{N_x+1}^q\rangle &=\dfrac{1}{\Delta_t}\langle\mu_{-}(q+1, N_{N_x+1}, e_{N_x+1}) {n}_{N_x+1}^{q+1}\rangle + \dfrac{1}{\Delta_t}\langle\mu_{+}(q-1, N_{N_x+1}, e_{N_x+1}) {n}_{N_x+1}^{q-1}\rangle\nonumber \\ &\quad-\dfrac{1}{\Delta_t}\langle\mu_{-}(q, N_{N_x+1}, e_{N_x+1}) {n}_{N_x+1}^{q}\rangle -\dfrac{1}{\Delta_t}\langle\mu_{+}(q, N_{N_x+1}, e_{N_x+1}) {n}_{N_x+1}^{q}\rangle\nonumber \\ &\quad+  \dfrac{1}{\Delta_t}\langle\beta_{+}(q, N_{N_x+1},e_{N_x+1}) {n}_{N_x}^{q}\rangle -\dfrac{1}{\Delta_t}\langle\beta_{-}(q, {N_{N_x},e_{N_x})} {n}_{N_x+1}^{q}\rangle \nonumber \\ &\quad+\dfrac{1}{\Delta_t}  \langle\gamma(q, N_{N_x+1},e_{N_x+1})n_{N_x+1}^q\rangle . \label{SIeq:BC_xNx}
\end{align}

Now we want to find the continuum equivalent of this equation. 
Recalling the {{continuum equivalents of the dependent variables}} whilst employing Taylor series expansions around $x=X_{\text{max}}$,  we can rewrite Eq.~\eqref{SIeq:BC_xNx} as (dropping the dependent variables for simplicity)
\begin{align*}
    \Delta_t \dfrac{\partial n}{\partial t} &= \left(\mu_{-}+\Delta_y\dfrac{\partial \mu_{-}}{\partial y}+\dfrac{\Delta_y^2}{2}\dfrac{\partial^2 \mu_{-}}{\partial y^2}+\dots\right)\left(n+\Delta_y\dfrac{\partial n}{\partial y}+\dfrac{\Delta_y^2}{2}\dfrac{\partial^2 n}{\partial y^2}+\dots\right) \nonumber \\ 
    &\quad +\left(\mu_{+}+\Delta_y\dfrac{\partial \mu_{+}}{\partial y}+\dfrac{\Delta_y^2}{2}\dfrac{\partial^2 \mu_{+}}{\partial y^2}+\dots\right)\left(n+\Delta_y\dfrac{\partial n}{\partial y}+\dfrac{\Delta_y^2}{2}\dfrac{\partial^2 n}{\partial y^2}+\dots\right) \nonumber \\ 
    &\quad -\mu_{+}n-\mu_{-}n +\gamma n + \beta_{+} \left(n+\Delta_x\dfrac{\partial n}{\partial x}+\dfrac{\Delta_x^2}{2}\dfrac{\partial^2 n}{\partial x^2}+\dots\right) \nonumber \\ 
    &\quad -n\left(\beta_{-}+\Delta_x\dfrac{\partial \beta_{-}}{\partial x}+\dfrac{\Delta_x^2}{2}\dfrac{\partial^2 \beta_{-}}{\partial x^2}+\dots\right),
\end{align*}
at $x=X_{\text{max}}$ so that 
\begin{align*}
    \dfrac{\partial n}{\partial t}&= \dfrac{\Delta_y}{\Delta_t}\dfrac{\partial}{\partial y}\left(\left(\mu_{-}-\mu_{+}\right)n\right)+\dfrac{\Delta_y^2}{2\Delta_t}\dfrac{\partial^2}{\partial y^2}\left(\left(\mu_{-}+\mu_{+}\right)n\right) - \dfrac{1}{\Delta_t}n\left(\beta_{-}-\beta_{+}\right) \nonumber \\ &\quad +\dfrac{\Delta_x}{\Delta_t} \left(\beta_{+} \dfrac{\partial n}{\partial x} - n \dfrac{\partial \beta_{-}}{\partial x}\right) +\dfrac{\Delta_x^2}{2\Delta_t}\left( \beta_{+} \dfrac{\partial^2 n}{\partial x^2}- n \dfrac{\partial^2 \beta_{-}}{\partial x^2}\right)+\dfrac{1}{\Delta_t}\gamma n .
\end{align*}
Recalling that
{\small{
\begin{align*}
    \lim_{\Delta_x,  \Delta_t\rightarrow 0} \dfrac{\Delta_x}{\Delta_t} \Big(\beta_{-}(y, \rho(x,t), e(x,t))-\beta_{+}(y, \rho(x,t), e(x,t))\Big) &= v^m(y, \rho(x,t), e(x,t)), \\
    \lim_{\Delta_x,  \Delta_t\rightarrow 0} \dfrac{\Delta_x^2}{2\Delta_t} \Big(\beta_{-}(y, \rho(x,t), e(x,t))+\beta_{+}(y, \rho(x,t), e(x,t))\Big) &= D^m(y, \rho(x,t), e(x,t)), \\
    \lim_{\Delta_y,  \Delta_t\rightarrow 0} \dfrac{\Delta_y}{\Delta_t} \Big(\mu_{-}\left(y, \rho(x,t), e(x,t)\right)-\mu_{+}\left(y, \rho(x,t), e(x,t)\right)\Big) &= v^p(y, \rho(x,t), e(x,t)), \\
    \lim_{\Delta_y,  \Delta_t\rightarrow 0} \dfrac{\Delta_y^2}{2\Delta_t} \Big(\mu_{-}\left(y, \rho(x,t), e(x,t)\right)+\mu_{+}\left(y, \rho(x,t), e(x,t)\right)\Big) &= D^p(y, \rho(x,t), e(x,t)), \\
    \lim_{\Delta_t\rightarrow 0} \dfrac{1}{\Delta_t} \gamma\left(y, \rho(x,t), e(x,t)\right) &= r(y, \rho(x,t), e(x,t)),
\end{align*}
}}
such that 
\begin{align*}
    \beta_{\pm}&=\dfrac{D^m\Delta_t}{\Delta_x^2}\mp\dfrac{v^m\Delta_t}{2\Delta_x},
\end{align*}
then the equation can be rewritten as
\begin{align*}
    \dfrac{\partial n}{\partial t}&= \dfrac{\partial}{\partial y}\left(v^pn\right)+\dfrac{\partial^2}{\partial y^2}\left(D^pn\right) - \dfrac{1}{\Delta_x}v^mn +\dfrac{1}{2}v^m\dfrac{\partial n}{\partial x} + \dfrac{1}{\Delta_x}D^m\dfrac{\partial n}{\partial x} \nonumber \\ &\quad -\dfrac{1}{\Delta_x}n\dfrac{\partial}{\partial x}D^m-\dfrac{1}{2}n\dfrac{\partial}{\partial x}v^m+\dfrac{1}{2}D^m\dfrac{\partial^2n}{\partial x^2}-\dfrac{1}{2}n\dfrac{\partial^2}{\partial x^2}D^m+rn .
\end{align*}
In order to prevent blow-up of terms in the limit $\Delta_x\rightarrow 0$, we require
\begin{equation*}
    -v^mn +D^m\dfrac{\partial n}{\partial x}-n\dfrac{\partial}{\partial x}D^m = 0  \qquad\qquad \text{at} \quad x=X_{\text{max}}.
\end{equation*}
As such, we have no flux of cells out of the physical space boundary at $x=X_{\text{max}}.$

%%%%%%%%%%%%%%%%%%%%%%%%%%
\paragraph{Boundary condition at $y=Y_{\text{min}}$.}
Returning to Eq.~\eqref{SIeq:master}, we seek an equation for the evolution of the cell number on the left most lattice site in phenotype space, \textit{i.e.}, at $y_1$. 
To find this, we multiply Eq.~\eqref{SIeq:master} by $n_{s}^{1}$ and sum over all possible states ${\bf{n}}$ and ${\bf{e}}$:
{\small{
\begin{align}
    &\Delta_t \sum_{\bf{n}}\sum_{\bf{e}}n_{s}^{1} \dfrac{\partial}{\partial t} p({\bf{n}}, {\bf{e}},  t_h)+O(\Delta_t^2) \nonumber \\ 
    &\quad= \sum_{\bf{n}}\sum_{\bf{e}}n_{s}^{1} \sum_{i=1}^{N_x+1}\sum_{j=1}^{N_y}\mu_{-}(j+1, N_i, e_i)\times \nonumber \\ 
    &\qquad\qquad\qquad\qquad\qquad\qquad\left\{ (n_i^{j+1}+1)p(U_{i,j}^{\text{p}}{\bf{n}}, {\bf{e}},  t_h)-n_i^{j+1}p({\bf{n}}, {\bf{e}},  t_h)\right\} \nonumber \\ 
    &\qquad + \sum_{\bf{n}}\sum_{\bf{e}}n_{s}^{1} \sum_{i=1}^{N_x+1}\sum_{j=2}^{N_y+1}\mu_{+}(j-1, N_i, e_i)\times \nonumber \\ 
    &\qquad\qquad\qquad\qquad\qquad\qquad\left\{ (n_i^{j-1}+1)p(D_{i,j}^{\text{p}}{\bf{n}}, {\bf{e}},  t_h)-n_i^{j-1}p({\bf{n}}, {\bf{e}},  t_h)\right\} \nonumber \\
    &\qquad + \sum_{\bf{n}}\sum_{\bf{e}}n_{s}^{1}\sum_{i=1}^{N_x} \sum_{j=1}^{N_y+1}\beta_{-}(j, {N_i,e_i)}\left\{ (n_{i+1}^{j}+1)p(R_{i,j}^{\text{m}}{\bf{n}}, {\bf{e}},  t_h)-n_{i+1}^{j}p({\bf{n}}, {\bf{e}},  t_h)\right\} \nonumber \\ 
    &\qquad + \sum_{\bf{n}}\sum_{\bf{e}}n_{s}^{1}\sum_{i=2}^{N_x+1} \sum_{j=1}^{N_y+1}\beta_{+}(j, N_i,e_i)\left\{ (n_{i-1}^{j}+1)p(L_{i,j}^{\text{m}}{\bf{n}}, {\bf{e}},  t_h)-n_{i-1}^{j}p({\bf{n}}, {\bf{e}},  t_h)\right\}\nonumber \\ 
    &\qquad + \sum_{\bf{n}}\sum_{\bf{e}}n_{s}^{1}\sum_{i=1}^{N_x+1} \sum_{j=1}^{N_y+1}\Big\{\gamma(j, N_i-1,e_i)(n_i^j-1)p(G_{i,j}{\bf{n}}, {\bf{e}},  t_h)\nonumber \\ &\qquad\qquad\qquad\qquad\qquad\qquad\qquad\qquad\qquad\qquad\qquad\qquad-\gamma(j, N_i,e_i)n_i^jp({\bf{n}}, {\bf{e}},  t_h)\Big\} .\label{SIeq:master_mult_y1}
\end{align}
}}

Using the same methods as on the boundaries in physical space, we can change variables in each term to find an equation for evolution of cell number. 
Consider the first term on the right-hand side. 
The non-zero contributions come from when $j=1$ and $i=s$.
In the second term, the non-zero terms are for $j=2$ and $i=s$. 
The third term gives non-zero contributions when $j=1$ and $i=s$ or $i=s-1$. 
In the fourth term, there are non-zero contributions when $j=1$ and $i=s$ or $i=s+1$.
The final term produces non-zero contributions only when $i=s$ and $j=1.$
Employing this knowledge, Eq.~\eqref{SIeq:master_mult_y1} becomes
\begin{align}
    &\Delta_t \sum_{\bf{n}}\sum_{\bf{e}}n_{s}^{1} \dfrac{\partial}{\partial t} p({\bf{n}}, {\bf{e}},  t_h) +O(\Delta_t^2)  \nonumber \\ 
    &\quad=\sum_{\bf{n}}\sum_{\bf{e}}n_{s}^{1} \mu_{-}(2, N_{s}, e_s)\left\{ (n_s^{2}+1)p(U_{s, 1}^{\text{p}}{\bf{n}}, {\bf{e}},  t_h)-n_s^{2}p({\bf{n}}, {\bf{e}},  t_h)\right\} \nonumber \\ 
    &\qquad + \sum_{\bf{n}}\sum_{\bf{e}}n_{s}^{1} \mu_{+}(1, N_{s}, e_s)\left\{ (n_s^{1}+1)p(D_{s,2}^{\text{p}}{\bf{n}}, {\bf{e}},  t_h)-n_s^{1}p({\bf{n}}, {\bf{e}},  t_h)\right\} \nonumber \\
    &\qquad + \sum_{\bf{n}}\sum_{\bf{e}}n_{s}^{1}\sum_{i=s,  s-1}\beta_{-}(1, {N_i,e_i)}\left\{ (n_{i+1}^{1}+1)p(R_{i,1}^{\text{m}}{\bf{n}}, {\bf{e}},  t_h)-n_{i+1}^{1}p({\bf{n}}, {\bf{e}},  t_h)\right\} \nonumber \\ 
    &\qquad + \sum_{\bf{n}}\sum_{\bf{e}}n_{s}^{1}\sum_{i=s, s+1}\beta_{+}(1, N_i,e_i)\left\{ (n_{i-1}^{1}+1)p(L_{i,1}^{\text{m}}{\bf{n}}, {\bf{e}},  t_h)-n_{i-1}^{1}p({\bf{n}}, {\bf{e}},  t_h)\right\}\nonumber \\ 
    &\qquad + \sum_{\bf{n}}\sum_{\bf{e}}n_{s}^{1}\gamma(1, N_{s},e_s)\left\{n_s^{1}p({\bf{n}}, {\bf{e}},  t_h)-(n_s^{1}+1)p(G_{s,1}{\bf{n}}, {\bf{e}},  t_h)\right\}.\label{SIeq:master_mult_y12}
\end{align}
The first term can be rewritten using the change of variables $\bar{n}_s^{2}=n_s^{2}+1$ and $\bar{n}_s^{1}=n_s^{1}-1$
\begin{align*}
    &\sum_{\bf{n}}\sum_{\bf{e}}n_{s}^{1} \mu_{-}(2, N_{s}, e_s)\left\{ (n_s^{2}+1)p(U_{s, 1}^{\text{p}}{\bf{n}}, {\bf{e}},  t_h)-n_s^{N_y+1}p({\bf{n}}, {\bf{e}},  t_h)\right\} \nonumber \\ 
    &\quad= \sum_{\bf{n}}\sum_{\bf{e}}n_{s}^{1} \mu_{-}(2, N_{s}, e_s)\times \nonumber \\ 
    &\qquad\qquad\qquad\qquad\left\{ (n_s^{2}+1)p([n_s^1-1, n_{s}^{2}+1, \dots n_{s}^{N_y+1}], {\bf{e}},  t_h)-n_s^{2}p({\bf{n}}, {\bf{e}},  t_h)\right\} \nonumber \\ 
    &\quad=  \sum_{\bf{n}}\sum_{\bf{e}}\mu_{-}(2, N_{s}, e_s)\times \nonumber \\ 
    &\qquad\qquad\qquad\qquad\left\{ \bar{n}_s^{2}(\bar{n}_{s}^{1} +1)p([\bar{n}_s^1, \bar{n}_s^2, \dots, \bar{n}_{s}^{N_y+1}], {\bf{e}},  t_h)-n_s^{2}n_{s}^{1} p({\bf{n}}, {\bf{e}},  t_h)\right\} \nonumber \\ 
    &\quad=  \sum_{\bf{n}}\sum_{\bf{e}}\mu_{-}(2, N_{s}, e_s)\left\{ {n}_s^{2}({n}_{s}^{1} +1)p({\bf{n}}, {\bf{e}},  t_h)-n_s^{1}n_{s}^{2} p({\bf{n}}, {\bf{e}},  t_h)\right\} \nonumber \\ 
    &\quad=  \sum_{\bf{n}}\sum_{\bf{e}}\mu_{-}(2, N_{s}, e_s) {n}_s^{2}p({\bf{n}}, {\bf{e}},  t_h) \nonumber \\ 
    &\quad=  \langle\mu_{-}(2, N_{s}, e_s) {n}_s^{2}\rangle.
\end{align*}
The second term in the right-hand side of Eq.~\eqref{SIeq:master_mult_y12} can then be rewritten as the following (using $\bar{n}_s^{2}=n_s^{2}-1$ and $\bar{n}_s^{1}=n_s^{1}+1$):
\begin{align*}
    &\sum_{\bf{n}}\sum_{\bf{e}}n_{s}^{{{1}}} \mu_{+}({1}, N_{s}, e_s)\left\{ (n_s^{{1}}+1)p(D_{s,1}^{\text{p}}{\bf{n}}, {\bf{e}},  t_h)-n_s^{{1}}p({\bf{n}}, {\bf{e}},  t_h)\right\} \nonumber \\ 
    &\quad= \sum_{\bf{n}}\sum_{\bf{e}}n_{s}^{{1}} \mu_{+}({1}, N_{s}, e_s)\times \nonumber \\ 
    &\qquad\qquad\qquad\qquad\left\{ (n_s^{{1}}+1)p([n_{s}^{{1}}+1, n_{s}^{{2}}-1, \dots, n_s^{N_y+1}], {\bf{e}},  t_h)-n_s^{{1}}p({\bf{n}}, {\bf{e}},  t_h)\right\} \nonumber \\ 
    &\quad=  \sum_{\bf{n}}\sum_{\bf{e}}\mu_{+}({1}, N_{s}, e_s)\times \nonumber \\ 
    &\qquad\qquad\qquad\qquad\left\{ \bar{n}_s^{{1}}(\bar{n}_{s}^{1}-1 )p([\bar{n}_s^1, \bar{n}_s^2, \dots, {n}_{s}^{{N_y}}], {\bf{e}},  t_h)-n_s^{{1}}n_{s}^{{1}} p({\bf{n}}, {\bf{e}},  t_h)\right\} \nonumber \\ 
    &\quad=  \sum_{\bf{n}}\sum_{\bf{e}}\mu_{+}({1}, N_{s}, e_s)\left\{ {n}_s^{{1}}({n}_{s}^{{2}} +1)p({\bf{n}}, {\bf{e}},  t_h)-n_s^{{1}}n_{s}^{{1}} p({\bf{n}}, {\bf{e}},  t_h)\right\} \nonumber \\ 
    &\quad=  -\sum_{\bf{n}}\sum_{\bf{e}}\mu_{+}({1}, N_{s}, e_s) {n}_s^{{1}}p({\bf{n}}, {\bf{e}},  t_h) \nonumber \\ 
    &\quad=  -\langle\mu_{+}({1}, N_{s}, e_s) {n}_s^{{1}}\rangle.
\end{align*}
The contributions from the terms describing movement in physical space are the same as in Eq.~\eqref{SIeq:full_IBM_prolif_ECM} with $j=1$
and the growth terms are also the same as those in Eq.~\eqref{SIeq:full_IBM_prolif_ECM} with $i=s$ and $j=1.$
Putting these together, we get 
\begin{align}
    \dfrac{\partial}{\partial t}\langle n_s^{1}\rangle &=\dfrac{1}{\Delta_t} \langle\mu_{-}(2, N_{s}, e_s) {n}_s^{2}\rangle- \dfrac{1}{\Delta_t}\langle\mu_{+}({1}, N_{s}, e_s) {n}_s^{{1}}\rangle \nonumber \\ &\quad+ \dfrac{1}{\Delta_t}\langle\beta_{+}({1}, N_{s},e_s)n_{s-1}^{1}\rangle +\dfrac{1}{\Delta_t} \langle\beta_{-}({1}, N_{s},e_s)n_{s+1}^{1}\rangle \nonumber \\ 
    &\quad-\dfrac{1}{\Delta_t} \langle\beta_{-}({1}, N_{s-1},e_{s-1})n_{s}^{1}\rangle -\dfrac{1}{\Delta_t}\langle\beta_{+}({1}, N_{s+1},e_{s+1})n_{s}^{1}\rangle \nonumber \\ &\quad+\dfrac{1}{\Delta_t}  \langle\gamma({1}, N_{s},e_{s})n_{s}^{1}\rangle . \label{SIeq:BC_y1}
\end{align}

Then, in the limit $\Delta_x,  \Delta_y, \Delta_t \rightarrow 0$, the {{continuum equivalents of the dependent variables}}, Eq.~\eqref{SIeq:BC_y1} can be rewritten at $y=Y_{\text{min}}$ as follows:
\begin{align*}
    \dfrac{\partial}{\partial t} n(x, y, t) &= \dfrac{1}{\Delta_t}\beta_{+}(y, \rho(x,  t),e(x,  t))n(x-\Delta_x,  y,  t)\nonumber \\ &\quad +\dfrac{1}{\Delta_t} \beta_{-}(y, \rho(x,  t),e(x,  t))n (x+\Delta_x,  y,  t) \nonumber \\ 
    &\quad-\dfrac{1}{\Delta_t} \beta_{-}(y, \rho(x-\Delta_x,  t),e(x-\Delta_x,  t))n(x,y,t) \nonumber \\ &\quad-\dfrac{1}{\Delta_t}\beta_{+}(y, \rho(x+\Delta_x,  t),e(x+\Delta_x,  t))n(x,y,t) \nonumber \\ 
    &\quad+\dfrac{1}{\Delta_t}\mu_{-}(y+\Delta_y, \rho(x,  t),e(x,  t))n(x,  y+\Delta_y,  t)\nonumber \\ &\quad -\dfrac{1}{\Delta_t}\mu_{+}(y, \rho(x, t),e(x,  t))n(x,y,t)\nonumber\\
    &\quad+\dfrac{1}{\Delta_t} \gamma(y, \rho(x,  t),e(x,  t))n(x,y,t). 
\end{align*}
Then, employing the aforementioned Taylor expansions, we find 
\begin{align*}
    \dfrac{\partial}{\partial t} n(x, y, t) &= \dfrac{1}{\Delta_t}\beta_{+}(y, \rho(x,  t),e(x,  t))\times\nonumber\\ &\qquad\qquad\qquad\left[n(x,y,t)-\Delta_x\dfrac{\partial}{\partial x}n(x,y,t)+\dfrac{\Delta_x^2}{2}\dfrac{\partial^2}{\partial x^2}n(x,y,t)\right]\nonumber \\ &\quad +\dfrac{1}{\Delta_t} \beta_{-}(y, \rho(x,  t),e(x,  t))\times\nonumber\\ &\qquad\qquad\qquad\left[n(x,y,t)+\Delta_x\dfrac{\partial}{\partial x}n(x,y,t)+\dfrac{\Delta_x^2}{2}\dfrac{\partial^2}{\partial x^2}n(x,y,t)\right]\nonumber \\ 
    &\quad-\dfrac{1}{\Delta_t}n(x,y,t) \times\nonumber\\ &\qquad\qquad\qquad\Bigg[\beta_{-}(y, \rho(x, t),e(x, t))- \Delta_x\dfrac{\partial}{\partial x}\beta_{-}(y, \rho(x,t), e(x,t)\nonumber \\ &\qquad\qquad\qquad\qquad\qquad\qquad\qquad+\dfrac{\Delta_x^2}{2}\dfrac{\partial^2}{\partial x^2}\beta_{-}(y, \rho(x,t), e(x,t)\Bigg]\nonumber \\ &\quad-\dfrac{1}{\Delta_t}n(x,y,t)\times\nonumber\\ &\qquad\qquad\qquad\Bigg[\beta_{+}(y, \rho(x, t),e(x, t))+ \Delta_x\dfrac{\partial}{\partial x}\beta_{+}(y, \rho(x,t), e(x,t)\nonumber \\ &\qquad\qquad\qquad\qquad\qquad\qquad\qquad+\dfrac{\Delta_x^2}{2}\dfrac{\partial^2}{\partial x^2}\beta_{+}(y, \rho(x,t), e(x,t)\Bigg]\nonumber \\ 
    &\quad+\dfrac{1}{\Delta_t}\left[n(x,y,t)+\Delta_y\dfrac{\partial}{\partial y}n(x,y,t)+\dfrac{\Delta_y^2}{2}\dfrac{\partial^2}{\partial y^2}n(x,y,t)\right]\times\nonumber \\ &\qquad\qquad\qquad\Bigg[\mu_{-}(y, \rho(x, t),e(x, t))+ \Delta_y\dfrac{\partial}{\partial y}\mu_{-}(y, \rho(x,t), e(x,t)\nonumber \\ &\qquad\qquad\qquad\qquad\qquad\qquad\qquad+\dfrac{\Delta_y^2}{2}\dfrac{\partial^2}{\partial y^2}\mu_{-}(y, \rho(x,t), e(x,t)\Bigg]\nonumber \\ &\quad -\dfrac{1}{\Delta_t}\mu_{+}(y, \rho(x, t),e(x,  t))n(x,y,t)\nonumber\\
    &\quad+\dfrac{1}{\Delta_t} \gamma(y, \rho(x,  t),e(x,  t))n(x,y,t). 
\end{align*}
This can be rewritten as (dropping the dependent variables for simplicity)
\begin{align}
    \dfrac{\partial n}{\partial t} 
    &= \dfrac{1}{\Delta_t }\left(\mu_{-}-\mu_{+}\right)n+\dfrac{\Delta_y}{\Delta_t}\dfrac{\partial}{\partial y}\left(\mu_{-}n\right)+\dfrac{\Delta_y^2}{2\Delta_t}\dfrac{\partial^2}{\partial y^2}\left(\mu_{-}n\right) +\dfrac{\Delta_x}{\Delta_t}\dfrac{\partial}{\partial x}\Big(\left(\beta_{-}-\beta_{+}\right)n\Big) \nonumber \\ &\qquad+\dfrac{\Delta_x^2}{2\Delta_t}\dfrac{\partial}{\partial x}\left( (\beta_{-}+\beta_{+}) \dfrac{\partial n}{\partial x}- n \dfrac{\partial }{\partial x}(\beta_{-}+\beta_{+}) \right)+\dfrac{1}{\Delta_t}\gamma n, \label{eq:SIhi}
\end{align}
at $y=Y_{\text{min}}.$
Recalling that
{\small{
\begin{align*}
    \lim_{\Delta_x,  \Delta_t\rightarrow 0} \dfrac{\Delta_x}{\Delta_t} \Big(\beta_{-}(y, \rho(x,t), e(x,t))-\beta_{+}(y, \rho(x,t), e(x,t))\Big) &= v^m(y, \rho(x,t), e(x,t)), \\
    \lim_{\Delta_x,  \Delta_t\rightarrow 0} \dfrac{\Delta_x^2}{2\Delta_t} \Big(\beta_{-}(y, \rho(x,t), e(x,t))+\beta_{+}(y, \rho(x,t), e(x,t))\Big) &= D^m(y, \rho(x,t), e(x,t)), \\
    \lim_{\Delta_y,  \Delta_t\rightarrow 0} \dfrac{\Delta_y}{\Delta_t} \Big(\mu_{-}\left(y, \rho(x,t), e(x,t)\right)-\mu_{+}\left(y, \rho(x,t), e(x,t)\right)\Big) &= v^p(y, \rho(x,t), e(x,t)), \\
    \lim_{\Delta_y,  \Delta_t\rightarrow 0} \dfrac{\Delta_y^2}{2\Delta_t} \Big(\mu_{-}\left(y, \rho(x,t), e(x,t)\right)+\mu_{+}\left(y, \rho(x,t), e(x,t)\right)\Big) &= D^p(y, \rho(x,t), e(x,t)), \\
    \lim_{\Delta_t\rightarrow 0} \dfrac{1}{\Delta_t} \gamma\left(y, \rho(x,t), e(x,t)\right) &= r(y, \rho(x,t), e(x,t)),
\end{align*}
}}
such that 
\begin{align*}
    \mu_{\pm}&=\dfrac{D^p\Delta_t}{\Delta_y^2}\mp\dfrac{v^p\Delta_t}{2\Delta_y},
\end{align*}
then {{Eq.~\eqref{eq:SIhi}}} can be rewritten as
\begin{align*}
    \dfrac{\partial n}{\partial t}&= \dfrac{1}{\Delta_y}v^pn + \dfrac{\partial}{\partial y}\left(\dfrac{1}{2}v^pn+\dfrac{1}{\Delta_y}D^pn\right)+\dfrac{1}{2}\dfrac{\partial^2}{\partial y^2}\left(D^pn\right)  +\dfrac{\partial }{\partial x}(v^mn) \nonumber \\ &\quad +\dfrac{\partial}{\partial x}\left( D^m \dfrac{\partial n}{\partial x}- n \dfrac{\partial }{\partial x}D^m\right)+rn .
\end{align*}
Therefore, in order to prevent blow-up of terms at $y=Y_{\text{min}}$, we require that 
\begin{equation*}
    v^pn +\dfrac{\partial }{\partial y}(D^pn)= 0  \qquad\qquad \text{at} \quad y=Y_{\text{min}}.
\end{equation*}

%%%%%%%%%%%%%%%%%%%%%%%%%
\paragraph{Boundary condition at $y=Y_{\text{max}}$.}
Returning to the Eq.~\eqref{SIeq:master}, we seek an equation for the evolution of the cell number on the upper most lattice site in phenotype space, corresponding to the site $y_{N_y+1}$. 
To find this, we multiply Eq.~\eqref{SIeq:master} by $n_{s}^{N_y+1}$ and sum over possible states ${\bf{n}}$:
{\small{
\begin{align}
    &\Delta_t \sum_{\bf{n}}\sum_{\bf{e}}n_{s}^{N_y+1} \dfrac{\partial}{\partial t} p({\bf{n}}, {\bf{e}},  t_h) = \nonumber \\ 
    &\quad\sum_{\bf{n}}\sum_{\bf{e}}n_{s}^{N_y+1} \sum_{i=1}^{N_x+1}\sum_{j=1}^{N_y}\mu_{-}(j+1, N_i, e_i)\left\{ (n_i^{j+1}+1)p(U_{i,j}^{\text{p}}{\bf{n}}, {\bf{e}},  t_h)-n_i^{j+1}p({\bf{n}}, {\bf{e}},  t_h)\right\} \nonumber \\ 
    &\quad + \sum_{\bf{n}}\sum_{\bf{e}}n_{s}^{N_y+1} \sum_{i=1}^{N_x+1}\sum_{j=2}^{N_y+1}\mu_{+}(j-1, N_i, e_i)\left\{ (n_i^{j-1}+1)p(D_{i,j}^{\text{p}}{\bf{n}}, {\bf{e}},  t_h)-n_i^{j-1}p({\bf{n}}, {\bf{e}},  t_h)\right\} \nonumber \\
    &\quad + \sum_{\bf{n}}\sum_{\bf{e}}n_{s}^{N_y+1}\sum_{i=1}^{N_x} \sum_{j=1}^{N_y+1}\beta_{-}(j, {N_i,e_i)}\left\{ (n_{i+1}^{j}+1)p(R_{i,j}^{\text{m}}{\bf{n}}, {\bf{e}},  t_h)-n_{i+1}^{j}p({\bf{n}}, {\bf{e}},  t_h)\right\} \nonumber \\ 
    &\quad + \sum_{\bf{n}}\sum_{\bf{e}}n_{s}^{N_y+1}\sum_{i=2}^{N_x+1} \sum_{j=1}^{N_y+1}\beta_{+}(j, N_i,e_i)\left\{ (n_{i-1}^{j}+1)p(L_{i,j}^{\text{m}}{\bf{n}}, {\bf{e}},  t_h)-n_{i-1}^{j}p({\bf{n}}, {\bf{e}},  t_h)\right\}\nonumber \\ 
    &\quad + \sum_{\bf{n}}\sum_{\bf{e}}n_{s}^{N_y+1}\sum_{i=1}^{N_x+1} \sum_{j=1}^{N_y+1}\gamma(j, N_i,e_i)\Big\{\gamma(j, N_i-1,e_i)(n_i^j-1)p(G_{i,j}{\bf{n}}, {\bf{e}},  t_h)\nonumber \\ &\qquad\qquad\qquad\qquad\qquad\qquad\qquad\qquad\qquad\qquad\qquad\qquad-\gamma(j, N_i,e_i)n_i^jp({\bf{n}}, {\bf{e}},  t_h)\Big\} .\label{SIeq:master_mult_yNp}
\end{align}
}}
Using the same methods as on the boundaries in physical space, we can change variables in each term to find an equation for evolution of cell number. 
Consider first the first term on the right-hand side. 
From previous analysis, we know that the only non-zero contributions come from when $j=N_y$ and $i=s$.
In the second term, the non-zero terms are $j=N_y+1$ and $i=s$. 
The third term gives contributions when $j=N_y+1$ and $i=s$ or $i=s-1$. 
In the fourth term, there are non-zero contributions when $j=N_y+1$ and $i=s$ or $i=s+1$.
The final term produces non-zero contributions only when $i=s$ and $j=N_y+1.$
Thus, Eq.~\eqref{SIeq:master_mult_yNp} can be written as
{\small{
\begin{align}
    &\Delta_t \sum_{\bf{n}}\sum_{\bf{e}}n_{s}^{N_y+1} \dfrac{\partial}{\partial t} p({\bf{n}}, {\bf{e}},  t_h) = \nonumber \\ 
    &\quad\sum_{\bf{n}}\sum_{\bf{e}}n_{s}^{N_y+1} \mu_{-}(N_y+1, N_{s}, e_s)\left\{ (n_s^{N_y+1}+1)p(U_{s, N_y}^{\text{p}}{\bf{n}}, {\bf{e}},  t_h)-n_s^{N_y+1}p({\bf{n}}, {\bf{e}},  t_h)\right\} \nonumber \\ 
    &\quad + \sum_{\bf{n}}\sum_{\bf{e}}n_{s}^{N_y+1} \mu_{+}(N_y, N_{s}, e_s)\left\{ (n_s^{N_y}+1)p(D_{s,N_y+1}^{\text{p}}{\bf{n}}, {\bf{e}},  t_h)-n_s^{N_y}p({\bf{n}}, {\bf{e}},  t_h)\right\} \nonumber \\
    &\quad + \sum_{\bf{n}}\sum_{\bf{e}}n_{s}^{N_y+1}\sum_{i=s,  s-1}\beta_{-}(N_y+1, {N_i,e_i)}\Big\{ (n_{i+1}^{N_y+1}+1)p(R_{i, N_y+1}^{\text{m}}{\bf{n}}, {\bf{e}},  t_h)\nonumber \\ &\qquad\qquad\qquad\qquad\qquad\qquad\qquad\qquad\qquad\qquad\qquad\qquad-n_{i+1}^{N_y+1}p({\bf{n}}, {\bf{e}},  t_h)\Big\} \nonumber \\ 
    &\quad + \sum_{\bf{n}}\sum_{\bf{e}}n_{s}^{N_y+1}\sum_{i=s, s+1}\beta_{+}(N_y+1, N_i,e_i)\Big\{ (n_{i-1}^{N_y+1}+1)p(L_{i,N_y+1}^{\text{m}}{\bf{n}}, {\bf{e}},  t_h)\nonumber \\ &\qquad\qquad\qquad\qquad\qquad\qquad\qquad\qquad\qquad\qquad\qquad\qquad-n_{i-1}^{N_y+1}p({\bf{n}}, {\bf{e}},  t_h)\Big\}\nonumber \\ 
    &\quad + \sum_{\bf{n}}\sum_{\bf{e}}n_{s}^{N_y+1}\Big\{\gamma(N_y+1, N_s-1,e_s)(n_s^{N_y+1}-1)p(G_{s,N_y+1}{\bf{n}}, {\bf{e}},  t_h)\nonumber \\ &\qquad\qquad\qquad\qquad\qquad\qquad\qquad\qquad-\gamma(N_y+1, N_s,e_s)n_s^{N_y+1}p({\bf{n}}, {\bf{e}},  t_h)\Big\}.\label{SIeq:master_mult_yNp2}
\end{align}
}}
The first term in Eq.~\eqref{SIeq:master_mult_yNp2} can be rewritten using the change of variables ($\bar{n}_s^{N_y+1}=n_s^{N_y+1}+1$ and $\bar{n}_s^{N_y}=n_s^{N_y}-1$) in the following way:
\begin{align*}
    &\sum_{\bf{n}}\sum_{\bf{e}}n_{s}^{N_y+1} \mu_{-}(N_y+1, N_{s}, e_s)\left\{ (n_s^{N_y+1}+1)p(U_{s, N_y}^{\text{p}}{\bf{n}}, {\bf{e}},  t_h)-n_s^{N_y+1}p({\bf{n}}, {\bf{e}},  t_h)\right\} \nonumber \\ 
    &\quad= \sum_{\bf{n}}\sum_{\bf{e}}n_{s}^{N_y+1} \mu_{-}(N_y+1, N_{s}, e_s)\times \nonumber \\ 
    &\qquad\qquad\left\{ (n_s^{N_y+1}+1)p([n_s^1, \dots, n_{s}^{N_y}-1, n_{s}^{N_y+1}-1], {\bf{e}},  t_h)-n_s^{N_y+1}p({\bf{n}}, {\bf{e}},  t_h)\right\} \nonumber \\ 
    &\quad=  \sum_{\bf{n}}\sum_{\bf{e}}\mu_{-}(N_y+1, N_{s}, e_s)\times \nonumber \\ 
    &\qquad\qquad\left\{ \bar{n}_s^{N_y+1}(\bar{n}_{s}^{N_y+1} -1)p([n_s^1, \dots, \bar{n}_{s}^{N_y}, \bar{n}_{s}^{N_y+1}], {\bf{e}},  t_h)-n_s^{N_y+1}n_{s}^{N_y+1} p({\bf{n}}, {\bf{e}},  t_h)\right\} \nonumber \\ 
    &\quad=  \sum_{\bf{n}}\sum_{\bf{e}}\mu_{-}(N_y+1, N_{s}, e_s)\left\{ {n}_s^{N_y+1}({n}_{s}^{N_y+1} -1)p({\bf{n}}, {\bf{e}},  t_h)-n_s^{N_y+1}n_{s}^{N_y+1} p({\bf{n}}, {\bf{e}},  t_h)\right\} \nonumber \\ 
    &\quad= - \sum_{\bf{n}}\sum_{\bf{e}}\mu_{-}(N_y+1, N_{s}, e_s) {n}_s^{N_y+1}p({\bf{n}}, {\bf{e}},  t_h) \nonumber \\ 
    &\quad= - \langle\mu_{-}(N_y+1, N_{s}, e_s) {n}_s^{N_y+1}\rangle.
\end{align*}
The second term in Eq.~\eqref{SIeq:master_mult_yNp2} can then be rewritten as the following (using $\bar{n}_s^{N_y+1}=n_s^{N_y+1}-1$ and $\bar{n}_s^{N_y}=n_s^{N_y}+1$):
\begin{align*}
    &\sum_{\bf{n}}\sum_{\bf{e}}n_{s}^{{{N_y+1}}} \mu_{+}({N_y}, N_{s}, e_s)\left\{ (n_s^{{N_y}}+1)p(D_{s,N_y+1}^{\text{p}}{\bf{n}}, {\bf{e}},  t_h)-n_s^{{N_y}}p({\bf{n}}, {\bf{e}},  t_h)\right\} \nonumber \\ 
    &\quad= \sum_{\bf{n}}\sum_{\bf{e}}n_{s}^{{N_y+1}} \mu_{+}({N_y}, N_{s}, e_s)\times \nonumber \\ 
    &\qquad\qquad\left\{ (n_s^{{N_y}}+1)p([n_s^1, \dots, n_{s}^{{N_y}}+1, n_{s}^{{N_y+1}}-1], {\bf{e}},  t_h)-n_s^{{N_y}}p({\bf{n}}, {\bf{e}},  t_h)\right\} \nonumber \\ 
    &\quad=  \sum_{\bf{n}}\sum_{\bf{e}}\mu_{+}({N_y}, N_{s}, e_s)\times \nonumber \\ 
    &\qquad\qquad\left\{ \bar{n}_s^{{N_y}}(\bar{n}_{s}^{N_y+1}+1 )p([n_s^1, \dots, \bar{n}_{s}^{{N_y}}, \bar{n}_{s}^{{N_y+1}}], {\bf{e}},  t_h)-n_s^{{N_y}}n_{s}^{{N_y+1}} p({\bf{n}}, {\bf{e}},  t_h)\right\} \nonumber \\ 
    &\quad=  \sum_{\bf{n}}\sum_{\bf{e}}\mu_{+}({N_y}, N_{s}, e_s)\times \nonumber \\ 
    &\qquad\qquad\left\{ {n}_s^{{N_y}}({n}_{s}^{{N_y+1}} +1)p({\bf{n}}, {\bf{e}},  t_h)-n_s^{{N_y}}n_{s}^{{N_y+1}} p({\bf{n}}, {\bf{e}},  t_h)\right\} \nonumber \\ 
    &\quad=  \sum_{\bf{n}}\sum_{\bf{e}}\mu_{+}({N_y}, N_{s}, e_s) {n}_s^{{N_y}}p({\bf{n}}, {\bf{e}},  t_h) \nonumber \\ 
    &\quad=  \langle\mu_{+}({N_y}, N_{s}, e_s) {n}_s^{{N_y}}\rangle.
\end{align*}
The contributions from the terms describing movement in physical space are the same as in the main body Eq.~\eqref{SIeq:full_IBM_prolif_ECM} where $j=N_y.$
This is also true for the growth terms with $i=s$ and $j=N_y.$
Putting these together, we get 
\begin{align}
    \dfrac{\partial}{\partial t}\langle n_s^{N_y+1}\rangle &=\dfrac{1}{\Delta_t}\langle\mu_{+}({N_y}, N_{s}, e_s) {n}_s^{{N_y}}\rangle- \dfrac{1}{\Delta_t}\langle\mu_{-}(N_y+1, N_{s}, e_s) {n}_s^{N_y+1}\rangle \nonumber \\ &\quad+ \dfrac{1}{\Delta_t}\langle\beta_{+}({N_y+1}, N_{s},e_s)n_{s-1}^{N_y+1}\rangle +\dfrac{1}{\Delta_t} \langle\beta_{-}({N_y+1}, N_{s},e_s)n_{s+1}^{N_y+1}\rangle \nonumber \\ 
    &\quad-\dfrac{1}{\Delta_t} \langle\beta_{-}({N_y+1}, N_{s-1},e_{s-1})n_{s}^{N_y+1}\rangle -\dfrac{1}{\Delta_t}\langle\beta_{+}({N_y+1}, N_{s+1},e_{s+1})n_{s}^{N_y+1}\rangle \nonumber \\ &\quad+\dfrac{1}{\Delta_t}  \langle\gamma({N_y+1}, N_{s},e_{s})n_{s}^{N_y+1}\rangle \label{SIeq:BC_yNy}. 
\end{align}
We can take the limit $\Delta_x,  \Delta_y, \Delta_t \rightarrow 0$, such that Eq.~\eqref{SIeq:BC_yNy} can be rewritten as the following at $y=Y_{\text{max}}$:
\begin{align*}
    \dfrac{\partial}{\partial t} n(x, y, t) &= \dfrac{1}{\Delta_t}\beta_{+}(y, \rho(x,  t),e(x,  t))n(x-\Delta_x,  y,  t)\nonumber \\ &\quad +\dfrac{1}{\Delta_t} \beta_{-}(y, \rho(x,  t),e(x,  t))n (x+\Delta_x,  y,  t) \nonumber \\ 
    &\quad-\dfrac{1}{\Delta_t} \beta_{-}(y, \rho(x-\Delta_x,  t),e(x-\Delta_x,  t))n(x,y,t) \nonumber \\ &\quad-\dfrac{1}{\Delta_t}\beta_{+}(y, \rho(x+\Delta_x,  t),e(x+\Delta_x,  t))n(x,y,t) \nonumber \\ 
    &\quad+\dfrac{1}{\Delta_t}\mu_{+}(y-\Delta_y, \rho(x,  t),e(x,  t))n(x,  y-\Delta_y,  t)\nonumber \\ &\quad -\dfrac{1}{\Delta_t}\mu_{-}(y, \rho(x, t),e(x,  t))n(x,y,t)\nonumber\\
    &\quad+\dfrac{1}{\Delta_t} \gamma(y, \rho(x,  t),e(x,  t))n(x,y,t). 
\end{align*}
Using Taylor series expansions, we find 
\begin{align*}
    \dfrac{\partial}{\partial t} n(x, y, t) &= \dfrac{1}{\Delta_t}\beta_{+}(y, \rho(x,  t),e(x,  t))\times\nonumber\\ &\qquad\qquad\qquad\left[n(x,y,t)-\Delta_x\dfrac{\partial}{\partial x}n(x,y,t)+\dfrac{\Delta_x^2}{2}\dfrac{\partial^2}{\partial x^2}n(x,y,t)\right]\nonumber \\ &\quad +\dfrac{1}{\Delta_t} \beta_{-}(y, \rho(x,  t),e(x,  t))\times\nonumber\\ &\qquad\qquad\qquad\left[n(x,y,t)+\Delta_x\dfrac{\partial}{\partial x}n(x,y,t)+\dfrac{\Delta_x^2}{2}\dfrac{\partial^2}{\partial x^2}n(x,y,t)\right]\nonumber \\ 
    &\quad-\dfrac{1}{\Delta_t}n(x,y,t) \times\nonumber\\ &\qquad\qquad\qquad\Bigg[\beta_{-}(y, \rho(x, t),e(x, t))- \Delta_x\dfrac{\partial}{\partial x}\beta_{-}(y, \rho(x,t), e(x,t)\nonumber \\ &\qquad\qquad\qquad\qquad\qquad\qquad\qquad+\dfrac{\Delta_x^2}{2}\dfrac{\partial^2}{\partial x^2}\beta_{-}(y, \rho(x,t), e(x,t)\Bigg]\nonumber \\ &\quad-\dfrac{1}{\Delta_t}n(x,y,t)\times\nonumber\\ &\qquad\qquad\qquad\Bigg[\beta_{+}(y, \rho(x, t),e(x, t))+ \Delta_x\dfrac{\partial}{\partial x}\beta_{+}(y, \rho(x,t), e(x,t)\nonumber \\ &\qquad\qquad\qquad\qquad\qquad\qquad\qquad+\dfrac{\Delta_x^2}{2}\dfrac{\partial^2}{\partial x^2}\beta_{+}(y, \rho(x,t), e(x,t)\Bigg]\nonumber \\ 
    &\quad+\dfrac{1}{\Delta_t}\left[n(x,y,t)-\Delta_y\dfrac{\partial}{\partial y}n(x,y,t)+\dfrac{\Delta_y^2}{2}\dfrac{\partial^2}{\partial y^2}n(x,y,t)\right]\times\nonumber \\ &\qquad\qquad\qquad\Bigg[\mu_{+}(y, \rho(x, t),e(x, t))- \Delta_y\dfrac{\partial}{\partial y}\mu_{+}(y, \rho(x,t), e(x,t)\nonumber \\ &\qquad\qquad\qquad\qquad\qquad\qquad\qquad+\dfrac{\Delta_y^2}{2}\dfrac{\partial^2}{\partial y^2}\mu_{+}(y, \rho(x,t), e(x,t)\Bigg]\nonumber \\ &\quad -\dfrac{1}{\Delta_t}\mu_{-}(y, \rho(x, t),e(x,  t))n(x,y,t)\nonumber\\
    &\quad+\dfrac{1}{\Delta_t} \gamma(y, \rho(x,  t),e(x,  t))n(x,y,t). 
\end{align*}
Now, dropping the dependent variables for simplicity, we find
\begin{align*}
    \dfrac{\partial n}{\partial t} 
    &= -\dfrac{1}{\Delta_t}(\mu_{-}-\mu_{+})n-\dfrac{\Delta_y}{\Delta_t}\dfrac{\partial}{\partial y}\left(\mu_{+}n\right)+\dfrac{\Delta_y^2}{2\Delta_t}\dfrac{\partial^2}{\partial y^2}\left(\mu_{+}n\right)  +\dfrac{\Delta_x}{\Delta_t}\dfrac{\partial}{\partial x}\Big(\left(\beta_{-}-\beta_{+}\right)n\Big) \nonumber \\ &\qquad+\dfrac{\Delta_x^2}{2\Delta_t}\dfrac{\partial}{\partial x}\left( (\beta_{-}+\beta_{+}) \dfrac{\partial n}{\partial x}- n \dfrac{\partial }{\partial x}(\beta_{-}+\beta_{+}) \right)+\dfrac{1}{\Delta_t}\gamma n.
\end{align*}
at $y=Y_{\text{max}}.$
Recalling that
{\small{
\begin{align*}
    \lim_{\Delta_x,  \Delta_t\rightarrow 0} \dfrac{\Delta_x}{\Delta_t} \Big(\beta_{-}(y, \rho(x,t), e(x,t))-\beta_{+}(y, \rho(x,t), e(x,t))\Big) &= v^m(y, \rho(x,t), e(x,t)), \\
    \lim_{\Delta_x,  \Delta_t\rightarrow 0} \dfrac{\Delta_x^2}{2\Delta_t} \Big(\beta_{-}(y, \rho(x,t), e(x,t))+\beta_{+}(y, \rho(x,t), e(x,t))\Big) &= D^m(y, \rho(x,t), e(x,t)), \\
    \lim_{\Delta_y,  \Delta_t\rightarrow 0} \dfrac{\Delta_y}{\Delta_t} \Big(\mu_{-}\left(y, \rho(x,t), e(x,t)\right)-\mu_{+}\left(y, \rho(x,t), e(x,t)\right)\Big) &= v^p(y, \rho(x,t), e(x,t)), \\
    \lim_{\Delta_y,  \Delta_t\rightarrow 0} \dfrac{\Delta_y^2}{2\Delta_t} \Big(\mu_{-}\left(y, \rho(x,t), e(x,t)\right)+\mu_{+}\left(y, \rho(x,t), e(x,t)\right)\Big) &= D^p(y, \rho(x,t), e(x,t)), \\
    \lim_{\Delta_t\rightarrow 0} \dfrac{1}{\Delta_t} \gamma\left(y, \rho(x,t), e(x,t)\right) &= r(y, \rho(x,t), e(x,t)),
\end{align*}
}}
such that 
\begin{align*}
    \mu_{\pm}&=\dfrac{D^p\Delta_t}{\Delta_y^2}\mp\dfrac{v^p\Delta_t}{2\Delta_y},
\end{align*}
then the equation can be rewritten as
\begin{align*}
    \dfrac{\partial n}{\partial t}&= -\dfrac{1}{\Delta_y}v^pn + \dfrac{\partial}{\partial y}\left(-\dfrac{1}{2}v^pn+\dfrac{1}{\Delta_y}D^pn\right)+\dfrac{1}{2}\dfrac{\partial^2}{\partial y^2}\left(D^pn\right)  +\dfrac{\partial }{\partial x}(v^mn) \nonumber \\ &\quad +\dfrac{\partial}{\partial x}\left( D^m \dfrac{\partial n}{\partial x}- n \dfrac{\partial }{\partial x}D^m\right)+rn .
\end{align*}
Therefore, in order to prevent blow-up of terms at $y=Y_{\text{max}}$, we require that 
\begin{equation*}
    -v^pn +\dfrac{\partial }{\partial y}(D^pn)= 0  \qquad\qquad \text{at} \quad y=Y_{\text{max}}.
\end{equation*}

\subsection{Equation for the density of the local environment}
Following the assumptions outlined in Sec.~\ref{sec:IBM} and methodology above, we can {{review the}} master equation~\eqref{SIeq:master} describing the evolution of the number of elements of the local environment at position $x_i$, denoted $e_i$, {{and 
multiply}} by $e_s$ and sum over all possible states ${\bf{e}}$ and  ${\bf{n}}$ to get 
{\small{
\begin{align*}
    &\sum_{\bf{e}}\sum_{\bf{n}} e_s \Delta_t \dfrac{\partial }{\partial t} p({\bf{n}}, {\bf{e}},  t_h) \nonumber \\ &= \sum_{\bf{e}}\sum_{\bf{n}} e_s \sum_{i=1}^{N_x+1} \sum_{j=1}^{N_y+1}\lambda(j, n_i^j)\left\{(e_i+1)p({\bf{n}},  H_i{\bf{e}},  t_h) - e_i p({\bf{n}},  {\bf{e}},  t_h) \right\} \nonumber \\ 
    &= \sum_{\bf{e}}\sum_{\bf{n}} e_s \sum_{i=1}^{N_x+1} \sum_{j=1}^{N_y+1} \lambda(j, n_i^j)\big\{(e_i+1)p({\bf{n}},  
    [e_1, \dots, e_i +1, \dots, e_{N_x+1}],  t_h)  \nonumber \\ 
    & \qquad\qquad\qquad\qquad\qquad\qquad\qquad\qquad\qquad\qquad\qquad\qquad\qquad\qquad\qquad- e_i p({\bf{n}},  {\bf{e}},  t_h) \big\}, 
\end{align*}
}}
{{recalling that contributions from the terms describing cell dynamics only sum to zero.}}
Now we consider two cases: $i=s$ and $i\neq s$. 
First, consider $i \neq s$ and use the change of variables $\bar{e}_i = e_i+1$ in the second term, and then drop the bar:
{\small{
\begin{align*}
     &\sum_{\bf{e}} \sum_{\bf{n}} e_s \sum_{\substack{i=1, \\ i\neq s}}^{N_x+1} \sum_{j=1}^{N_y+1}\lambda(j, n_i^j)\left\{(e_i+1)p({\bf{n}},  
    [e_1, \dots, e_i +1, \dots, e_{N_x+1}],  t_h)  - e_i p({\bf{n}},  {\bf{e}},  t_h)\right\} \nonumber \\
    &=\sum_{\bf{e}}\sum_{\bf{n}} e_s \sum_{\substack{i=1, \\ i\neq s}}^{N_x+1} \sum_{j=1}^{N_y+1}\lambda(j, n_i^j)\left\{\bar{e}_ip({\bf{n}},  
    [e_1, \dots, \bar{e}_i, \dots, e_{N_x+1}],  t_h) - e_i p({\bf{n}},  {\bf{e}},  t_h) \right\} \nonumber \\
    &=\sum_{\bf{e}} \sum_{\bf{n}}e_s \sum_{\substack{i=1, \\ i\neq s}}^{N_x+1} \sum_{j=1}^{N_y+1} \lambda(j, n_i^j)\left\{ e_i p({\bf{n}},  {\bf{e}},  t_h)-{e}_ip({\bf{n}},  {\bf{e}},  t_h)\right\} \nonumber \\
    &= 0. 
\end{align*}
}}
Now consider the case when $i=s$ and use the change of variables $\bar{e}_s=e_s+1$:
{\small{
\begin{align*}
     &\sum_{\bf{e}}\sum_{\bf{n}} \sum_{j=1}^{N_y+1}e_s \lambda(j, n_s^j)\left\{  (e_s+1)p({\bf{n}}, 
    [e_1, \dots, e_s +1, \dots, e_{N_x+1}],  t_h) - e_s p({\bf{n}},  {\bf{e}},  t_h)\right\} \nonumber \\
    &=\sum_{\bf{e}} \sum_{\bf{n}}\sum_{j=1}^{N_y+1}\lambda(j, n_s^j)\left\{ \bar{e}_s(\bar{e}_s-1)p({\bf{n}},  
    [e_1, \dots, \bar{e}_s, \dots, e_{N_x+1}], {\bf{e}},  t_h) - e_s^2 p({\bf{n}},  {\bf{e}},  t_h) \right\} \nonumber \\
    &=\sum_{\bf{e}}\sum_{\bf{n}} \sum_{j=1}^{N_y+1} e_s \lambda(j, n_s^j)\left\{ (e_s-1) p({\bf{n}},  {\bf{e}},  t_h) - {e}_s p({\bf{n}},  {\bf{e}},  t_h)  \right\} \nonumber \\
    &= -\sum_{\bf{e}}\sum_{\bf{n}}\sum_{j=1}^{N_y+1} \lambda(j,  n_s^j) e_s p({\bf{n}},  {\bf{e}},  t_h) \nonumber \\ 
    &= -\sum_{j=1}^{N_y+1}\langle \lambda(j,  n_s^j) e_s \rangle.
\end{align*}
}}
Putting this together, we get 
\begin{align*}
     \Delta_t \dfrac{\partial}{\partial t} \langle e_s\rangle&=- \sum_{j=1}^{N_y+1}\langle\lambda (j, n_s^j)e_s\rangle.
\end{align*}
Defining 
\begin{equation*}
    \lim_{\Delta_t\rightarrow0}\dfrac{1}{\Delta_t} \lambda(y, n(x,y,t), e(x,t))= \nu(y, n(x, y, t), e(x,t)),
\end{equation*}
which we can substitute into {{the}} equation, rearrange and take limits as $\Delta_x, \Delta_y, \Delta_t \rightarrow 0$, to find that the differential equation for the density of the local environment, $e(x,t)$, is given by
\begin{equation*}
       \dfrac{\partial }{\partial t} e(x,t) = - \int_{y=Y_{\text{min}}}^{y=Y_{\text{max}}} \nu(y, n(x,  y,  t)) e(x,  t) \mathrm{d}y.
\end{equation*}
No boundary conditions are required for this equation.

%%=============================================%%
%% For submissions to Nature Portfolio Journals %%
%% please use the heading ``Extended Data''.   %%
%%=============================================%%

\newpage
{{
\section{Individual-based model functions}\label{app:IBMfcts}
\subsection{Phenotypic structuring during range expansion}\label{app:PSfct}
As per the rules described in Sec.~\ref{sec:IBM}, we can write functions to describe the movement and growth of cells over time. 
In particular, noting that in this case we have homogeneous cells, with constant random movement in all directions, then we can write 
$$\beta_{\pm}(j, N_{i\pm1}, e_{i\pm1})=k=\text{constant}.$$
Furthermore, when considering KPP type invasion, we know that cells grow faster in areas with higher amounts of available space, which {{can be modelled as}} 
\begin{equation*}
    \gamma_K(j, N_i) = 1-\dfrac{N_i}{\kappa},
\end{equation*}
whereas, with the addition of the Allee effect, we instead have
\begin{equation*}
    \gamma_A(j, N_i) = \left(1-\dfrac{N_i}{\kappa}\right)(N_i-p^{*}), 
\end{equation*} 
with $p^{*}\in(0, 1/2)$ and $\kappa>0$ describing the maximum total number of cells that can fit in any single site.
Implementing these functional forms during the coarse-graining process described in Sec.~\ref{app:deriv}, absorbing constants in the continuum limit and rescaling as appropriate, the resulting continuum equation is given by Eq.~\eqref{eq:struct} with functions~\eqref{eq:KPP}~and~\eqref{eq:Allee}.

\subsection{A go-or-grow model of cells invading the extracellular matrix (ECM)}\label{app:GGfct}
When describing cells moving into the extracellular matrix (ECM), we know that volume filling constraints will {{a}}ffect the movement in physical space. 
In fact, as space decreases, cells {{have less space in which to move}}. 
Furthermore, we implement a continuum of cell phenotypes in this case, such that cells in phenotypic state $j=Y_{\text{max}}$ are the most proliferative, but least motile and degrading cells. 
As such, the individual-based function{{s}} describing movement in physical space can be written as 
\begin{equation*}
    \beta_{\pm}(j, N_{i\pm1}, e_{i\pm1}) = (1-j)\left(1-\dfrac{N_{i\pm1}+e_{i\pm1}}{\kappa}\right), %\dfrac{1}{2}
\end{equation*}
where $\kappa>0$ is the total number of available sites for cells and ECM elements, known as the carrying capacity.

Cells {{are able to proliferate more rapidly when there is}} a larger amount of available space, and when they occupy a phenotypic state with higher values.
As such, we have that
\begin{equation*}
    \gamma(j, N_{i}, e_{i}) = j\left(1-\dfrac{N_{i}+e_{i}}{\kappa}\right).
\end{equation*}
Alternatively, it is cells in a lower phenotypic state, $j$, that degrade the surrounding ECM at a higher rate. 
The corresponding function to describe this is given by
\begin{equation*}
    \lambda(j,  n_i^j) = (1-j)n_i^j.
\end{equation*}

In Sec.~\ref{sec:gg}, we consider a number of different functions to describe movement in phenotypic space.
The first phenotypic drift term we consider is cell-dependent drift, where cells transition into a phenotypic state with lower values at an increasing rate in regions with more cells present. 
The second phenotypic drift term we evaluate considers the role of the ECM in determining phenotypic transitions. 
In this case, cells transition to phenotypic states with lower values as the number of ECM elements in the same physical site increases.
Finally we consider space-dependent phenotype transitions such that cells move into phenotypic sites with higher values at an increasing rate as the available space in the same physical site increases. 
These options are all described in Table~\ref{tab:PD_IBM}.
\begin{table}[htbp]
\begin{center}
{\renewcommand{\arraystretch}{2.2} % Adjust for spacing without making rows massive
\begin{tabular}{|>{\centering\arraybackslash}m{2.8cm}|
                >{\centering\arraybackslash}m{2.8cm}|
                >{\centering\arraybackslash}m{2.8cm}|}
\hline
Phenotypic drift & $\mu_{-}(j, N_i, e_i)$ & $\mu_{+}(j, N_i, e_i)$       \\
\hline
\hline
Cell-dependent           & $ \dfrac{N_i}{\kappa}$           & $1-\dfrac{N_i}{\kappa}$                                \\
\hline
ECM-dependent            & $ \dfrac{e_i}{\kappa}$              & $1-\dfrac{e_i}{\kappa}$                               \\
\hline
Space-dependent          & $ \dfrac{N_i+e_i}{\kappa}$         & $ 1-\dfrac{N_i+e_i}{\kappa}$     \\
\hline
\end{tabular}
}
\end{center}
\caption{Table listing the individual-based functions used during coarse-graining, that correspond to those continuum equivalents described in Table~\ref{tab:PD}. The functions shown describe the probabilities of transitions up and down the phenotype space, $\mu_{+}(j, N_i, e_i)$ and $\mu_{-}(j, N_i, e_i)$, respectively.}
\label{tab:PD_IBM}
\end{table}

Implementing the individual-based functions described above during the coarse-graining process, absorbing parameters and rescaling gives the resulting continuum equations and functions as stated in Sec.~\ref{sec:gg} of the main text.

\subsection{T cell exhaustion}\label{app:Tfct}
In the case where we consider T cell exhaustion, we are simulating both T cells and tumour cells. 
Both the T cells and tumour cells can move and grow, whilst T cells have a further variable, exhaustion, attached to them, and they are able to kill off the tumour cells. 

The movement of both the tumour cells and T cells is subject to volume exclusion, but also depends on the exhaustion level of the cells for the T cells. 
As such, we can write that the individual-based function{{s}} describing the movement of the T cells is given by
$$\beta_{\pm}(j, N_{i\pm1}, e_{i\pm1}) = j \bigg(1-\dfrac{N_{i\pm1}+e_{i\pm1}}{\kappa}\bigg),$$
where $\kappa>0$ is the carrying capacity of each physical site, $i$.
Alternatively, the {{movement of the}} tumour cells does not depend on the phenotype of the cells, and can be given as 
$$r_{\pm}(N_{i\pm1}, e_{i\pm1})=\left(1-\dfrac{N_{i\pm1}+e_{i\pm1}}{\kappa}\right).$$

T cells are able to {{divide}} and produce a daughter cell in the same phenotypic and physical site at a rate described by $\gamma_1\geq0$.
However the rate of reproduction depends on available space and is also greater for less exhausted cells, in a phenotypic state with higher values.
T cells can also die at a rate $\gamma_0\geq0$ which increases as they exhaust. 
As such, the function describing the net growth of the T cells at an individual-level is given by 
$$\gamma(j, N_i, e_i) = \gamma_1 j\left(1-\dfrac{N_i+e_i}{\kappa}\right)-\gamma_0(1-j).$$
Concurrently, the probability of tumour cell growth increases in more available space. As such, we write
$$b(N_i, e_i)=1-\dfrac{N_i+e_i}{\kappa}.$$

T cells exhaust as a result of interactions with (being in the same site as) tumour cells. They also naturally exhaust. 
Both of these occur faster when a cell is less exhausted (in a higher phenotypic state). 
To implement this, we write that the probability of cells moving up or down in phenotype space can be written as 
\begin{align*}
    \mu_{\pm}(j, e_i)=\dfrac{1}{2}\left(1\pm jk_1\pm jk_2e_i\right),
\end{align*}
% which{{, for $\hat{v^p}(j, N_i, e_i)=\mu_{-}(j, N_i, e_i)-\mu_{+}(j, N_i, e_i)$, and $\hat{D^p}(j, N_i, e_i)=\mu_{-}(j, N_i, e_i)+\mu_{+}(j, N_i, e_i)$,}} results in
% $${{\hat{v^p}}}(j, e_i) = j(k_1+k_2e_i),$$ and $${{\hat{D^p}}}(y, \rho(x,t), C(x,t))=1,$$ 
where $k_1, k_2 \geq0$ describe the exhaustion rate of the T cells as a result of movement and growth, and as a result of interactions with the tumour cells, respectively.

Finally, tumour cells die (and are removed from their site) as a result of interactions with T cells in the same physical site. 
The individual-based description of this is given by
$$\lambda(j, n_i^j) = j n_i^j.$$ 

Using these functions in a coarse-graining process similar to that in Sec.~\ref{app:deriv} and rescaling as appropriate, we find that the resulting system of equations is given by Eqs.~\eqref{eq:Tcell-T}~and~\eqref{eq:Tcell-C}, with continuum functions as described in Sec.~\ref{sec:Tcell}.
}}

\newpage
{{
\section{Numerical methods}\label{sec:SI-NM}
The deterministic, continuum counterpart of the individual-based model described in Sec.~\ref{sec:IBM} is given by the PDEs in Eqs.~\eqref{eq:cont-n}~and~\eqref{eq:cont-e}, with boundary conditions given in Eqs.~\eqref{eq:BC1}-\eqref{eq:BC4} and initial conditions given in Eqs.~\eqref{eq:ICn}~and~\eqref{eq:ICe}.

To solve this {{system}} numerically, we use an {{advection-diffusion-reaction}} (A-DR) scheme that discretises the spatial variable $x$ using a central finite difference stencil.
In the phenotypic axis, $y$, we use a finite volume scheme, which divides the axis into $N_y+1$ sites of equal width, controlled using the Koren limiter \citep{koren1993robust}.
The discretised equations which are solved numerically to produce the simulations take the following form: 
\begin{align*}
\dfrac{\mathrm{d} n_i^j}{\mathrm{d}t} &=
\dfrac{((D^m)_{i+1}^j + (D^m)_i^j)(n_{i+1}^j - n_i^j) - ((D^m)_i^j + (D^m)_{i-1}^j)(n_i^j - n_{i-1}^j)}{2\Delta x^2}
\nonumber \\
&\qquad+ \dfrac{n_i^{j+1} - 2n_i^j + n_i^{j-1}}{\Delta y^2} (D^p)_i^j
+ \mathcal{A}_{i,j}
+ n_i^j r(\bar{y_j}, N_i, e_i)
\end{align*}
where
\begin{align*}
N_i &= \sum_j n_i^j \Delta y, \\
(D^m)_i^j &= D^m(\bar{y_j}, N_i, e_i), \\
(D^p)_i^j &= D^p(\bar{y_j}, N_i, e_i), \\
\bar{y}_j &= \text{mean of } y_{j} \text{ and } y_{j+1}, \\
\mathcal{A}_{i,j} &= \text{flux-limited advection in } y.
\end{align*}
The advection term in the \( y \)-direction is discretised using a slope-limited upwind scheme and can be written as:
\[
\left( \frac{\partial}{\partial y}(v^p n) \right)_i^j \approx \frac{F_i^{j+1/2} - F_i^{j-1/2}}{\Delta y},
\]
where \( F_i^{j+1/2} \) is the numerical flux across the interface between phenotypic points \( y_j \) and \( y_{j+1} \), given by:
\[
F_i^{j+1/2} =
\begin{cases}
(v^p)_i^{j+1/2} n_{i}^{j,+} & \text{if } (v^p)_i^{j+1/2} > 0, \\
(v^p)_i^{j+1/2} n_{i}^{j+1,-} & \text{if } (v^p)_i^{j+1/2} \leq 0,
\end{cases}
\]
which is calculated using
\begin{align*}
R_i^j &= \frac{n_i^{j+1} - n_i^j}{n_i^j - n_i^{j-1}}, \\
n_i^{j,+} &= n_i^j + \frac{1}{2} \phi(R_i^j) (n_i^{j+1} - n_i^j), \\
n_i^{j+1,-} &= n_i^{j+1} - \frac{1}{2} \phi(1/R_i^j) (n_i^{j+1} - n_i^j).
\end{align*}

Alongside this, to simulate the evolution of the density of the local environment, as described by Eq.~\eqref{eq:cont-e}, we use the finite difference scheme, with a summation to approximate the integral, which can be written as
\begin{equation*}
\frac{d e_i}{dt} = 
- e_i \cdot \sum_j \nu(\bar{y_j}, n_i^j) \Delta_y.
\end{equation*}
This resulting system of ordinary differential equations are then integrated in time using python's in-built ordinary differential equation solver {\tt{scipy.integrate.solve\_ivp}} with the explicit Runge-Kutta integration method of order 5 and time step $\Delta_t=0.1$. 
The phenotype step is $\Delta_y=0.02$ and the spatial step is $\Delta_x=0.1$, both of which were chosen to be sufficiently small to ensure that we observed convergence in the solutions.   
}}

\newpage
\section{Supplementary figures}
\begin{figure}[htbp]
    \centering
    \includegraphics[width=0.925\linewidth]{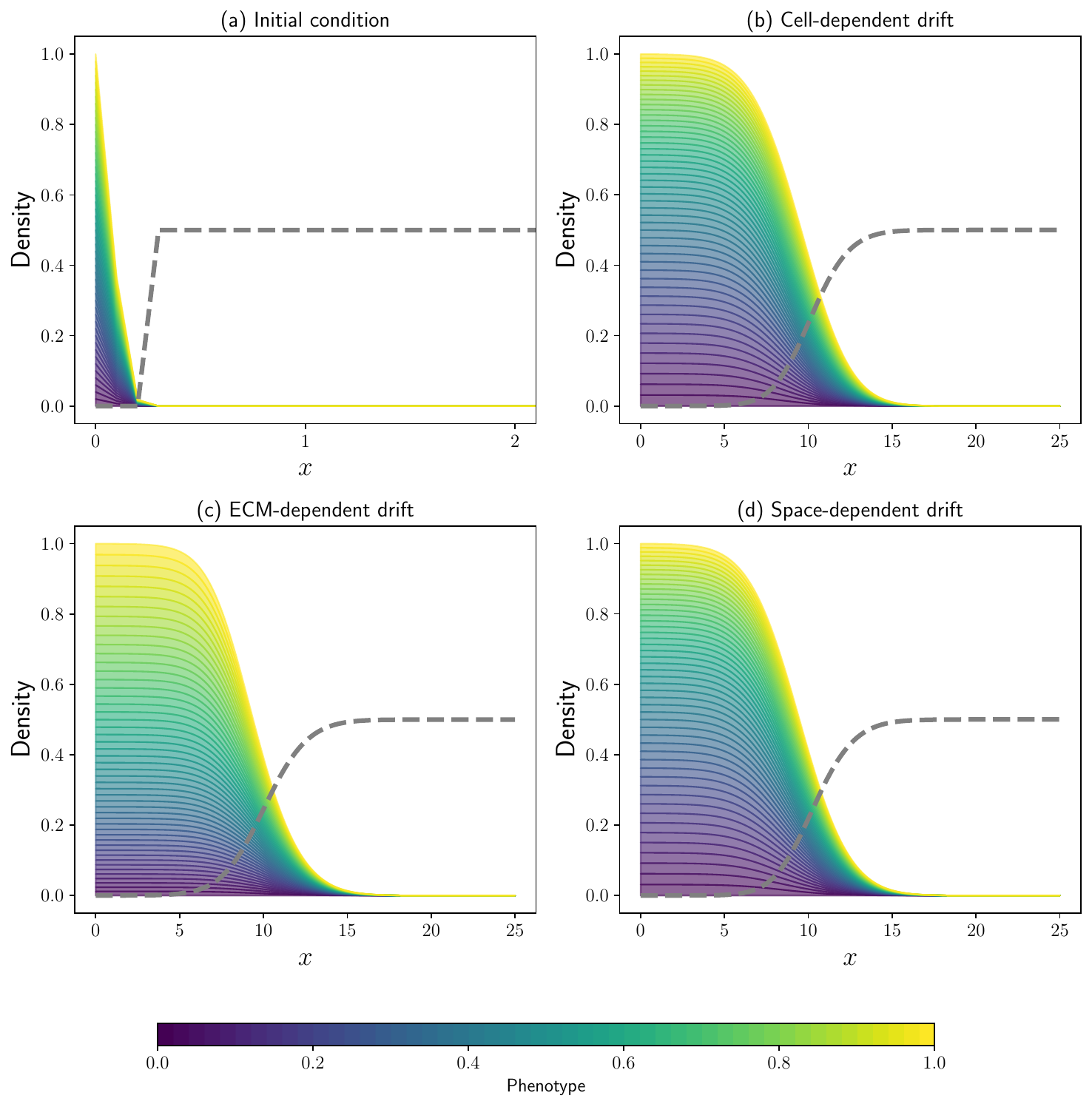}
    \caption{Evolution of the phenotypic structure of cells in Eqs.~\eqref{eq:cont-n}--\eqref{eq:cont-e} subject to various phenotypic drift terms, with the corresponding ECM density shown as a dashed grey line. (a) The initial distribution of the ECM and the cells with different phenotypes. (b) The spatial structure of the invading wave subject to cell-dependent phenotypic drift. (c)  The spatial structure of the invading wave subject to ECM-dependent phenotypic drift. (d) The spatial structure of the invading wave subject to space-dependent phenotypic drift. Results in (b), (c) and (d) are all plotted at time 30 wand simulations are carried out with $\kappa=1$. See Table~\ref{tab:PD} for explicit forms of the phenotypic drift terms.}
    \label{fig:x}
\end{figure}

\bibliographystyle{plain}
\bibliography{references}

\end{document}